\documentclass[11pt,a4paper]{article}

\usepackage[toc,page]{appendix}
\usepackage{graphicx}
\usepackage{dcolumn}
\usepackage{bm}

\usepackage{amsmath}
\usepackage{a4wide}
\usepackage{amssymb}
\usepackage{amsfonts}
\usepackage{ifpdf}
\usepackage{setspace}
\usepackage{color}
\usepackage{amsthm}
\usepackage{slashed}
\usepackage{afterpage}
\usepackage{epsfig}
\usepackage{epstopdf}
\usepackage{tikz}
\usepackage{graphicx}
\usepackage{epstopdf}
\usepackage{jheppub}

\DeclareMathOperator{\Tr}{Tr}
\newcommand*{\Comb}[2]{{}^{#1}C_{#2}}%

\title{Notes on Melonic $O(N)^{q-1}$ Tensor Models}

\author{Sayantan Choudhury$^a$\footnote{sayantan@theory.tifr.res.in}, \ Anshuman Dey$^a$\footnote{anshuman@theory.tifr.res.in}, \ Indranil Halder$^a$\footnote{indranil.halder@tifr.res.in}, \ Lavneet Janagal$^a$\footnote{lavneet@theory.tifr.res.in},
Shiraz Minwalla$^a$\footnote{minwalla@theory.tifr.res.in}, \  Rohan R. Poojary$^a$\footnote{ronp@theory.tifr.res.in}
\\[1.5mm]
\small{\emph{$^{a}$Department of Theoretical Physics,}}
\small{\emph{Tata Institute of Fundamental Research,}} \\
\small{\emph{Homi Bhabha Rd, Mumbai 400005, India}}
}

\abstract{
 It has recently been demonstrated that the large N limit of a model of 
fermions charged under the global/gauge symmetry group $O(N)^{q-1}$ 
agrees with the large $N$ limit of the SYK model. In these notes we 
investigate aspects of the dynamics of the $O(N)^{q-1}$ theories that 
differ from their SYK counterparts. We argue that the spectrum of 
fluctuations about the finite temperature saddle point in these theories
has $(q-1)\frac{N^2}{2}$ new light modes in addition to the light 
Schwarzian mode 
that exists even in the SYK model, suggesting that the bulk dual 
description of theories differ significantly if they both exist. We also
study the thermal partition function of a mass deformed version of the 
SYK model. At large mass we show that the effective entropy of this theory 
grows with energy like $E \ln E$ (i.e. faster than Hagedorn) 
up to energies of order $N^2$. The canonical partition function of 
the model displays a deconfinement or Hawking Page type phase transition 
at temperatures of order $1/\ln N$. We derive these results in the large mass 
limit but argue that they are qualitatively robust to small corrections in $J/m$.
}

\begin{document}
\preprint{{\Large\bf TIFR/TH/17-28}}
\maketitle

\section{Introduction}

It has recently been demonstrated that the dynamically rich Sachdev-Ye-Kitaev model - 
a quantum mechanical model of fermions interacting with random potentials - is solvable 
at large $N$ \cite{Sachdev:2010um, kitaev-talk, Maldacena:2016hyu}. This model is interesting partly because its 
thermal properties have several features in common with those of black holes. The SYK model 
self equilibrates over a time scale of order the inverse temperature and has a Lyapunov index 
that saturates the chaos bound \cite{Maldacena:2016hyu, kitaev-talk}. Moreover the long time behaviour of this model at finite temperature 
is governed by an effective action that has been reinterpreted as a particular theory of gravity 
expanded about $AdS_2$ background solution \cite{Sachdev:2010um, Almheiri:2014cka, Jensen:2016pah, Maldacena:2016upp,  Mandal:2017thl, Gross:2017hcz, Forste:2017kwy, Das:2017pif}. 

These facts have motivated the suggestion that the SYK model 
is the boundary dual of a highly curved bulk gravitational theory whose finite temperature behaviour is dominated by a black hole saddle point. 
If this suggestion turns out to be correct, the solvability of the SYK model at large $N$ - and its 
relative simplicity even at finite $N$- could allow one to probe old mysteries of black hole physics in a manner that is 
nonperturbative in $\frac{1}{N}$, the effective dual gravitational coupling (see e.g. \cite{Garcia-Garcia:2016mno,Cotler:2016fpe, Garcia-Garcia:2017pzl} 
for recent progress). 

There is, however, a potential fly in the ointment. While the SYK model  - defined as a theory with random couplings -
is an average over quantum systems, it is not a quantum system by itself. One cannot, for instance, associate the 
SYK model with a Hilbert space in any completely precise manner, or find a unitary operator that generates time evolution in this
model. As several of the deepest puzzles of black hole physics concern conflicts with unitarity, this feature 
of the SYK model is a concern. 

Of course any particular realisation of the couplings drawn from the SYK 
ensemble is a genuine quantum theory. It is plausible that several 
observables - like the partition function - have the same large $N$ limit 
when computed for any given typical member of the ensemble as they 
do for the SYK model defined by averaging over couplings 
\cite{Cotler:2016fpe, Stanford:2017thb, Belokurov:2017eit}. 
It might thus seem that 
{\it every} typical realization of random couplings is an inequivalent 
consistent quantization of classical large $N$ SYK system. As the number of 
such quantizations is very large, this would be an embarrassment of riches. The 
potential issue here is that if we work with any given realization of the 
SYK model, it appears inconsistent to restrict attention to averaged 
observables for any finite $N$ no matter how large. On the other hand 
correlators of individual $\psi_i$ operators (as opposed to their 
averaged counterparts) presumably do not have a universal large $N$ limit 
(and so are not exactly solvable even at large $N$). 
\footnote{We thank S. Sachdev for  discussion on this point.}

In order to address these concerns some authors have recently 
\cite{Witten:2016iux, Klebanov:2016xxf, Klebanov:2017nlk} (based on 
earlier work \cite{Gurau:2009tw, Gurau:2010ba, 
Gurau:2011aq, Gurau:2011xq, Bonzom:2011zz, Gurau:2011xp}) studied
a related class of models. 
These models are ordinary quantum mechanical systems; in fact they describe the global or gauged quantum 
mechanics of a collection of fermions in 0+1 dimensions. 
In this paper we will focus our attention on the model 
\begin{equation}\label{sykqm} \begin{split} 
&S= \int dt \sum_{a=1}^{N_F} [  {\bar \psi}_a D_0 \psi_a  
  - \left(g \ \psi_a^q + h.c. \right)  
],  \\
& D_0 = \partial_0 + i A_0 \ , \ g=\frac{J}{N^{\frac{(q-1)(q-2)}{4}}}, \\
\end{split}
\end{equation} 
that was first proposed - at least in the current context - in  
\cite{Klebanov:2016xxf} . 
In \eqref{sykqm} $\psi_a$ are a collection of complex gauged fermionic fields 
in $0+1$ dimensions that transform in the fundamental of each of the 
$q-1$ copies of $O(N)$. The index $a$ is a flavour index that runs from $1 \ldots N_F$. 
\footnote{For this simplest case $N_F=1$ this model was presented in Eq 3.23 
of  \cite{Klebanov:2016xxf}. }. $J$ is a coupling 
constant with dimensions of mass and $\psi^q$ is a schematic for a $q$ vertex
generalisation of a `tetrahedronal' interaction term between $q$ copies of 
the fermionic fields, whose gauge index contraction structure is explained 
in detail in \cite{Witten:2016iux, Klebanov:2016xxf} and will be 
elaborated on below. 

The tetrahedral structure of the interaction\cite{Witten:2016iux, Klebanov:2016xxf} is such that for any even number of fermions $q$ each fermion has $q-1$ indices each in a different $O(N)({\textrm or}\, U(N))$. The indices among the $q$ fermions are contracted such that every fermion is index contracted with an index of the same gauge group on one of the remaining fermions. Moreover, given any - and every - 2 fermions have a single 
index (of some gauge group) contracted between them. For $q=4$ it is easy to check that these words define a unique contraction structure 
which may be viewed as a tetrahedral contraction among the 4 fermions each with $q-1=3$ indices(legs) with every fermion(point or vertex of the tetrahedron) connected to 3 different coloured legs. For $q\geq6$ it is not clear that the words above define a unique contraction 
structure. In case the contraction structure is not unique, we pick one choice - for example the Round-Robin scheduling process 
to define our interaction \cite{Narayan:2017qtw, Yoon:2017nig}. \footnote{We would like to thank J. Yoon for explaining the 
Round Robin scheduling process to us and clearing up our misconceptions about uniqueness of the contraction structure for 
$q >4$.}

The connection between the quantum mechanical theories \eqref{sykqm} and the 
SYK model itself is the following; it has been demonstrated (subject 
to certain caveats) that sum over Feynman graphs of the theory \eqref{sykqm} 
coincides with the sum over Feynman graphs of the SYK model at at leading order at large $N$ (see \cite{Witten:2016iux} for the argument in a very similar model), even though these two sums differ at finite values of $N$ 
(see e.g. the recent paper \cite{Dartois:2017xoe} and references therein). 
It follows that the 
quantum mechanical models \eqref{sykqm} are exactly as solvable as the SYK model at large $N$; 
moreover they also inherit much of the dynamical 
richness of the SYK model. In other words the models \eqref{sykqm} are solvable at 
large $N$, are unitary and are potentially boundary duals of (highly curved) 
black hole physics.

Motivated by these considerations, in this note we study the effective theory 
that governs the long time dynamics of the model 
\eqref{sykqm} at finite temperature. We focus attention on dynamical aspects 
of \eqref{sykqm} that have no counterpart in the already well studied dynamics 
of the original SYK model. 
\footnote{See \cite{Nishinaka:2016nxg, Peng:2016mxj, Krishnan:2016bvg, Ferrari:2017ryl, Gurau:2017xhf,  Bonzom:2017pqs, Krishnan:2017ztz, Narayan:2017qtw, Chaudhuri:2017vrv, Azeyanagi:2017drg, Giombi:2017dtl} 
for other recent work on the model \eqref{sykqm} and 
its close relatives.}

In the rest of this introduction we will explain and describe our principal observations and results.

\subsection{New light modes}
\label{Newlightmodes}

The thermal behaviour of both the theory \eqref{sykqm} and 
the original SYK model is determined by the path integral of these theories 
on a circle of circumference $\beta$. 

It was demonstrated in 
\cite{kitaev-talk, Maldacena:2016hyu} that, in the case of the original 
SYK model, this path integral is dominated by a saddle point of an 
effective action whose fields are the two point function and self energy
of the fermions. An extremization of this effective action determines 
both the fermionic two point function at finite temperature as well as 
the free energy  of the system at leading order at large $N$. 

In a similar manner, the thermal behaviour of the quantum mechanical systems
\eqref{sykqm} is dominated by a saddle point 
at large $N$. Under appropriate assumptions it may be shown that 
resultant effective action has the same minimum as that 
of the original SYK theory \cite{Witten:2016iux}.
\footnote{A potential subtlety  is that  path integral of the quantum 
mechanical system \eqref{sykqm} has a degree of freedom that is absent 
in the original SYK model, namely the holonomy of the gauge group 
$O(N)^{q-1}$. As for the SYK model, integrating out 
the fermions leads to an effective action - proportional to $N^{q-1}$ 
- whose fields are a two point function of the fermions, a self 
energy and the holonomy of the gauge group. 
As in the case of the original SYK model, at leading order in the large 
$N$ limit the free energy of the system is captured by the saddle point 
of this effectively classical action. If we work at temperatures that 
are held fixed as $N \to \infty$ it is highly plausible that 
this effective action is minimised when the holonomy is the identity 
matrix (see \ref{sec3} below). Under this assumption the saddle point of the quantum mechanical system 
coincides with that of the SYK model.}. Specialising to the case $N_F=1$, 
the leading order fermionic two point function of the quantum mechanical system is given by 
\begin{equation}\label{tpf} 
\langle {\bar \psi}^a (t) \psi_b(t') \rangle
= \delta^a_b G^{SYK}(t-t'),
\end{equation} 
where $a$ and $b$ denote the (collection of) vector indices for the fermions and $G^{SYK}(t)$ is the 
thermal propagator of the original SYK model. \footnote{\eqref{tpf} applies both the the case 
that the group $O(N)^{q-1}$ is global and local. In the latter case this equation applies in the 
gauge $\partial_0 A_0=0$. Assuming that the holonomy degree of freedom is frozen to identity at 
large $N$, the gauged and global model coincide.}

While the thermal behaviour of the model \eqref{sykqm} is thus 
indistinguishable from that of the SYK model at leading order in the large 
$N$ limit, the dynamics of the quantum mechanical model \eqref{sykqm} differs from 
that of the SYK model at subleading orders in $1/N$. The first 
correction to leading large $N$ thermal behaviour may be  obtained 
by  performing a one loop path integral over quadratic fluctuations 
around the saddle point.
In the long time limit, correlators are dominated by the lightest 
fluctuations around the saddle point.

Recall that in the UV (i.e. as $\beta J \rightarrow 0$ )
the fermions of \eqref{sykqm} have dimension zero. The term proportional 
to $\psi^q$ in \eqref{sykqm} represents a dimension zero relevant deformation 
of this UV fixed point. The resultant RG flow ends in the deep IR in 
a conformal field theory in which the fermions have dimension $\frac{1}{q}$. 
\cite{kitaev-talk, Maldacena:2016hyu}. In this IR limit 
(relevant to thermodynamics when $\beta J \to \infty$) 
 $\psi^q$ is marginal while the kinetic term in 
\eqref{sykqm}  is irrelevant \cite{kitaev-talk, Maldacena:2016hyu}. The 
fact that the kinetic term is irrelevant in the IR - and so can effectively 
be ignored in analysing the symmetries of \eqref{sykqm} at large $\beta J$ 
- has important implications for the structure of light fluctuations about 
the thermal saddle point.  

The first implication of the irrelevance of the kinetic term 
occurs already in the SYK model and was explored 
in detail in \cite{kitaev-talk, Maldacena:2016hyu, Maldacena:2016upp}.
The main point is that the action \eqref{sykqm}, with the kinetic 
term omitted,  enjoys invariance under conformal diffeomorphisms 
(i.e. diffeomorphisms together with a Weyl transformation). 
 However the saddle point solution for the Greens function 
$G^{\text{SYK}}(t)$ is not invariant under conformal diffeomorphisms. It  
follows immediately that the action of infinitesimal conformal diffeomorphisms 
on this solution generates zero modes in the extreme low energy limit.

At any finite temperature, no matter 
how small, the kinetic term in \eqref{sykqm} cannot completely be ignored 
and conformal invariance is broken; the action of conformal 
diffeomorphisms on the SYK saddle point consequently 
produces anomalously light (rather than exactly zero) modes. 
The action for these modes was computed in 
\cite{kitaev-talk, Maldacena:2016hyu, Maldacena:2016upp} and takes the form 
of the Schwarzian for the conformal diffeomorphisms.

A very similar line of reasoning leads to the conclusion that the model 
\eqref{sykqm} has $(q-1) \frac{N^2}{2}$ additional light modes in the 
large $\beta J$ limit, as we now explain. 
Let us continue to work in the gauge $A_0=0$. In this gauge the action 
\eqref{sykqm} is obviously invariant under the global rotations 
$\psi \rightarrow V \psi $, $ {\bar \psi} \rightarrow {\bar \psi} V ^{\dagger}$
where $V$ is an arbitrary time independent 
$O(N)^{q-1}$ rotation. In the global model \eqref{sykqm}
the rotation by $V$ is the action of a global symmetry. In gauged model 
on the other hand, these rotations are part of the gauge group and do not generate global 
symmetries of our model; the Gauss law in the theory ensures that all 
physical states are uncharged under this symmetry. 

Let us now consider the transformation 
$\psi \rightarrow V(t) \psi$ together with 
$ {\bar \psi} \rightarrow {\bar \psi} V(t)^{\dagger}$ where $V(t)$ is an 
arbitrary time {\it dependent}
$O(N)^{q-1}$ rotation. In the case of the gauged 
models, this transformation is not 
accompanied by a change in $A_0$ ($A_0=const$ throughout) so 
the rotation is not a gauge transformation. 

At finite $\beta J$ the rotation by a time dependent $V(t)$ is not a 
symmetry of the action \eqref{sykqm} in either the global or the 
gauged theory as the kinetic term in \eqref{sykqm} is not left invariant 
by this transformation. 
As we have explained above, however, the kinetic term is irrelevant 
in the low temperature limit $\beta J \to \infty$. It follows that 
the time dependent transformation is an effective symmetry of 
dynamics this strict low temperature limit. 

However the saddle point solution \eqref{tpf} is clearly not invariant 
under the time dependent rotations by $V(t)$. It follows that, as in 
the discussion for conformal diffeomorphisms above, the 
action of $V(t)$ on \eqref{tpf} generates exact zero modes in the strict 
limit $\beta J \to \infty$ and anomalously light modes at any finite 
$\beta J$. 
We emphasise that this discussion applies both to the global model where 
$O(N)^{q-1}$ is a global symmetry, and the gauged model where it is not. 

In section \ref{EffectiveActionSection} below we argue that the dynamics of our new light modes is governed by the 
effective sigma model on the group manifold  
\begin{equation}
	\label{effectiveaction}
	\begin{aligned}
		S=-\mathcal{A} \ \frac{N^{q-2}}{|J|}   \int dt \ \sum_{l=1}^{q-1} \ Tr \ \left[ \left( V_l^{-1}(t)\ \frac{\partial}{\partial t} V_l(t) \right)^2 \right],
	\end{aligned}
\end{equation}
where $V_l(t)$ is an arbitrary element of the group $O(N)$ and $\mathcal{A} $ is a number of order 
unity that we have not been able to determine. 

The formula \eqref{effectiveaction} has appeared before in a closely related context. 
The authors of \cite{Sachdev:2017mar} (see also \cite{Stanford:2017thb})
studied the a complex version of the SYK model.
Their model had an exact $U(1)$ symmetry at all energies, which - using the arguments 
presented in the previous paragraphs -  was approximately enhanced to a 
local $U(1)$ symmetry at low energies. The authors of \cite{Sachdev:2017mar} argued the 
long distance dynamics of the new light modes is governed by a sigma model 
on  the group manifold $U(1)$.
\footnote{They also argued for some mixing between the diffeomorphism and $U(1)$ long distance modes.} 
Given these results, the appearance of a low energy sigma model 
in the large $\beta J$ finite temperature dynamics of 
the theory \eqref{sykqm} seems natural. 

We would, however,  like to emphasise two qualitative differences between 
the sigma model \eqref{effectiveaction} and the model that appeared in  
\cite{Sachdev:2017mar}. First \eqref{effectiveaction} is a sigma model 
for a group $O(N)^{q-1}$ whose dimensionality 
goes to infinity in the large $N$ limit, $N \to \infty$. Second that we 
find the new light modes of the action 
even of the gauged model \eqref{sykqm} even though $O(N)^{q-1}$ is not a global symmetry of this theory.  

 The new modes governed by \eqref{effectiveaction} are 
approximately as light - and so potentially as important to long time 
dynamics - as the conformal diffeomorphisms described above.  
Note, however, that there are $(q-1) \frac{N^2}{2}$ light time dependent 
$O(N)^{q-1}$ modes but (as far as we can tell) only one conformal 
diffeomorphism.

We have already remarked above that the light diffeomorphism degree of freedom
described above has been given an interpretation as a particular gravitational
action in an $AdS_2$ background. It seems likely to us that the 
effective action \eqref{effectiveaction} will, in a similar way, admit 
a bulk interpretation as a gauge field propagating in $AdS_2$. The Yang 
Mills coupling of this gauge field - like Newton's constant for the 
gravitational mode - will be of order $\frac{1}{N^{q-1}}$ (this is simply 
a reflection of the fact that our model has $N^{q-1}$ degrees of freedom). 
This means that the t' Hooft coupling of all the gauge fields in 
the bulk will be of order $g_{YM}^2 N \sim \frac{1}{N^{q-2}}$. 
The fact that this coupling goes to zero in the large $N$ limit
implies that the bulk gauge fields are classical even though there are 
so many of them. \footnote{We would like to thank J. Maldacena for 
a discussion of this point.} 

It has been established that the light diffeomorphism degree of freedom 
has a qualitatively important effect on out of time ordered thermal 
correlators; it leads to exponential growth in such correlators at a 
rate that saturates the chaos bound $G \sim e^{ 2 \pi Tt}$. When we include 
the contribution of the new light modes described in this subsection, we 
expect this growth formula to be modified to 
\begin{equation}\label{mgf}
G(t) \sim \left( e^{2 \pi T t} + N^2 f(t) \right).
\end{equation} 
\footnote{See \cite{Yoon:2017nig} for related work.}
The factor of $N^2$ is a reflection of the fact that our new 
modes are $N^2$ in number, whereas - as far as we can tell - there 
is only a single light mode corresponding to conformal diffeomorphisms.

 Given that the solutions of the equations of 
motion to the Sigma model \eqref{effectiveaction} grow no faster than 
linearly in time, we expect $f(t)$ to grow at most polynomial 
in time. This suggests it that the light modes \eqref{effectiveaction} will 
dominate correlators  up to a time of order 
$\frac{1}{\pi T} \ln N $. At later times the exponentially growing 
diffeomorphism mode will dominate, leading to exponential growth and a 
 Lyapunov index that saturates the chaos bound. 

To end this subsection let us return to a slightly subtle point in our 
discussion. 
 In order to derive the effective action for $V(t)$ we worked in the 
gauge $A_0=0$. As our theory is on a thermal circle, in the case of the gauged 
model \eqref{sykqm} 
we have missed a degree of freedom - the gauge holonomy - by working in the 
gauge $A_0=0$. This,  however,  is easily corrected for. Even in the presence of 
a holonomy, we can set the gauge field $A_0$ to zero by a gauge 
transformation provided we allow ourselves to work with gauge transformations
that are not single valued on the circle. The net effect of working with 
such a gauge transformation is that the matter fields are no longer periodic
around the thermal circle but obey the boundary conditions 
\begin{equation} \label{bcV}
\psi(\beta)= -U \psi(0),
\end{equation}
where $U$ is the holonomy around the thermal circle. For the fields of the 
low energy effective action \eqref{effectiveaction} this implies the 
boundary conditions 
\begin{equation} \label{lea}
V(\beta)= U V(0) U^{-1}.
\end{equation} 
Recall we are instructed to integrate over all values of the holonomy $U$.
Consequently we must integrate over the boundary conditions \eqref{lea} 
with the Haar measure. See section \ref{sec5} for some discussion of this 
point.

In summary, the discussion of this subsection suggests that the 
bulk low energy effective action `dual' to the gauged/global quantum mechanics 
of \eqref{sykqm} differs from the low energy effective action `dual'
to the SYK model in an important way; in addition to the gravitational 
field it contains gauge fields of a gauge group whose rank is a positive
fractional power of the inverse Newton (and Yang Mills) coupling constant 
of the theory. In the classical limit in which Newton's constant is taken
to zero, the rank of the low energy gauge fields also diverges. Nonetheless
the limits are taken in such a way that the effective bulk theory remains 
classical.

\subsection{Holonomy dynamics and the spectrum at large mass}

Our discussion up to this point has applied equally to the `global' and 
`gauged' quantum mechanical models \eqref{sykqm}. In the rest of this 
introduction we focus attention on the gauged models, i.e. the models in 
which the $O(N)^{q-1}$ symmetry algebra is gauged. 
In this case the thermal path integral of our system includes an integral 
over gauge holonomies over the thermal circle. We wish to study the effect 
of this holonomy integral on the dynamics of our system. 

In order to do this in the simplest and clearest possible way we deform the model 
\eqref{sykqm} in a way that trivializes the dynamics of all non holonomy 
modes in the theory. This is accomplished by adding a mass to the fermions. For 
concreteness we work with the $O(N)^{q-1}$ model  
\begin{equation}\label{sykqmmd} \begin{split} 
&S= \int dt \sum_{a=1}^{N_F} \left[ \left(  {\bar \psi}_a D_0 \psi_a  
+ m {\bar \psi}_a \psi_a \right)  -\left(  g \
\psi_a^q  + h.c.  \right) \right],  \\
& D_0 = \partial_0 + i A_0 \ , \ g=\frac{J}{N^{\frac{(q-1)(q-2)}{4}}},\\
\end{split}
\end{equation} 
where $m$, the mass of the fermion is taken to be positive. \footnote{
In the case that the mass is negative, most of our formulae below go 
through once under the replacement $ m \rightarrow |m|$.}
We work the large mass limit, i.e. the limit $\frac{m}{J} \gg 1$.
The effective interaction between fermions in \eqref{sykqmmd}, 
$\frac{J}{m}$, is small in this limit and can be handled perturbatively. 
In the strict $m \to \infty$ limit the 
only interaction that survives in the system 
is that between the (otherwise free) matter fields and the holonomy 
$U$. \footnote{We emphasize that, in the limit under consideration, 
modes corresponding to diffeomorphisms or $V(t)$
are no longer light - and so are irrelevant. However the holonomy continues 
to be potentially important.}

Let us first work in the strict limit $\frac{m}{J} \rightarrow \infty$. 
In this limit the dynamics of the holonomy field $U$ in this theory is governed 
by an effective action obtained by integrating out the matter fields 
at one loop. \footnote{For orientation, we remind the reader that the 
integral over the holonomy is the device the 
path integral uses to ensure that
the partition function only counts those states that obey the $A_0$ equation
of motion, i.e. the Gauss law constraint. Restated, the integral 
over holonomies ensures that the partition function only counts 
those states in the matter Hilbert space that are singlets under 
the gauge group. }
The resultant effective action is easily obtained and is 
given by (\cite{Aharony:2003sx})
\begin{equation}\label{freesykmd} \begin{split}
&Z= \Tr x^{\frac{H}{m}} = \int  \prod_{i=1}^{q-1}dU_i  
\exp (- S_{\text{eff}} (U_i)), \\
&S_{\text{eff}}(U_i)= -N_F \sum_{n=1}^\infty  \frac{ (-x)^n \left( \prod_{i=1}^{q-1} \Tr U_i^n \right) 
}{n},\\
&x= e^{- \beta |m|},\\
\end{split}
\end{equation}
where $H$ is the Hamiltonian of our theory. 
\footnote{The generalization of these 
results to a model with $N_B$ bosons and $N_F$ fermions yields the 
holonomy effective action 
\begin{equation}\label{fneq}
S_{\text{eff}}(U_i)=\sum_{n=1}^\infty (N_B+ (-1)^{n+1} N_F) \ x^n \frac{ \left( \prod_{i=1}^{q-1} \Tr U_i^n 
\right) }{n} .
\end{equation}
As we will see below, in the scaling limit of interest to this paper, only the term with $n=1$ 
is important. In the strictly free limit it follows that most the results presented above 
apply also to a theory with $N_F$ fermions and $N_B$ bosons once we make the 
replacement $N_F \rightarrow N_F+N_B$.}

Each $U_i$ is an $O(N)$ matrix that represents the holonomy in the $i^{th}$ factor in the 
gauge group $O(N)^{q-1}$. $dU$ is the Haar measure over the group $O(N)^{q-1}$ normalized so that 
the total group volume is unity. 

Notice that when $x$ is of order unity, $S_{\text{eff}} \sim N^{q-1}$ in 
\eqref{freesykmd}. On the other hand the contribution of the group 
measure to the `effective' action is of order $N^2$. The integral in 
\eqref{freesykmd} is interesting when these two contributions are comparable. 
This is the case if we scale temperatures so that 
\begin{equation}\label{tempscal} 
  x=e^{- \beta |m|} =\frac{\alpha}{N_F N^{q-3}},
\end{equation}
with $\alpha$ held fixed as $N$ is taken to infinity. 
In this limit the terms in the second of \eqref{freesykmd} 
with $n>1$ are subleading and can be ignored. Effectively 
\begin{equation}\label{pfotss}
 \begin{split}
&Z(x)= \int  \prod_{i=1}^{q-1}d U_i 
\exp (- S_{\text{eff}} (U_i)) \\
&S_{\text{eff}} = \frac{\alpha}{N^{q-3}}  
\left(\prod_{i=1}^{q-1}\Tr U_i \right). 
\end{split} 
\end{equation} 

In the large $N$ limit the matrix integral \eqref{pfotss} is equivalent - as we 
show below - to the well known Gross Witten Wadia 
model and is easily solved. The solution - presented in detail below - has the following features
\begin{itemize}
\item{1.} In the canonical ensemble, the partition function undergoes a 
deconfinement type phase transition at $\alpha=\alpha_{1pt}$ where the 
value of $\alpha_{1pt}$ is given in \eqref{alpha1pt}. At smaller values of $\alpha$ 
the system is dominated by the `confining' saddle point in which $U$ is 
the clock matrix. At larger values of $\alpha_{1pt}$ the system is dominated 
by a more complicated `deconfined' or black hole 
saddle point. The phase transition is reminiscent of the transitions 
described in \cite{Witten:1998apr, Aharony:2005bq}. \footnote{We note that the 
first order phase transitions described in \cite{Aharony:2005bq}
were strongly first order (i.e. not on the edge between first and second 
order) only after turning on gauge interactions. In the current context, 
in contrast, the phase transition in our system is strongly first order
even in the absence of interactions.}
\item{2.} In the microcanonical ensemble, the scaling limit described above
captures the density of states of the system at energies less than or 
of order $N^2$. Over the range of energies $1 \ll E < \frac{N^2}{4} $, the entropy $S$ is given by the simple formula
\begin{equation}\label{leent} 
S(E) = (q-3)\left[ \frac{E}{2} \ln \left( \frac{E}{2} \right) -\frac{E}{2} \right]+E \ \log N_F + (q-3)\frac{E}{2}\ln(2).
\end{equation}
The saddle point that governs the density of states of the theory
changes in a non analytic manner at $E= \frac{N^2}{4}$. For $E > \frac{N^2}{4}$ 
the formula for the entropy is more complicated. For energies 
$E \gg (q-2) \frac{N^2}{4} $, however, the entropy 
simplifies to the formula for $ n_B N^{q-1}$ complex bosonic and 
$n_F N^{q-1}$ free complex fermionic harmonic oscillators 
\begin{equation}\label{eent} 
S(E) = E \left[1-\log \left(\frac{E}{pN^{q-1}}\right)\right].
\end{equation}
The complicated formula that interpolates between these special results is presented in \eqref{eneg}.
\end{itemize}

The formula \eqref{leent} suggests that if a dual bulk interpretation 
of the theory \eqref{freesykmd} exists, it is given in terms of a collection 
of bulk fields whose number grows faster than exponentially with energy.
It would be fascinating to find a bulk theory with this unusual behaviour.

Moreover the existence of a Hawking Page type phase transition in this model
- and in particular the existence of a subdominant saddle point even 
at temperatures at which the dominant phase is a black hole phase - 
opens the possibility of the subdominant phase playing a role in effectively 
unitarizing correlators about the black hole saddle point by putting 
a floor on the decay of the amplitude of 
correlators as in \cite{Maldacena:2001kr}. 

The results presented above apply only in the limit $\frac{m}{J}\to \infty$. 
We have also investigated how these results are modified at very weak 
(rather than zero) coupling. We continue to work at low temperatures, 
in a manner we now describe in more detail. It turns that $S_{\text{eff}}(U)$ takes 
the schematic form 
\begin{equation}\label{pfsf}
S_{\text{eff}}(U) = \sum_{a=1}^\infty  x^a f_a(\beta, U).
\end{equation}
Working to any given order in perturbation theory, the functions $f_a(\beta)$ 
are all polynomials of bounded degree in $\beta$. We work at temperatures 
low enough so that we can truncate \eqref{pfsf} to its first term. 
In other words the terms we keep are all proportional to $x$ multiplied 
by a polynomial dressing in $\beta$. 

We demonstrate below that within this approximation the partition function 
\eqref{pfsf} takes the form 
\begin{equation}\label{sfai}
-S_{\text{eff}}(U)= N^{q-1} x \left(\prod_{m=1}^{q-1} \rho^1_m \right) 
\left( \sum_{k=0}^\infty \left( \frac{J}{m} \right)^{2k} 
{\tilde H}_k(\frac{J^2 \beta}{m}) \right).
\end{equation}
Note that \eqref{sfai} asserts that the interacting effective action has the 
same dependence on $x$ and $U$ as its free counterpart did. The only difference
between the interacting and free effective action is a prefactor 
which is a function of the two effective couplings  $\frac{J}{m}$ and 
$\frac{J^2 \beta}{m}$. Below we have summed an infinite class of graphs and 
determined the function ${\tilde H}_0$. Working at $N_F=1$ we find 
\begin{equation}\label{parfmoi}
\begin{aligned}
	{\tilde H}_0=& 2 
  \left[ \frac{1}{2}+2\gamma(q)\ \frac{(-\beta)}{m}|J|^2  e^{\gamma(q)\ \frac{(-\beta)}{m}|J|^2}-\frac{1}{2}e^{2 \gamma(q)\ \frac{(-\beta)}{m}|J|^2} -\frac{(-1)^{q/2}}{2} q \ \beta \frac{|J|^2}{m}  \right],
\end{aligned}
\end{equation}
where $\gamma(q)$ is defined in \eqref{gde}.

\eqref{sfai} and \eqref{parfmoi} determine the effective action of our system
whenever the terms proportional to ${\tilde H}_m$ $(m=1, 2 \ldots)$ 
in the second line of \eqref{sfa} can be ignored compared to the term 
proportional to ${\tilde H}_0$. This is always the case at weak enough 
coupling; the precise condition on the coupling when this is the case 
depends on the nature of the 
as yet unknown large argument behaviour of the functions 
${\tilde H}_m$ .

The partition function that follows from the action 
\eqref{sfai} is identical to the free partition function described above 
under the replacement $ \alpha \rightarrow \alpha {\tilde H}_0$.
It follows that the interacting partition function is essentially identical 
to the free one in the canonical ensemble. The $\beta$ dependence of 
the effective value of $\alpha$ leads to some differences in the 
micorcanonical ensemble that turn out not to impact  the main qualitative 
conclusions of the analysis of the free theory. For instance the super hagedorn 
growth of the entropy persists upon  including the effects of 
interaction.
\newline
\newline
{\it Note Added}: `We have recently become aware of the preprint
 \cite{Bulycheva:2017ilt} which overlaps with this paper in multiple ways. We hope it will prove possible to combine the results of this paper with the methods of \cite{Bulycheva:2017ilt} to better understand the new light modes discussed earlier in this introduction'.

\section{Light thermal modes of the Gurau-Witten-Klebanov-Tarnopolsky
 models} \label{sec2}

In this section we consider the Gurau-Witten-Klebanov-Tarnopolsky model at 
finite temperature. The Lagrangians for the specific theories we 
study was listed in \eqref{sykqm}. 
As we have explained in the introduction, this model has 
a new set of light modes parameterized by $V(t)$, an arbitrary group element
as a function of time, where $V$ belongs to $O(N)^{q-1}$. 
In this section we will present an 
argument that suggests that the dynamics of these light modes is governed 
by a (quantum mechanical) sigma model on the group manifold. We will also 
present an estimate for the coupling constant of this sigma model. 

That the dynamics of $V(t)$ should be governed by a sigma model is very 
plausible on general grounds. Recall that in the formal IR limit, 
$V(t)$ is an exact zero mode of dynamics. It follows that $V(t)$ picks up 
dynamics only because of corrections to extreme low energy dynamics. From 
the point of view of the low energy theory these corrections are UV effects, 
and so should lead to a local action for $V(t)$. The resultant action 
must be invariant under global shifts $V(t ) \rightarrow V_0 V(t)$. 
We are interested in the term in the action that will dominate long time 
physics, i.e. the action with this property that has the smallest number of 
time derivatives. Baring a dynamical coincidence (that sets the coefficient 
of an apparently allowed term to zero) the action will be that of the 
sigma model. 

In the rest of this section we will put some equations to these words. 
We would like to emphasise that the `derivation' of the sigma model 
action presented in this section is intuitive rather than rigorous - 
and should be taken to be an argument that makes our result highly 
plausible rather than certain.

\subsection{Classical effective action}
\label{ClassicalEffectiveAction}

In \cite{Maldacena:2016hyu} the effective large $N$ dynamics of the SYK model was recast 
as the classical dynamics of two effective fields; the Greens function 
$G(t)$ and the self energy $\Sigma(t)$. The action for $\Sigma$ and 
$G$ derived in \cite{Maldacena:2016hyu} was given by 
\begin{equation} \label{msee} 
	\begin{aligned}
		S=
N^{q-1} \left( 
-\log Pf(\partial_t-\tilde{\Sigma} )   + \int dt_1 \ dt_2 \ 
\left[ - \tilde{\Sigma}(t_1,t_2)\tilde{G}(t_2,t_1)-
\frac{J^2}{q}\tilde{G}^q(t_1,t_2)  \right] \right). 
	\end{aligned}
\end{equation}
The utility of the action \eqref{msee} was twofold. First, the 
solutions to the equations of motion that follow from varying \eqref{msee}
are the saddle point that govern thermal physics of the SYK model. Second, 
an integral over the fluctuations in \eqref{msee} also correctly captures 
the leading order (in $\frac{1}{N}$) correction to this saddle point 
result. In order to obtain these 
corrections, one simply integrates over the quadratic fluctuations about 
this saddle point. In particular the action \eqref{msee} was used to 
determine the action for the lightest fluctuations about the 
saddle point \eqref{msee}, namely conformal diffeomorphism
\cite{Maldacena:2016hyu}.

In this section we wish to imitate the analysis of \cite{Maldacena:2016hyu} to determine 
the action for fluctuations of the new zero modes - associated 
with time dependent $O(N)^{q-1}$ rotations - described in the introduction. 
 The action \eqref{msee} is not sufficient for this purpose. 
As explained in the introduction, the low energy fluctuations we wish to 
study are obtained by acting on the saddle point Greens function with 
time dependent $O(N)^{q-1}$ rotations; however the fields 
$G$ and $\Sigma$ that appear in \eqref{msee} have no indices and so cannot 
be rotated. 

As the first step in our analysis we proceed to generalise the effective 
action \eqref{msee} to an action whose variables are the matrices 
$G_a^b$ and $\Sigma_a^b$. The indices $a$ and and $b$ are both fundamental
indices of the group $O(N)^{q-1}$. Our generalised action is given by 
\begin{equation} \label{genact}
	\begin{aligned}
		S=-\log Pf(D_0-\tilde{\Sigma} )   + \int dt_1 \ dt_2 \ \left[ - \tilde{\Sigma}_a^{\ b}(t_1,t_2)\tilde{G}_b^{\ a}(t_2,t_1)-\frac{|g|^2}{q}\tilde{G}^q(t_1,t_2)  \right].
	\end{aligned}
\end{equation}

In this action, the expression $\tilde{G}^q$ is a product of $q$ copies of 
$\tilde{G}^{a}_{b}$ where all gauge indices are contracted in a manner we now 
describe. Recall that $a$ and $b$ are fundamental indices for the group 
$O(N)^{q-1}$. Each of these indices may be thought of as a collection of $q-1$ 
fundamental indices
$$a=(a_1 a_2 \ldots a_{q-1}), ~~~b=(b_1 b_2 \ldots b_{q-1}),$$
where $a_i$ and $b_i$ are fundamental indices in the ($i^{th}$ factor of) 
$O(N)$. In the contraction $\tilde{G}^q$, $a$ type indices are contracted with 
each other while $b$ type indices are also contracted with each other - 
there is no cross contraction between $a$ and $b$ type indices. The 
structure of contractions is as follows; the $a$ indices of precisely 
one of the $O(N)$ factors of the gauge group are contracted between 
any two (and every two)  $Gs$ and, simultaneously, the $b$ indices 
of the same two $O(N)$ factors are also contracted between the 
same two $\tilde{G}s$. \footnote{These rules have their origin 
in the generalized `tentrahedronal' contraction structure described 
in the introduction. For values of $q$ at which the basic interaction 
structure has an ambiguity, we make one choice; for instance we 
adopt the `Round Robin' scheme to fix the ambiguities. As far as we 
can tell, none of our results depend on the details of the choice we make.}

As a quick check note that 
the total number of contraction of $a$ (or $b$) indices, according to 
our rule, is the number of ways of choosing two objects from a group of 
$q$, or, $\frac{q(q-1)}{2}$. As each pair hit two indices, we see that 
the pairing rule described in this paragraph saturates the indices 
present $q$ copies of $\tilde{G}$ (there are a total of $q(q-1)$ 
$a$ type indices).  

The contraction structure described for $a$ type indices in the previous 
paragraph is precisely the contraction structure for the interaction term 
$\psi^{q}$ in the action \eqref{sykqm}. 

We regard \eqref{genact} as a phenomenological action with the following 
desirable properties. First it is manifestly invariant under global 
$O(N)^{q-1}$ transformations. Second if we make the substitutions 
$\tilde{G}^a_b \to \tilde{G} \delta^a_b$, $\tilde{\Sigma}^a_b \to \tilde{\Sigma} \delta^a_b$
into \eqref{genact} we recover the action \eqref{msee}. It follows 
in particular that, if $G$ and $\Sigma$ denote the saddle point values 
of \eqref{msee} then 
\begin{equation}\label{sadptsol}
G^a_b= \delta^a_b G, ~~~\Sigma^a_b= \delta^a_b \Sigma,
\end{equation}
are saddle points of \eqref{genact}. This point can also be verified 
directly from the equations of motion that follow from varying \eqref{genact}, 
i.e.  
\begin{equation}
\label{LargeNeq}
	\begin{aligned}
		G_a^{\ b}(t_1,t_2)=& ((D_0-\Sigma )^{-1})_a^{\ b}(t_1,t_2),\\
		\Sigma_a^{\ b}(t_1,t_2)=&|g|^2 \ (G^{q-1})_a^{\ b}(t_1,t_2).
	\end{aligned}
\end{equation}
While \eqref{genact} correctly reproduces finite temperature 
saddle point of the the model \eqref{sykqm}, it does not give us a 
weakly coupled description of arbitrary fluctuations about this saddle 
point. The fact that \eqref{genact} has $N^{2(q-1)}$ fields makes the action 
very strongly coupled. The key assumption in this section - for which we will 
offer no detailed justification beyond its general plausibility - is that 
the action \eqref{genact} can, however, be reliably used to obtain the 
effective action for the very special manifold of configurations described 
in the introduction, namely 
\begin{equation} \label{symo}
	\begin{aligned}
		\ \tilde{G}_b^{\ a}(t_1,t_2)  =&   
V_b^{\ b'}(t_1) G(t_1, t_2) V_{b'}^{\ a}(t_2),\\
	\tilde{\Sigma}_b^{\ a}(t_1,t_2) =& V_b^{\ b'}(t_1) 
\Sigma(t_1,t_2)V_{b'}^{\ a}(t_2),
	\end{aligned}
\end{equation}
where the index free functions $G(t_, t_2)$ and $V(t_1, t_2)$ are the 
solutions to the SYK gap equations and 
$V(t)$ is an arbitrary $O(N)^{q-1}$ group element. The RHS
in \eqref{symo} is the result of performing a time dependent $O(N)^{q-1}$ 
rotation on the saddle point solution \eqref{sadptsol}.

The fact that 
we have only $(q-1)\frac{N^2}{2}$ fields ($V(t)$) on this manifold of 
solutions - at least formally makes the action restricted to this special 
manifold weakly coupled, as we will see below.

In the rest of this section we will use the action \eqref{genact} 
to determine the effective action that controls the dynamics of 
the matrices $V(t)$ at leading order in the long wavelength limit.

\subsection{Effective action}\label{EffectiveActionSection}

In order to study quadratic fluctuations about \eqref{sadptsol}, 
we follow \cite{Maldacena:2016hyu} to  insert the expansion  
\footnote{Note that we have scaled $G$ fluctuations and $\Sigma$ 
fluctuations with factors that are inverses of each other 
ensures that our change of variables does not change the path integral 
measure. The scalings of fluctuations in \eqref{sov} are chosen to 
ensure that the second line of \eqref{gsqa} takes the schematic 
form $gg$ rather than $g K' G$ where $K'$ is an appropriate Kernel. 
We emphasise that the scaling factor $|G(t_1,t_2)|^{\pm \frac{q-2}{2}}$ in 
\eqref{sov} represents the power of  a function; no matrices are involved.}
\begin{equation} \label{sov}
	\begin{aligned}
		\tilde{G}_a^{\ b}(t_1,t_2)=&G_a^{\ b}(t_1,t_2)+|G(t_1,t_2)|^{\frac{q-2}{2}}g_a^{\ b} (t_1,t_2),\\
		\tilde{\Sigma}_a^{\ b}(t_1,t_2)=&\Sigma_a^{\ b}(t_1,t_2)+|G(t_1,t_2)|^{\frac{2-q}{2}}\sigma_a^{\ b} (t_1,t_2),
	\end{aligned}
\end{equation}
into \eqref{genact} and work to quadratic order in 
$g_a^{\ b} (t_1,t_2)$ and $\sigma_a^{\ b} (t_1,t_2)$. Integrating out $\sigma_a^{\ b} (t_1,t_2)$ using the linear equations of motion, we find an effective 
action of the general structure
\begin{equation} \label{gsqa}
	\begin{aligned}
		S(\tilde{G},\tilde{\Sigma})= S(G,\Sigma)&+\frac{1}{2}\int dt_1..dt_4 \ g_a^{\ b}(t_1,t_2)\tilde{K}^{-1}(t_1,t_2;t_3,t_4)g_b^{\ a}(t_3,t_4)\\
		& - \frac{|g|^2}{q} \ \frac{q}{2} N^{\frac{1}{2}(q-1)(q-4)+1} \int dt_1 \ dt_2 \ g(t_1,t_2) g(t_1,t_2).
	\end{aligned}
\end{equation}
The expression in the first line of \eqref{gsqa} results from varying the 
first two terms in \eqref{genact}, while the second line is the variation 
of the $\tilde{G}^q$ term  in \eqref{genact}. This term denotes the 
a sum of different contraction of indices between the two $gs$ 
\begin{equation}\label{ggf}
g(t_1,t_2) g(t_1,t_2)
= \sum_{k=1}^{q-1} g^{c_1 c_2 ...c_{k-1} a_k c_{k+1}....c_{q-1}}_{c_1 c_2 ...c_{k-1} b_k c_{k+1}....c_{q-1}} g^{d_1 d_2 ...d_{k-1} a_k d_{k+1}....d_{q-1}}_{d_1 d_2 ...d_{k-1} b_k d_{k+1}....d_{q-1}}.
\end{equation}
In the special case that the fluctuation fields $g$ are taken to be of the form 
$g^a_b= \delta^a_b g$, the matrix contractions in \eqref{gsqa} give appropriate 
powers of $N$, and \eqref{gsqa} reduces to the effective action for $g$ 
presented in \cite{Maldacena:2016hyu}.

It was demonstrated in \cite{Maldacena:2016hyu} that 
\begin{equation} \label{kform}
	\begin{aligned}
		\tilde{K}(t_1,t_2;t_3,t_4)=-|G(t_1,t_2)|^{\frac{q-2}{2}}G(t_1,t_3)G(t_2,t_4)|G(t_3,t_4)|^{\frac{q-2}{2}}.
	\end{aligned}
\end{equation}
In the long distance limit the Greens function can be expanded as 
\begin{equation}\label{gfe}
\begin{split}
G=&G_c + \delta G + ...,\\
\delta G(t_1,t_2) & \equiv G_c(t_1,t_2) \ f_0(t_1,t_2),\\
\end{split}
\end{equation}
where $G_c$ is the Greens function in the conformal limit and $\delta G$ is the first correction to $G_c$ in a derivative expansion. It follows that $f_0$ is an even function of the time difference, an approximate form of which is given in \cite{Maldacena:2016hyu}. Plugging this expansion into \eqref{kform} it follows that 
${\hat K}$ can be expanded as 
\begin{equation} \label{kexp}
{\tilde K}= {\tilde K}_c + \delta \tilde{K}  + ...,
\end{equation} 
where \cite{Maldacena:2016hyu}
\begin{equation} \label{rru}
	\begin{aligned}
		\delta \tilde{K}(t_1,t_2;t_3,t_4)=  \ \tilde{K}_c(t_1,t_2;t_3,t_4)  \left[\frac{q-2}{2}(f_0(t_1,t_2)+f_0(t_3,t_4))+ f_0(t_1,t_3)+f_0(t_2,t_4)\right].
	\end{aligned}
\end{equation}
The first two contributions have their origin in the factors of 
$G^\frac{q-2}{2}$ in \eqref{kform}  and were called rung contributions
in \cite{Maldacena:2016hyu} \eqref{kform}. The remaining two contributions have their origin in the factors of $G$ in \eqref{kform} and were called 
rail contributions in \cite{Maldacena:2016hyu}. We note that for rung contributions $f_0$ appears with either first two times or last two times of the kernel.
On the other hand the two times in rail contributions are one from the first 
set and one from the second.

Our discussion so far has applied to  general fluctuations about the 
saddle point, and has largely been a review of the general  results of 
\cite{Maldacena:2016hyu} with a few extra indices sprinkled in. 
In the rest of this subsection we now focus attention on the specific 
fluctuations of interest to us, namely those generated by the linearized form of \eqref{symo} around conformal solution
\begin{equation} \label{spcmodes}
(g_c)_a^{\ b} (t_1,t_2)= |G_c(t_1,t_2)|^{\frac{q-2}{2}}G_c (t_1,t_2)\left[H_a^{\ b} (t_1)- H_a^{\ b} (t_2)\right].
\end{equation}
Notice that the fluctuations \eqref{spcmodes} represent the change of the 
propagator under a time dependent $O(N)^{q-1}$ rotation. The form 
of \eqref{spcmodes} is similar in some respects to the variation of the 
propagator under diffeomorphisms, studied in \cite{Maldacena:2016hyu}, with 
one important difference; the factors of $H^b_a(t_1)$ and $H^b_a(t_2)$ 
appear with a relative negative sign in \eqref{spcmodes}, whereas the 
infinitesimal diffeomorphism fields in the light fluctuations of 
 \cite{Maldacena:2016hyu} appeared with a relative positive sign in 
\cite{Maldacena:2016hyu}. The fact that our fluctuations are `antisymmetric' 
rather than` symmetric' will play an important role below. 

Specialising to this particular fluctuation, It can be shown (see Appendix \ref{ConformalKernel}) that \textit{$g_c$ is an eigenfunction of $\tilde{K}_c^{-1}$ with eigenvalue $|J|^2$} more clearly
\begin{equation} \label{eval}
	\begin{aligned}
	\int dt_3 \ dt_4 \ 	\tilde{K_c}^{-1}(t_1,t_2;t_3,t_4)(g_c)_a^{\ b}(t_3,t_4)=|J|^2 \ (g_c)_a^{\ b}(t_1,t_2).
	\end{aligned}
\end{equation}
It follows immediately from \eqref{eval} that 
\begin{equation}
\label{conformaleq}
	\begin{aligned}
		\frac{1}{2} \ g_c \ \tilde{K}_c^{-1} \ g_c=\frac{|g|^2}{q} \ g_c \  g_c.
	\end{aligned}
\end{equation}
Using this equation it may be verified that for the 
for the particular fluctuations under study- the second line of 
\eqref{gsqa} simply cancels the part of the term in the first line 
obtained by replacing ${\tilde K}$ with ${\tilde K}_c$. 

It follows that the action \eqref{gsqa} evaluated on the 
modes \eqref{spcmodes} is nonzero only because $K^{-1}$ differs from 
$K_c^{-1}$. Recall $K= K_c+ \delta K$ (see \eqref{kexp}). Using  $\delta K ^{-1}=-K \delta K K^{-1}$ that the action for our special modes evaluates at quadratic
order to 
\begin{equation}
	\begin{aligned}
		S_{\text{eff}}=-\frac{1}{2}g_c \ \tilde{K}_c^{-1} \ \delta \tilde{K} \ \tilde{K}_c^{-1} \ g_c.
	\end{aligned}
\end{equation}
Using the fact that $\tilde{K}^{-1}$ is hermitian (\cite{Maldacena:2016hyu}) 
and the eigenvalue equation \eqref{eval}, the action simplifies to 
\begin{equation}
\label{actionsimplified}
	\begin{aligned}
		S_{\text{eff}}=-\frac{1}{2}|J|^4 \int dt_1..dt_4 \ (g_c)_a^{\ b} (t_1,t_2) \ \delta \tilde{K}(t_1,t_2;t_3,t_4) \ \ (g_c)_b^{\ a}(t_3,t_4).
	\end{aligned}
\end{equation}
Plugging the specific form of our fluctuations \eqref{spcmodes} into this 
expression we find  \footnote{Here factors of $N$ comes from trace over other colour index $\delta$-functions that multiply $H_l$ of any colour. }
\begin{equation} \label{corre}
	\begin{aligned}
		S_{\text{eff}}=-\frac{1}{2}  N^{q-2} \sum_{l=1}^{q-1} \sum_{(i,k)  \textit{ pair}}(-1)^{i-k} \int dt_i \ dt_k \ (H_l)_a^{\ b}(t_i)(H_l)_b^{\ a}(t_k)L_{ik}(t_i,t_k),
	\end{aligned}
\end{equation}
where $i \in (1,2)$, $k \in (3,4)$ and
\begin{eqnarray} \label{dem}
		L_{ik}(t_i,t_k)&=&\int A(t_1,..,t_4)\prod_{m \neq i, m \neq k}dt_m, \nonumber\\
		A(t_1,..t_4)&=&|J|^4 \ G_c(t_1,t_2) |G_c(t_1,t_2)|^{\frac{q-2}{2}}  
		\delta \tilde{K}(t_1,t_2;t_3,t_4)
		    |G_c(t_3,t_4)|^{\frac{q-2}{2}}G_c(t_3,t_4).~~~~~ 
\end{eqnarray}
The expression \eqref{corre} is not yet completely explicit, as $L_{ik}$ in 
\eqref{dem} is given in terms of $\delta K$ which is given in terms of 
the first correction to the conformal propagator $G_c$ which, in turn,
is not explicitly known. Luckily $\delta G$ can be eliminated from 
\eqref{corre} as we now demonstrate. \footnote{
Using the fact that $g_c$ is an eigenfunction of $\tilde{K}_c$ with eigenvalue $\frac{1}{|J|^2}$ rung contributions can easily be summed up to 
\begin{equation}
	\begin{aligned}
		S_{\text{eff}}^{\text{rung}}=-\frac{1}{2}  (q-2)\frac{1}{|J^2|}\int (g_c)_a^{\ b} (t_1,t_2) \ f_0(t_1,t_2)\ (g_c)_b^{\ a}(t_1,t_2) \ dt_1 dt_2.
	\end{aligned}
\end{equation}
This expression is not by itself useful as the integral that appears in it has a  $\log$ divergence once numerically determined form of $f_0(\tau_1,\tau_2) \xrightarrow[|\tau_1-\tau_2|\rightarrow 0]{}\frac{1}{|\tau_1-\tau_2|}$ (from \cite{Maldacena:2016hyu}) is used; follows from 
\begin{equation}
g_c(\tau_1,\tau_2) = |G_c(\tau_1,\tau_2)|^{\frac{q-2}{2}} G_c(\tau_1,\tau_2)[H(\tau_1)-H(\tau_2)] 
\xrightarrow[|\tau_1-\tau_2|\rightarrow 0]{} \frac{\textrm{sgn}(\tau_1-\tau_2) }{|\tau_1-\tau_2|}H'(\tau_1)(\tau_1-\tau_2)\sim O(|\tau_1-\tau_2|^0).
\end{equation}
 }

While we do not know the explicit form of the correction to the 
conformal two-point function $\delta G(t_1,t_2)$, we know that it 
satisfies the equation 
 \begin{equation}
 \label{perturbationeq}
 	\begin{aligned}
 		\Sigma_c*\delta G+\delta \Sigma*G_c+s*G_c=0.
 	\end{aligned}
 \end{equation}
This is simply the gap equation expanded around the conformal point.  
Here $s(t_1,t_2)=-\frac{\partial}{\partial t_1}\delta(t_1-t_2)$ is a 
local differential operator.

In order to make the expression \eqref{corre} explicit we first simplify the formulae \eqref{dem} for $L_{ij}$. Plugging the expansion $G= G_c+ \delta G$ into \eqref{kform}, and using properties of conformal solutions, it may be verified after some algebra that for odd $i-k$ \footnote{Here overall factor of 2 comes from symmetry of the integrations and $\frac{q-2}{2}$ comes from rung part.}
\begin{equation} \label{ole}
	\begin{aligned}
		L_{ik}(t_i,t_k)=2 \ \delta(t_i-t_k)\left[\frac{q-2}{2}G_c* \frac{\delta \Sigma}{q-1} +\Sigma_c*\delta G\right](t_i,t_k).
	\end{aligned}
\end{equation}
 The fact that $L_{ik}$ is proportional 
to a $\delta$ function establishes that the contribution of terms with 
odd $i-k$ to the action is local. \eqref{ole} may be further simplified using the relation
\begin{equation}\label{DeltaIdentity}
	\begin{aligned}
		\delta(t_i-t_k)G_c* \frac{\delta \Sigma}{q-1}(t_i,t_k)= \delta(t_i-t_k)\Sigma_c* \delta G(t_i,t_k),
	\end{aligned}
\end{equation}
and to give
\begin{equation} \label{ole1}	
		L_{ik}(t_i,t_k)=q\delta(t_i-t_k)\Sigma_c*\delta G (t_i,t_k).
\end{equation}
Multiplying $\delta$-function on both sides of \eqref{perturbationeq} and using \eqref{DeltaIdentity},  we find
\begin{equation} \label{ole2}	
		L_{ik}(t_i,t_k)=-\delta(t_i-t_k)s*G_c (t_i,t_k)= \delta(t_i-t_k)\frac{\partial}{\partial t_i} G_c(t_i,t_k).
\end{equation}
On the other hand when  $i-k$ is even, using properties of conformal solutions \footnote{As before $\frac{q-2}{2}\times 2$ comes from rung part.}
\begin{equation}\label{lnex}
L_{ik}(t_i,t_k) = -\left[\frac{q-2}{2}\times 2 +1\right]\Sigma_c(t_i,t_k) \delta G(t_i,t_k)+(\Sigma_c*\delta G*\Sigma_c)(t_i,t_k)G_c(t_i,t_k).
\end{equation}
\eqref{lnex} can be further simplified by substituting 
\begin{equation}
	\begin{aligned}
		\Sigma_c*\delta G*\Sigma_c=\delta \Sigma +s,
	\end{aligned}
\end{equation}
and then using the linearized form of the gap equation 
\begin{equation}\label{sma}
\delta \Sigma \ G_c = (q-1) \delta G \ \Sigma_c,
\end{equation} 
to give
\begin{equation}
	\begin{aligned}
		L_{ik}(t_i,t_k)=- G_c(t_i,t_k)\frac{\partial}{\partial t_i}\delta(t_i-t_k).
	\end{aligned}
\end{equation}
Adding together the contributions of $i-k$ even and $i-k$ odd we have a manifestly local effective action, whose structure accounts for the fact that we have 
worked beyond the purely conformal limit (recall that in the purely conformal 
limit our fluctuation action simply vanished) even though the final expression
makes no reference to the explicit form of the correction $\delta G$ to the 
conformal propagator $G_c$.
\begin{equation} \label{seffaf}
	\begin{aligned}
		S_{\text{eff}}=&-N^{q-2}\sum_{l=1}^{q-1}\int dt_i \ dt_k G_c(t_i-t_k)\delta(t_i-t_k) \Tr \ \left( \frac{\partial}{\partial t_i} H_l(t_i)H_l(t_k)\right)\\
	\end{aligned}
\end{equation}
Expanding $H_l(t_k)$ in a Taylor series expansion about $t_i$ 
$$H_l(t_k)= \sum_{n=0}^\infty \frac{\partial^n}{\partial t^n} H_l(t_i)
\frac{(t_k-t_i)^n}{n!}$$
allows us to recast \eqref{seffaf} into the form 
\begin{equation} \label{effactaaf} 
S_{\text{eff}}=-N^{q-2}\int dt \ \sum_{l=1}^{q-1} \sum_{n=0}^{\infty}C_n \ Tr \ \left( \frac{\partial}{\partial t} H_l(t) \ \frac{\partial^n}{\partial t^n} H_l(t) \right).
\end{equation}
where
\begin{equation} \label{cn}
	\begin{aligned}
		C_n=\frac{1}{n!}\int dt \  G_c(t)\delta(t)t^n.
	\end{aligned}
\end{equation}
 The term in the sum \eqref{effactaaf} with 
$n=0$ is a total derivative and so can be ignored. It follows that 
\begin{equation} \label{effact} 
S_{\text{eff}}=-\int dt \ \sum_{l=1}^{q-1} \sum_{n=1}^{\infty}C_n \Tr \ \left( \frac{\partial}{\partial t} H_l(t) \ \frac{\partial^n}{\partial t^n} H_l(t) \right).
\end{equation}
 Our final result \eqref{effact} for the effective action, has now been 
arranged as an expansion over terms with increasing numbers of derivatives. 

Recall that all the results of this section have been obtained after expanding 
the Greens function
\begin{equation}\label{expgf}
G(t_1, t_2)= G_c(t_1, t_2)+ \delta G(t_1, t_2),
\end{equation}
and assumed that $\delta G \ll G_c$. This assumption is only valid when 
$t_1 - t_2 \gg \frac{1}{J}$, but are not valid for 
$t_1 -t_2 \sim \frac{1}{J}$. All potential non localities in the 
effective action for $H$ presumably have their origin 
in regions where our approximations are valid. It thus seems plausible 
that the central result of this section - namely the absence of nonlocalities in the effective action on length scales large compared to $\frac{1}{J}$
- which therefore takes the form \eqref{effact} - is a reliable result.

On the other hand the precise expressions for the coefficient functions $C_n$ 
involve integrals over a function - namely the delta function - which varies 
over arbitrarily small distances - and so is not reliable (it uses our 
approximations in a regime where they are not valid). We would expect 
the correct versions of \eqref{cn} to be given by smeared out versions of 
the integrals in \eqref{cn}. On general dimensional grounds it follows that 
\begin{equation}\label{rcn}
C_n \rightarrow \frac{A_n}{|J|^n}.
\end{equation}
We will make the replacement \eqref{rcn} in what follows. The numbers 
$A_n$ could presumably be computed by studying four point correlators 
of appropriate operators at finite temperature. We will not attempt this 
exercise in this paper. 

For the purposes of long time physics we are interested only in the term with 
the leading number of derivatives, i.e. with the term with $n=1$ in 
\eqref{effact}. The coefficient of our action in this case is proportional 
to $A_1\equiv {\cal A}$. \footnote{Note that 
\begin{equation} \label{expct} 
C_1=  \int dt \ \delta (t) G_c(t) \ t. 
\end{equation} 
Plugging the formula  
\begin{equation} \label{gc}
G_c= b\frac{ {\rm sgn} (t)}{ |Jt |^{\frac{2}{q}}},\\
\end{equation}
into \eqref{expct} we find, formally, that 
\begin{equation}  
C_1 \propto \int dt |t|^{1-\frac{2}{q}} \delta(t)=0,  
\end{equation} 
(where we have used the fact that $q>2$).
As explained above, we expect that the 
vanishing of $C_1$ is not a physical result but rather is
a consequence of inappropriate use of approximations. We 
assume that $C_1 \rightarrow \frac{\cal A}{|J|}$ in what follows
where ${\cal A}$ is an unknown dimensionless number.}
and the effective action of our theory at leading order in the 
derivative expansion takes the form
\begin{equation}
	\label{effectiveactioninfinitesimal}
	\begin{aligned}
		S=-\mathcal{A} \ \frac{N^{q-2}}{|J|}   \int dt \ \sum_{l=1}^{q-1} \ \Tr \ \left( \frac{\partial}{\partial t} H_l(t) \ \frac{\partial}{\partial t} H_l(t) \right).
	\end{aligned}
\end{equation}

In the analysis presented so far we have determined the form of the effective 
action for infinitesimal group rotations $H$. The group invariant extension 
of our result to finite group rotations is the sigma model action  
\begin{equation}
	\label{effectiveactiont}
	\begin{aligned}
		S=-\mathcal{A} \ \frac{1}{|J|}   \int dt \ \  \sum_{l=1}^{q-1} \Tr \ \left[ \left( V_l^{-1}(t)\ \frac{\partial}{\partial t} V_l(t) \right)^2 \right],
	\end{aligned}
\end{equation}
where $V_l \in SU(N)$ whose infinitesimal form is $V_l=1+H_l+\mathcal{O}(H_l^2)$. 
\eqref{sma} is simply the action for a free particle moving on the group 
manifold $O(N)^{q-1}$ \footnote{Non-trivial holonomy can be turned on for these new light modes, details of contribution of these light modes to effective action for holonomy is presented in Appendix  \ref{sec5}.}. As explained in the introduction, 
the structure of this action could have been anticipated on general grounds. 
The fact that the action is proportional to $\frac{1}{J}$ follows largely 
on grounds of dimensional analysis. 

As we have already seen in the introduction,  once we have established that 
the action for $V(t)$ is local the form of the low energy effective action 
\eqref{effectiveaction} for our system is almost inevitable using the general 
principles of effective field theory. The main accomplishment of the algebra
presented in this section is the demonstration that the effective action 
for $V(t)$ is, indeed, local.

Note that the Sigma model action \eqref{sma} has an 
$O(N)^{q-1} \times O(N)^{q-1}$ global symmetry under which 
\begin{equation} \label{gsym}
V_l \rightarrow A V_l B,
\end{equation}
where $A$ and $B$ both belong to $O(N)^{q-1}$. The rotations by $A$ are 
simply the global symmetry that the microscopic SYK model possesses. Rotations 
by $B$ are an emergent symmetry of the low energy effective action. 
The corresponding conserved quantities are 
$L_l=\dot{ V_l} V_l^{-1}$, and $R_l=V_l^{-1} \dot{ V_l}$ 
\footnote{A dot over a quantity indicates derivative with respect to time.}. 
Choosing a basis $(T_a)$ \footnote{It is assumed in what follows that this basis puts the Killing form in a form proportional to identity.} 
of Lie algebra $\mathcal{O}(N)$ it can be shown that Hamiltonian vector fields corresponding to group functions $L_{l,a}=Tr \ (T_a L_l) $, $R_{l,a}=Tr \ (T_a R_l) $ give two copies of $\mathcal{O}(N)$ (at both classical and quantum level), 
both of which commutes with the Hamiltonian which is the quadratic 
Casimir of the algebra.

\section{Holonomy dynamics and density of states at large mass}\label{sec3}

We now switch gears; in this section and next we discuss 
a the mass deformed SYK theory \eqref{sykqmmd} in the large mass 
limit. We work with the theory based on the $O(N)^{q-1}$ symmetry 
where this symmetry is gauged. The large mass limit is of interest because
it allows us to focus on the dynamics of the holonomy at finite temperature, 
and also allows us to compute the growth of states in the theory as a 
function of energy in a very simple setting. 

\subsection{Scaling limit}

As explained in the introduction, in this section we will compute the finite
temperature partition function 
$$ Z= \Tr \ x^{\frac{H}{m}},$$
for the mass deformed gauged $O(N)^{q-1}$ melonic theory \eqref{sykqmmd}.

In the large mass limit all fields in \eqref{sykqmmd} except the holonomies 
of the gauge group can be integrated out at quadratic order. The result of 
this integration is easily obtained using the formulae of 
\cite{Aharony:2003sx}, and is given by \eqref{freesykmd}.

Notice that the effective action $S_{\text{eff}}(U_i)$ presented in \eqref{freesykmd} 
is invariant under the global `gauge transformations'  
$U_i \rightarrow V_i U_i V_i^{-1}$ for arbitrary orthogonal 
matrices $V_i$. This invariance may be used to diagonalize each $U_i$. 
The integral in \eqref{freesykmd} 
may then be recast as an integral over the eigenvalues of each of the 
holonomy matrices $U_i$ with the appropriate measure. As $U_m$ are each 
unitary, their eigenvalues take the form $e^{i \theta^n_m}$ where $n$ runs 
from $1$ to $N$. We define the eigen value density functions 
\begin{equation} \label{evdf}
\rho_m(\theta)= \frac{1}{N} \sum_{n=1}^N \delta( \theta- \theta_m^n).
\end{equation} 
As we are dealing with orthogonal matrices, the eigenvalues of our matrix
occurs in equal and opposite pairs $(\theta_a, -\theta_a)$ and 
so the eigenvalue density function defined in \eqref{evdf} is an even 
function.

As usual the rather singular looking sum over delta functions in 
\eqref{evdf} morphs into an effectively smooth function at large $N$  
as the individual eigenvalues merge into a continuum.  Note that 
\begin{equation} \label{evdfmd}
\frac{ \Tr U_m^n}{N}= \frac{ \sum_{j=1}^N e^{i n \theta_m^j}}{N}= 
\int \rho_i(\theta)e^{i n \theta}  \equiv \rho_m^n,
\end{equation} 
where the last equality defines the symbol $\rho_i^n$. 
Note that the subscript $m$ on $\rho$ runs from $ 1 \ldots q-1$ and 
labels the $O(N)$ factor under study, while the superscript $n$ runs 
from $1 \ldots \infty$ and labels the Fourier mode of the eigenvalue 
distribution. Using the fact that $\rho_i(\theta)=\rho_i(-\theta)$ it follows that 
\begin{equation} \label{altrin}
\rho_i^n = \int d \theta \rho(\theta) \cos n \theta. 
\end{equation}
It follows that $\rho_i^n$ are all real numbers and that $\rho_i^n=\rho_i^{-n}$.

In the large $N$ limit the  integral over the eigenvalues 
$\theta_m^n$ may be recast, in the large $N$ limit into a path 
integral over the eigenvalue functions $\rho_m(\theta)$ given by    
\begin{equation} \label{pfotn}
Z(x)= 
\int \prod_{i=1}^{q-1}D \rho_i 
\exp \left[\frac{1}{2}\sum_{n=1}^\infty\left(    -N^2 \sum_{m=1}^{q-1} \frac{|\rho_m^n|^2}{n}   -2 N_F N^{q-1}   (-x)^n \frac{ \left( \prod_{m=1}^{q-1} \rho^n_m 
\right)  }{n} 
\right) \right], 
\end{equation}  
\footnote{Let us focus on the special case $N_F=1$. In this case the Hilbert 
space of our quantum mechanical problem is simply the sum of $q$ forms of 
the group $O(N^3)$ with $q$ running from $1$ to $N^3$. The exponential in 
\eqref{pfotn} is the character of this Hilbert space w.r.t the subgroup 
$O(N)^3$, with representations coming from $q$ forms in $O(N^3)$ graded by 
$x^q$. (In order to view the exponential as a character one must 
use \eqref{evdfmd}). The integral in \eqref{pfotn} projects onto the singlet 
subspace, and so counts the number of $O(N)^3$ singlets. Note that it was 
very important for this discussions that the fundamental fermions 
in this paper are complex. The 
case of real fermions was studied from this point of view in 
\cite{Krishnan:2017txw}. In this case the Hilbert space of the $N_F=1$ 
theory consists of spinors of $O(N^3)$, and the decomposition of this 
representation content into representations of $O(N)^3$ appears to be 
a very different problem; it was suggested in \cite{Krishnan:2017txw}
that this decomposition contains no singlets. We thank C. Krishnan for 
discusssions on this point. }
where the path integral is now taken over the eigenvalue density functions 
$\rho_m$ with a measure which descends from the flat integration measure 
for individual eigenvalues $\theta_m^j$. As we have only $(q-1)N$ eigenvalues, 
the Jacobian of this variable change is of order $N$ in the exponent and so 
is subleading at large $N$ and will not concern us. 

 Notice 
that the effective action in \eqref{pfotn} is a sum of two kinds of terms; 
those proportional to $N^2$ (we call these terms the contribution of the measure) and those proportional to $N^{q-1}$ (we call these terms the contribution 
of the energy). 
As $q \geq 4$ the energy overwhelms the measure at large $N$ if $x$ is taken 
to be of order unity. In order to work in a regime in which the measure 
and the energy compete with each other we define
\begin{equation}\label{scaling} 
x= \frac{\alpha}{p N^{q-3}},
\end{equation} 
where \footnote{As explained in the introduction, in the free limit we 
could as well study bosons coupled to the gauge field in which case 
we would have $p= N_B+N_F$ where $N_B$ is the number of bosons.} 
$$p= N_F,$$
 and take the limit $N \to \infty$ with $\alpha$ held fixed. In this limit 
the `energy' term with $n=1$ in \eqref{pfotn} is of order $N^2$ and so 
competes with the measure. All energy terms with $n > 1$ are, however, 
subleading compared to the measure and can be dropped at large $N$. In the 
limit under consideration, in other words, the effective action 
in \eqref{pfotn}
simplifies to 
\begin{equation}\label{pfots}
 \begin{split}
&Z(\alpha)= \int  \prod_{i=1}^{q-1}d U_i 
\exp (- S_{\text{eff}} (U_i)), \\
&S_{\text{eff}} = -\frac{\alpha}{N^{q-3}}  \left( 
\prod_{i=1}^{q-1}\Tr U_i \right).  
\end{split} 
\end{equation} 
We will now evaluate the integral \eqref{pfotn} at large $N$ 
with the effective action 
replaced by the simplified effective action \eqref{pfots}.  
In order to facilitate comparison with the matrix model literature, 
it is useful to note that the matrix integral \eqref{pfots} 
is closely related to the following integral over unitary matrices 
\begin{equation}\label{pfotsu}
 \begin{split}
&Z_{SU}(\alpha)= \int  \prod_{i=1}^{q-1}d U_i 
\exp (- S_{\text{eff}} (U_i)), \\
&S_{\text{eff}} = -\frac{\alpha}{N^{q-3}}  \left( 
\prod_{i=1}^{q-1} \Tr U_i  + \prod_{i=1}^{q-1} \Tr U_i^\dagger  \right).  
\end{split} 
\end{equation} 
Where the integral is now taken over unitary matrices. In the large 
$N$ limit the two matrix models have the same gap equation (see below) 
and 
\begin{equation}\label{suso}
\ln Z_{SU}(\alpha)= 2 \ln Z(\alpha).
\end{equation}

\subsection{Determination of saddle points} \label{detsad}

The matrix model \eqref{pfotsu} (and so \eqref{pfots})  is easily solved in the large $N$ limit using the usual 
saddle point method. In order to see how this can be done note that as far as the integral 
over the eigenvalues of $U_1$ are concerned, $\Tr U_2$, $\Tr U_3$ \ldots $\Tr U_{q-1}$ are all constants. 
Focusing only on the integral over $U_1$, \eqref{pfots} reduces to 
\begin{equation}\label{gww} \begin{split}
Z_{SU} & =\int d U_1 \exp \left( \frac{N}{g_1} \left( \Tr U_1 + \Tr U_1^\dagger \right) \right), \\
\frac{1}{g_1}& = \alpha \rho^1_2 \rho^1_3 \ldots \rho^1_{q-1} = \alpha u_2 u_3 \ldots u_{q-1},
\end{split}
\end{equation} 
where in order to lighten the notation we have defined 
\begin{equation}\label{urh}
\rho^1_m=u_m
\end{equation}
A similar statement applies to the integral over all $U_i$ for $i=1 \ldots q-1$. However 
\eqref{gww} is precisely the celebrated Gross Witten Wadia matrix integral 
\cite{Gross:1980he, Wadia:2012fr, Wadia:1980cp}. 
Recall that the saddle point that dominates the integral \eqref{gww} (and its counterparts for $U_2$ etc) is 
given by \cite{Gross:1980he, Wadia:2012fr, Wadia:1980cp}
\begin{equation} \label{trho} 
{\rho}_m( \theta ) = 
\begin{cases}
      \dfrac{1}{2\pi}\left[1+\dfrac{2}{g_m}\cos\theta \right], 
&  g_m \geq 2 ,\ \ |\theta|\le \pi \\
      \dfrac{2}{\pi g_m}\cos\dfrac{\theta}{2}\sqrt{\dfrac{g_m }{2} 
-\sin^2\dfrac{\theta}{2}}, & g_m < 2,\ \ |\theta| < 2\sin^{-1}\left(\dfrac{g_m}{2}\right)^{1/2},
    \end{cases}
\end{equation}
where\footnote{This eigenvalue densities produced above solve the GWW saddle point equations 
$$\frac{2 N}{g_m}  \sin \theta^n_m= \sum_{j \neq n} \cot\left( \frac{\theta^j_m - \theta^n_m}{2}\right),$$
in the large $N$ limit. 
}
\begin{equation} \label{speqsre} \begin{split} 
&\frac{1}{g_m}= \alpha \prod_{j \neq m} u_j.
\end{split}
\end{equation}
Taking the Fourier transform of \eqref{trho} it follows that 
\begin{equation}\label{uexpr}
 u_m =
    \begin{cases}
      \dfrac{1}{g_m}, &  g_m\ge 2  \\
     1- \dfrac{g_m}{4}, & g_m< 2. \\
    \end{cases}
\end{equation}
We refer to the solution $u_m= \frac{1}{g_m}$ as the wavy phase while 
the solution $u_m = 1- \dfrac{g_m}{4}$ as the gapped phase. 

\eqref{speqsre} and \eqref{uexpr} may be regarded as a set of $2(q-1)$ equations for the 
$2(q-1)$ variables $u_m$ and $g_m$. In order to complete the evaluation of our matrix 
integrals we will now solve these equations.

Let us first demonstrate that the variables $g_m$ are either all greater than 2 or 
all less than two simultaneously;  \eqref{speqsre} and \eqref{uexpr} admit no solutions 
in which some of the $g_m$ are greater than 2 while others are less than 2. 
\footnote{Equivalently $u_m$s are either all less than half or all greater than half. Equivalently 
the matrix models for $U_m$ are all simultaneously in the wavy phase or simultaneously in the 
gapped phase.} 

Let us assume that $g_m\geq 2$. It follows from \eqref{speqsre} and \eqref{uexpr} that
\begin{equation} \label{wavy}
\alpha u_1 u_2 \ldots u_{q-1} = \frac{u_m}{g_m} = \frac{1}{g_m^2} \leq \frac{1}{4}.
\end{equation}
On the other hand let us suppose that $g_k <2$ 
Then it follows from  \eqref{speqsre} and \eqref{uexpr} that 
\begin{equation} \label{gapped}
\alpha u_1 u_2 \ldots u_{q-1} = \frac{u_k}{g_k}= \frac{1}{g_k} -\frac{1}{4} > \frac{1}{4}.
\end{equation}
As \eqref{wavy} and \eqref{gapped} contradict each other it follows that either all $g_m \geq 2$ 
or all $g_m < 2$ as we wanted to show. Moreover it follows immediately from \eqref{gapped}
that when all $g_m\leq 2$ they are in fact all equal. Similarly it follows from \eqref{wavy} that
when all $g_m \geq 2$ then once again they are all equal. \footnote{Actually all solutions 
are equal up to sign - however saddle points that differ by sign assignments are actually 
essentially identical - they can be mapped to each other by $U \rightarrow -U$, so we 
ignore this issue.} It follows that in either case all $u_m$ and all $g_m$ are equal. 
Let us refer to the common saddle point value of $u_m$ as $u$. The saddle point 
equations \eqref{uexpr} now simplify to 
\begin{equation}\label{uexprs}
 u = 
    \begin{cases}
      \alpha u^{q-2} &  u \leq \frac{1}{2}  \\
     1- \dfrac{1}{4 \alpha u^{q-2}}, & u > \frac{1}{2}. \\
    \end{cases}
\end{equation}
Once we have determined the solution to \eqref{uexprs} value of the 
partition function \eqref{pfots}, in the large $N$ limit under consideration, 
is given by 
\begin{equation} \label{pffu} \begin{split} 
Z(\alpha)& = \exp \left( - \frac{N^2}{2} V(u) \right), \\
V(u)& = (q-1) f(u) -2 \alpha ~ u^{q-1}, \\
f(u)& = \begin{cases}
u^2, &  u\le \frac{1}{2}  \\
     \dfrac{1}{4} -\dfrac{1}{2}\ln\left[2(1-u)\right], & u> \frac{1}{2}.
\end{cases}
\end{split}
\end{equation}
\footnote{The factor of $\frac{1}{2}$ in the exponent of the first equation 
in \eqref{pfots} is a consequence of the fact that we are working with 
the orthogonal model. The analogous formula for the partition function of the 
unitary model, \eqref{pfotsu}, is the square of the partition function listed 
here and so does not have the factor of half in the exponential.}
Indeed the saddle point equation \eqref{uexprs} is simply the condition 
that the `potential' $V(u)$ in \eqref{pffu} is extremised. In other words
the saddle point solutions of our matrix integral are in one to one 
correspondence with the saddle points (or extrema) of $V(u)$; the contribution
of each saddle point to the matrix integral is simply given by 
$e^{-N^2 \frac{V(u)}{2}}$.

 \begin{figure}[!h]
\begin{center}
	\includegraphics[scale=0.6]{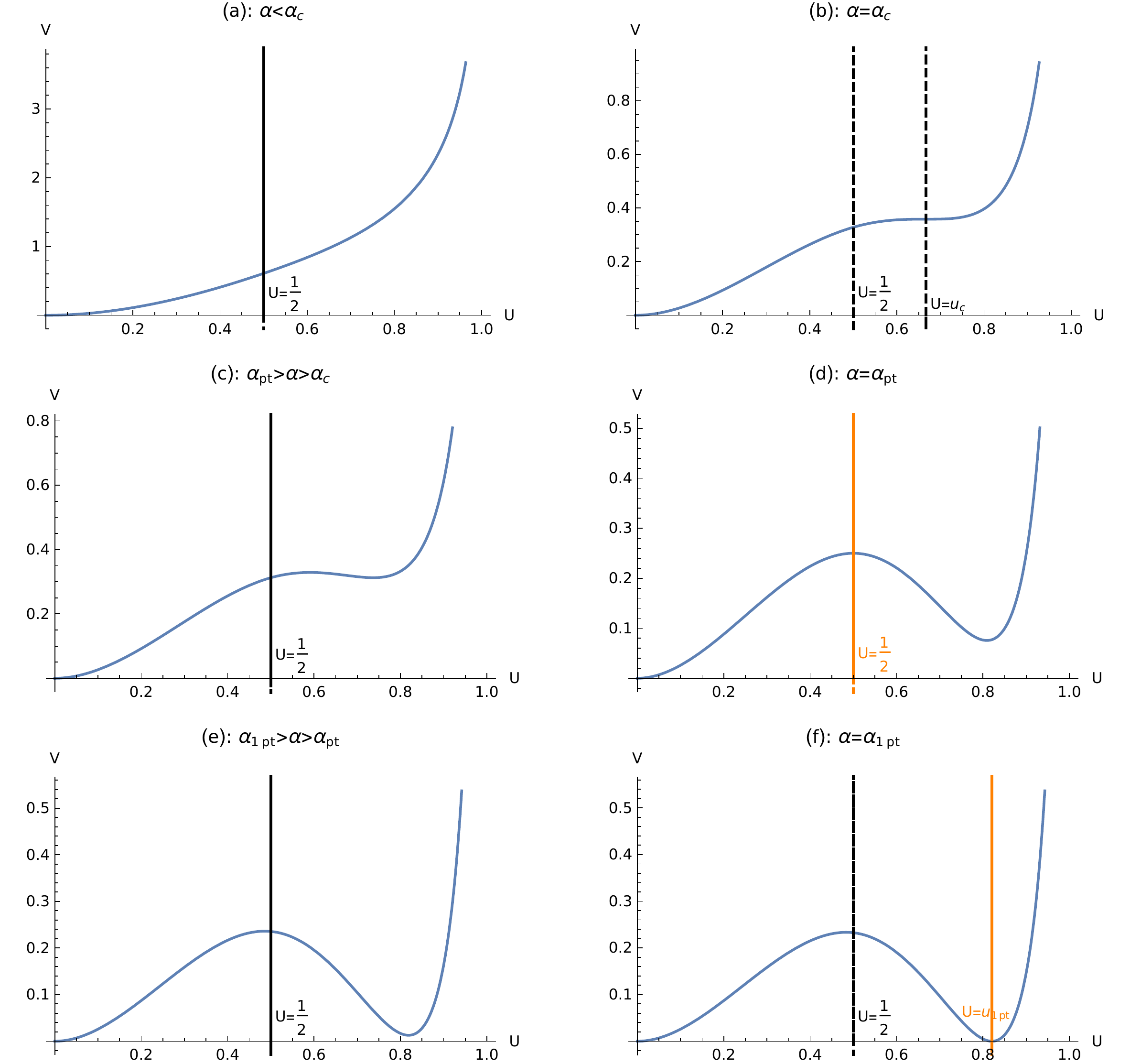}
\end{center}
    \caption{Effective potential for different values of temperature and associated phase transitions. The graphs are drawn for $q=4$.}
\label{vg}
   \end{figure}

At every positive value of $\alpha$, $V(u)=0$ when $u=0$ and $V(u)$ diverges 
as $u$ approaches unity from below. \footnote{Note that 
$u = \frac{ \Tr U}{N} \leq 1$.} However the qualitative behaviour 
of the function $V(u)$ for values between zero and unity depends sensitively 
on $\alpha$. 

It is easily verified that for $\alpha \leq \alpha_c=\frac{(q-1)^{q-1}}{4(q-2)^{q-2}}$ the function $V(u)$ increases monotonically as $u$ increases from $0$ 
to unity (see Fig \ref{vg} (a)). It follows that 
when $\alpha \leq \alpha_c$ the only saddle point lies at $u=0$. In this 
case the saddle point value of the partition function is $Z(x)=1$ 
(see below for a discussion of fluctuations about this saddle point value). 

At $\alpha=\alpha_c$ the potential $V(u)$ develops a point of inflection 
at $u=u_c= \frac{q-2}{q-1}$ (see Fig. \ref{vg} (b)). 
Note that $u_c > \frac{1}{2}$. At this value of $\alpha$ 
we have a new saddle point in the gapped phase. 

As $\alpha$ is increased above $\alpha_c$ the point of inflection at 
$u=u_c$ splits up into two saddle points; a local maximum at 
$u= u_{max} < u_c$ and a local minimum at $u=u_{min}> u_c$ (see Fig. \ref{vg} (c)). To start with 
both saddle points are in the gapped phase. We refer to the saddle point 
at $u_{max}$ as the upper saddle and the saddle point at $u_{min}$ as the 
lower saddle. 

As $\alpha$ is increased further the value of $u_{min}$ continues to 
increase while the value of $u_{max}$ continues to decrease. At 
$\alpha= \alpha_{pt} = 2^{q-3} > \alpha_c$, $u_{max}=\frac{1}{2}$. 
For $\alpha > \alpha_{pt}$,  $u_{max} <\frac{1}{2}$ and the upper saddle makes 
a Gross Witten Wadia phase transition into the wavy phase (see Fig \ref{vg} (d)).
\footnote{
The formula for 
$u_{max}, u_{min}$ as a function of $\alpha$ is complicated in general. However 
the formula simplifies at large $\alpha$ and we find 
\begin{equation}\label{largeu}
u_{max}= \left(\frac{1}{\alpha}\right)^{\frac{1}{q-3}} \ , \
u_{min}= 1+\frac{1}{-4 \alpha +q-2}+\frac{q^2-3 q+2}{2 (-4 \alpha +q-2)^3}+\ldots
\end{equation}}

Finally, when the new saddle point at $u=u_{c}$ is first nucleated, we 
have $V(u_c)>0$. As $\alpha$ is increased $V(u_{min})$ decreases 
below this value. At $\alpha= \alpha_{1pt} = $ we have $V(u_{min})=0$ 
(see Fig \ref{vg}(f)). 
For larger values of $\alpha$,  $V(u_{min})<0$ and our matrix model undergoes
a first order phase transition from the saddle at $u=0$ to the saddle 
at $u=u_{min}$. Note that at $\alpha=\alpha_{1pt}$ (i.e. at the `Hawking 
Page transition temperature') the saddle at $u=u_{max}$ is already in the 
the wavy phase when $q=4$ but is still in the gapped phase for $q>4$.

\subsection{Thermodynamics in the canonical ensemble}

The thermodynamics of our system in the canonical ensemble
follows immediately from the nature of the function $V(u)$ as a function 
of $\alpha$ described at the end of the last section. For convenience we 
discuss the phase diagram of our system as a function of $\alpha$ rather 
than temperature (recall that $\alpha$ is defined by the relations 
$e^{- \beta m}=x = \frac{\alpha}{p N^{q-3}}$). 

For $\alpha< \alpha_c$ the saddle at $u=0$ is the only saddle point in the 
theory (see Fig \ref{vg} (a)). For $\alpha_{c} < \alpha < \alpha_{pt}$ 
\footnote{In the text of this paragraph and the next we have assumed 
that $\alpha_{pt} < \alpha_{1pt}$ as is the case for $q=4$. For $q\geq 6$ 
the order above is reversed, and the discussion has obvious modifications.}  
there are two additional saddle points at $u=u_{min}$ and $u=u_{max}$ 
with $\frac{1}{2} < u_{max} < u_{min} <1$. The saddle point at $u=u_{max}$ 
is a local maximum and $V(u_{max})>0$ (see Fig \ref{vg} (c)).  The saddle 
point at $u=u_{min}$ is a local minimum and however $V(u_{min}) >0$. 
Both these saddles are subdominant compared 
to the flat saddle in this range of 
$\alpha$.

For $\alpha_{pt} < \alpha < \alpha_{1pt}$ the two new phases 
continue to be subdominant compared to the phase at $u=0$; in this range, 
however, the solution at $u=u_{max}< \frac{1}{2}$ is now in the wavy phase
(see Fig \ref{vg} (e)).

At $\alpha= \alpha_{1pt}$ we have $V(u_{min})=0$. For $\alpha> \alpha_{1pt}$
$V(u_{min})<0$, so the solution at $u=u_{min}$ is the dominant saddle point. 
Our system undergoes a phase transition at $\alpha=\alpha_{1pt}$ 
(see Fig \ref{vg} (e)). 
The value of $\alpha_{1pt}$ is given as a function of $q$ by 
\begin{equation}
\label{alpha1pt}
\begin{aligned}
	\alpha_{1pt}= \frac{1}{4} (q-1)w \left[1-\frac{1}{(q-1) w}\right]^{-(q-2)}, \ \  w=-W_{-1}\left[-\frac{2 \exp\left[{-\frac{(q+1)}{2(q-1)}}\right]}{q-1}\right],	
\end{aligned}
\end{equation}
where $W_{n}$ is the productlog function.

\subsection{Thermodynamics in the microcanonical ensemble} 

In this subsection we compute the density of states as a function of energy 
corresponding to each of the saddle points described in the previous 
subsection. In order to do this we use the thermodynamical relations 
\begin{equation}\label{thermody}
E( \alpha) = \alpha \partial_\alpha \ln Z(\alpha) \, ~~~
S( \alpha) = \left(\ln Z(\alpha) -  E(\alpha) \ln \frac{ \alpha}{N^{q-3} p}\right),
\end{equation}
where $E$ is the eigenvalue of $\frac{H}{m}$.
We invert the first of these equations to solve for $\alpha(E)$, and then 
plug this solution into the second equation to obtain $S=S(E)$. For the 
trivial saddle, the saddle value of $S(E)$ is trivial, so we include the 
contribution of fluctuations around this saddle. 

\subsubsection{The saddle at $u=0$}

The saddle point at $u=0$ exists at every value of $\alpha$. In this 
case the saddle point values of the energy and entropy both vanish 
so the first nontrivial contribution to the thermodynamics comes from the 
study of fluctuations about the saddle point. In this subsection - which is 
a bit of a deviation from the main flow of the (otherwise purely 
saddle point) computations of this paper we describe the relevant computations. 
For the purposes of this subsection - and this subsection only - we 
retreat away from the scaling limit \eqref{tempscal} and work with the full matrix model 
\eqref{freesykmd} - or more precisely with its generalisation 
\eqref{fneq} which allows for bosonic as well as fermionic harmonic 
oscillators. Working with this generalised model we compute the 
fluctuations around the trivial saddle point $\Tr U_m^n=0$, i.e. 
$\rho^n_m=0$. 

For the purposes of studying small fluctuations around this saddle point 
we work with the integral \eqref{pfotn}. The integral \eqref{pfotn} can be simplified by
 making the variable change
\begin{equation} \label{varchange}
\rho_m^n = \frac{\beta_m^n}{N} 
\end{equation} 
The point of the scaling \eqref{varchange} is that it eliminates all 
explicit factors of $N$ from the integral \eqref{pfotn}. It follows 
that - at least for the purposes of the perturbative Wick contraction 
evaluations we perform in this subsection - at any finite order in perturbation 
theory the integral over $\beta^m_n$ receives significant contributions only 
from values of $\beta^m_n$ of order unity. Note however that if $\beta^m_n$
are of order unity then $\rho^m_n$ are of order $\frac{1}{N}$ and so are 
very small. We can thus safely integrate over all values of $\beta^m_n$ 
without worrying about boundaries to the domain of integration. 
\footnote{More generally the variables $\rho^m_n$ are constrained 
by the requirement 
that the function  
$\rho_m(\theta)= \frac{1}{2 \pi } \sum_{n} \rho^n_m e^{-i n \theta},$
is positive for every value of $\theta$. This constraint is trivial when all 
$\rho^n_m$ as is effectively the case for the perturbative evaluations 
discussed above.} 
In other words \eqref{pfotn} may be rewritten in terms of these scaled 
variables 
\begin{equation} \label{fluc} \begin{split} 
&Z(x)   =  \prod_{n=1}^\infty F_n(x),\\
& F_n(x)  = 
\begin{cases} M_n\displaystyle  \int \prod_{m=1}^{q-1} d \beta_m^n  
\exp \left( -\sum_{m=1}^{q-1} \frac{|\beta_m^n|^2}{2n} + N_{F} x^n 
\frac{ \left( \prod_{m=1}^{q-1} \beta^n_m 
\right) + \text{c.c} }{n}  \right) & ~~~n ~~{\rm odd}\\
M_n \displaystyle\int \prod_{m=1}^{q-1} d \beta_m^n      
\exp \left( -\sum_{m=1}^{q-1} \frac{|\beta_m^n|^2}{2n} -N_{F} x^n 
\frac{ \left( \prod_{m=1}^{q-1} \beta^n_m 
\right) + \text{c.c} }{n}  \right) & ~~~n ~~{\rm even}.\\
\end{cases}
\end{split}
\end{equation} 
The expressions for $F_n$ above involve an integral with the usual measure 
$dz d{\bar z}$ for the complex variable $\beta^n_m$. The integral is taken 
over the whole complex plane\footnote{As mentioned above, the 
difference between this measure and $d \theta_m^{j}$ is sub-dominant in large $N$ limit.}.  The $x$ independent normalisation constant 
$M_n$ above are chosen to ensure that normalisation of Haar measure,i.e, $F_n(0)=1$ . 

The expressions for $F_n$ presented in \eqref{fluc} are formal as the 
integrals that define $F_n$ do not converge. However this fact does not 
bother us, as we are not really interested in the the expression for 
$Z(x)$ but only in the coefficients in of $x^{k}$ for each $k$ in that 
expression. Each of these coefficients is easily obtained (by Taylor 
expanding the non Gaussian terms in the integrands in the formulas for 
$F_n$ above and performing all integrals using Wicks theorem. We find 
\begin{equation}\label{fnres}
F_n(x)= 
\sum_{k=0}^\infty x^{2kn} \left( p^2 (2n)^{q-3}  \right)^k (k!)^{q-3}, ~~~ p \equiv N_F,
\end{equation}

Let $E$ denote the eigenvalues of $\frac{H}{m}$; in other words $E$ is the 
energy of our theory in units of the oscillator mass (or frequency). 
It follows from \eqref{fnres} that the functions $F_n(x)$ represent 
the partition function of a system whose entropy  as a function 
of energy is given by $S_n(E)$ where 
\begin{equation}\label{seoe}
e^{S_n(E)}=
\left( \frac{E}{2n}! \right)^{q-3} \left( p^2 (2n)^{q-3} \right)^\frac{E}{2n}.
\end{equation}
At large $E$ ( i.e. when $E \gg 2 n$) we may use Sterling's approximation 
to simplify \eqref{seoe} to obtain the asymptotic formula 
\begin{equation}\label{seoel}
S_n(E)= 
 (q-3)\frac{E}{2n} \ln \left( \frac{E}{2n} \right) + \frac{E}{2n}
\left( -(q-3) + 2 \ln p +(q-3) \ln(2n) \right). 
\end{equation}
Notice that the density of states
 grows faster than exponentially as a function of
energy, explaining the divergence of the integrals that define $F_n$ 
(or, equivalently, explaining why the sums in \eqref{fnres} are divergent 
at every $x$ no matter how small. 

As the partition function of our system is simply the product over the 
functions $F_n$, the entropy of our system at large energies is obtained 
by distributing the available energy $E$ among the various systems $S_n$ 
in such a way as to maximise the entropy. A glance at \eqref{seoel} is 
sufficient to convince oneself that the best one can do is to put all 
available energy into the `system' $S_1$. It follows that for $E \gg 1$, 
the contribution of the saddle point at $u_1=0$ to the entropy of the 
system is 
\begin{equation} \label{content}
S(E)=S_1(E)= (q-3)\frac{E}{2} \ln \left( \frac{E}{2} \right) + \frac{E}{2}
\left( (\ln(2)-1)(q-3) + 2 \ln p \right).
\end{equation}

The saddle at $u=0$ is exceptional in that it is trivial as a saddle point; 
in order to determine the thermodynamics of this `phase' we had to 
perform the one loop expansion about this saddle point. The remaining 
saddle points we will study in this section are nontrivial even at leading 
order, and so will be analysed only within the strict saddle point 
approximation. In the rest of this subsection we also return to the study 
of the strict scaling limit \eqref{tempscal}.

\subsubsection{The wavy phase}

In this subsection we study the thermodynamics of the wavy saddle, i.e. 
the saddle point at $u=u_{max}$ for $\alpha> \alpha_{pt}= 2^{q-3}$. 
The contribution of this saddle point to partition function is 
\begin{equation} \label{contwavy} 
\ln Z(\alpha)= - \frac{N^2}{2} (q-3) \alpha^{- \frac{2}{q-3}}.
\end{equation}
The energy of the corresponding phase is given by 
\begin{equation} \label{enphase1}
E( \alpha) = \alpha \partial_\alpha \ln Z(\alpha) =  \frac{N^2}{\alpha^{\frac{2}{q-3}}},
\end{equation}
Note that the energy is a decreasing function of $\alpha$ so that this 
phase has a negative specific heat. As this phase exists only for 
$\alpha > \alpha_{pt}$ it follows that the energy in this phase is bounded 
from above by 
\begin{equation} \label{Ept}
E_{pt} \equiv E(\alpha_{pt})= \frac{N^2}{4}. 
\end{equation} 
The entropy of this phase is given by 
\begin{equation} \label{entphase1}
S( \alpha) = \left(\ln Z(\alpha) - E(\alpha) \ln \frac{ \alpha}{N^{q-3} p}\right).
\end{equation}
Eliminating $\alpha$ between \eqref{enphase1} and \eqref{entphase1} we 
obtain 
\begin{equation} \label{enten1} 
S(E) = (q-3)\left[ \frac{E}{2} \ln \left( \frac{E}{2} \right) -\frac{E}{2} \right]+E \ \log p + (q-3)\frac{E}{2}\ln (2).
\end{equation} 
Note that \eqref{enten1} is in perfect agreement with \eqref{content}. 
This match strongly suggests that the formula \eqref{enten1} is correct 
for all values of $E$ in the range 
\begin{equation} \label{enrange} \
1 \ll E < \frac{N^2}{4}.
\end{equation}

\subsubsection{The gapped phase}

The analysis of this section applies to the saddle point at $u=u_{max}$ for 
$\alpha \leq \alpha_{pt}$ and to the saddle point at $u_{min}$. 
The partition function of this saddle is given by plugging the solution 
of the equation  
\begin{equation}\label{uppersaddle}
 u = 1- \dfrac{1}{4 \alpha u^{q-2}} \ , \quad u \geq \frac{1}{2}
\end{equation} 
into the formula 
\begin{equation} 
\ln Z= -\frac{N^2}{2} \left[ (q-1)  
\left(  \dfrac{1}{4} -\dfrac{1}{2}\ln\left[2(1-u)\right] 
\right) - 2 \alpha u^{q-1} \right].
\end{equation} 
 As we have explained above, for $\alpha< 
\alpha_c=\frac{(q-1)^{q-1}}{4 (q-2)^{q-2}}$ there are no legal solutions to 
\eqref{uppersaddle}. For $ \alpha_c< \alpha < \alpha_{pt}=2^{q-3}$ there 
are two legal solutions and for $\alpha> \alpha_{pt}$ there is a single 
legal solution to this equation. After the partition function is 
obtained one obtains the energy and entropy of the solution using the 
thermodynamical formulae 
\begin{equation}\label{thermodys}
E( \alpha) =\alpha \partial_\alpha \ln Z(\alpha) \ , ~~~
S( \alpha) = \left(\ln Z(\alpha) - E(\alpha) \ln \frac{ \alpha}{N^{q-3} p}\right).
\end{equation}
Eliminating $\alpha$ from the expressions obtained in \eqref{thermodys} we find 
the entropy $S$ as a function of the energy. This function $S=S(E)$ is 
difficult to find explicitly simply because \eqref{uppersaddle} is 
difficult to solve. The procedure described above, however, implicitly 
defines this function. It is not difficult to convince oneself that there 
is a single saddle point of this nature for every energy 
$E > \frac{N^2}{4}$ and that the function $S(E)$ is an analytic function of 
energy for every energy greater than $\frac{N^2}{4}$. 

While explicit formulae are difficult to obtain in general, they are easy 
to obtain in three special limits which we now describe 
\newline

\underline{\bf A.  The solutions with $\alpha$ near $\alpha_{pt}$ i.e. ($E$ near $E_{pt}$):}
\newline

At $\alpha=\alpha_{pt}$ \eqref{uppersaddle} admits the solution 
$u= \frac{1}{2}$. (This is a solution at $u=u_{max}$, i.e. the 
solutions that is a local maximum). It follows that at $\alpha= \alpha_{pt}- \delta \alpha$, \eqref{uppersaddle} 
admits a solution with $u = \frac{1}{2}+ \delta u$. Here $\delta u$ is solved order by order in $\delta \alpha$. A few lines of standard algebra gives:
\begin{equation} \label{ene}
\begin{split}
S\left( E_{pt} + \delta E\right) =& -\frac{1}{4} N^2 \left[ \log \left(\frac{2^{q-3} N^{3-q}}{p}\right)+\frac{q-3}{2}\right]
-\log \left[\frac{2^{(q-3)/2} N^{3-q}}{p}\right] \delta E \\
&~~~~~~+\frac{2(q-3)}{2 N^2}\left(\delta E \right)^2 +\frac{4(7-3 q)}{6 N^4}\left(\delta E\right)^3+\ldots
\end{split}
\end{equation} 
Comparing \eqref{ene} and \eqref{enten1}, it is easily verified that 
while $S(E)$, $S'(E)$ and $S''(E)$ are continuous at $E= \frac{N^2}{4}$, 
$S'''(E)$ is discontinuous. In that sense the function $S(E)$ has a 
third order phase transition' at $E= \frac{N^2}{4}$. Further taking the limit:
\begin{equation}\label{Sdis}
\lim_{\epsilon\to 0^{+}} S^{'''}\left(\frac{N^2}{4}-\epsilon\right) = \frac{4(6-2q)}{N^4}, \quad \lim_{\epsilon\to 0^{+}} S^{'''}\left(\frac{N^2}{4}+\epsilon\right) = \frac{4(7-3q)}{N^4}
\end{equation}
This discontinuity is a consequence
of the fact that the saddle point undergoes a Gross Witten Wadia 
transition at this energy. 
\newline\\
\underline{\bf B.  The solutions with $\alpha$ near $\alpha_c$ (i.e. $E$ near $E_c$):}
\newline

At $\alpha=\alpha_c$ \eqref{uppersaddle} admits the solution $u=\frac{q-2}{q-1} $.
For $\alpha= \alpha_c + \delta \alpha$ the \eqref{uppersaddle} admits 
two solutions near this critical solution at $u=u_{c} + \delta u$; these 
are the solutions at $u=u_{max}$ and $u=u_{min}$ respectively. A careful calculation shows $E,S$ as a function of $\alpha$ are different for this two branches but $S$ as a function of $E$ is same for both of them and given by:
\begin{equation} \label{enen}
\begin{split}
S\left( E_c + \delta E\right) =&  \frac{1}{4} N^2 \left[- (q-2) \log \left(\frac{(q-2)^{2-q} (q-1)^{q-1} N^{3-q}}{4p}\right)+ (q-1) \log \left(\frac{2}{q-1}\right)+\frac{(q-3)}{2}\right] \\
&-\log \left[\frac{(q-2)^{2-q} (q-1)^{q-1} N^{3-q}}{2^{(q+1)/2}p}\right]\left(\delta E\right) + \left[-\frac{8}{3 N^4 (q-2) (q-1)}\right]\left(\delta E\right)^3+\ldots
\end{split}
\end{equation} 
Note that \eqref{enen} is completely smooth around $E=E_c=\frac{1}{4} N^2 (q-2)$. 
\newline\\
\underline{\bf C. The solutions with $\alpha \gg 1$ (i.e. $E \gg \frac{N^2}{2} $):}
\newline

At $\alpha \gg 1$ \eqref{uppersaddle} admits the solution near $u=1$; 
this is the thermodynamically dominant saddle at $u=u_{max}$. 
Setting $u= 1- \delta u$, $\delta u$ is solved to give as series in $\frac{1}{\alpha}$:
\begin{equation}\label{dunn}
\delta u =  \left( \frac{1}{4}\right)\alpha^{-1}  +\left( \frac{q-2}{16}\right) \alpha^{-2}+ \ldots
\end{equation} 
It follows that: 
\begin{equation} \label{cnhnn} \begin{split}
\ln Z(\alpha) &=  N^2 \alpha+\left( -\frac{1}{4} N^2 (q-1)\right)  \log (\alpha)+\ldots,\\
E( \alpha) & = N^2 \alpha  +\left( -\frac{N^2 (q-2) (q-1)}{32  }\right)\alpha +\ldots, \\
S( \alpha) & = \left( - N^2\right) \alpha \log(\alpha)+\left(  N^2  \left(1+ \log \left(2 pN^{q-3}\right)\right)\right) \alpha+\ldots 
\end{split}
\end{equation}
which gives 
\begin{eqnarray} \label{eneg}
S(E) &=&  E-E\log \left( \frac{E}{pN^{q-1}}\right)- \frac{N^2}{2}\bigg[\frac{q-1}{2}\log \left(\frac{2 E}{N^2}\right)+\frac{3}{4}(q-1) +\frac{1}{8}(q-1)\left(\frac{2 E}{N^2}\right)^{-1}  \bigg] +\ldots~~~~~~~~~
\end{eqnarray}

\subsubsection{Entropy as a function of energy for $E \gg \frac{N^2}{2}$}

We have verified above that for $E \gg \frac{N^2}{2}$ the saddle point for the 
eigenvalue distribution function becomes very peaked and so is well 
approximated by a delta function. Whenever the eigenvalue distribution 
becomes so peaked effect of the holonomies on the partition function
of the system can be ignored. It follows that for energies much greater 
than $N^2$ the partition function of our system is simply that 
of  $N_F N^{q-1}$ complex fermionic oscillators.
The partition function for our system thus reduces to 
\begin{equation}\label{nbnf}
\ln Z(x) = N_F N^{q-1} \ln (1+x), 
\end{equation}
For $x \ll 1$ \eqref{nbnf} reduces to 
\begin{equation}\label{nbnfa}
\ln Z(x) = ~x~p~N^{q-1}.
\end{equation}
Substituting $x= \frac{\alpha}{N^{q-3} p}$ we find that \eqref{nbnfa}
agrees precisely with the leading term in the first line of \eqref{cnhnn}: \begin{equation} \label{contla} 
\ln Z(\alpha)= N^2 \alpha.
\end{equation}
The energy of the corresponding phase is given by 
\begin{equation} \label{enphase2}
E( \alpha) = \alpha \partial_\alpha \ln(Z(\alpha)) =  N^2 \alpha. 
\end{equation}
The entropy of this phase is given by 
\begin{equation} \label{entphase2}
S( \alpha) = \ln Z(\alpha) -E(\alpha) \ln \frac{ \alpha}{N^{q-3} p}
=  N^2\left(1-\log \left(\frac{\alpha}{pN^{q-3}}\right)\right)\alpha.
\end{equation}
Eliminating $\alpha$ between \eqref{enphase2} and \eqref{entphase2} we 
obtain 
\begin{equation} \label{enten2} 
S(E) = E \left(1-\log \left(\frac{E}{pN^{q-1}}\right)\right).
\end{equation} 
Note that \eqref{enten2} matches with the leading and 1st subleading term in \eqref{eneg}.

\section{The holonomy effective action with weak interactions} \label{sec4}

In the previous section we studied the free energy of the mass deformed 
SYK model in the zero coupling $\frac{J}{m} =0$. In this section 
we will study corrections to the results of the previous section in a 
power series expansion in the coupling constant. 
For simplicity we also study the special case $N_F=1$ in \eqref{sykqmmd} . 

In principle the leading large $N$ contribution to $S_{\text{eff}}$ is given as follows
(we restrict attention to the massless case for simplicity in this paragraph).
Consider the gap equation \eqref{LargeNeq}. 
We are instructed to solve this gap equation 
on a thermal circle, subject to the requirement that the solution respect 
the boundary conditions 
\begin{equation}\begin{split}\label{bcs}
&G\left(t_1+ \frac{\beta}{2}, t_2\right)=  -U G\left(t_1- \frac{\beta}{2}, t_2\right)\\
&G\left(t_1, t_2 + \frac{\beta}{2}\right)=   -G\left(t_1, t_2- \frac{\beta}{2}\right) U^{-1}\\
&\Sigma\left(t_1, t_2+ \frac{\beta}{2}\right)=   -U
\Sigma\left(t_1, t_2- \frac{\beta}{2}\right) \\
&\Sigma\left(t_1+ \frac{\beta}{2}, t_2\right)=  -\Sigma\left(t_1- \frac{\beta}{2}, t_2\right) U^{-1}\\
\end{split}
\end{equation}
We must then plug this solution into \eqref{genact} and the corresponding  
result is represented by $S_{\text{eff}}(U).$ While this prescription is clear it is rather difficult to implement 
in practice. In order to get some intuition for the effect of interactions
on $S_{\text{eff}}(U)$ present some perturbative results for this object. 

The thermal partition function of theory \eqref{sykqmmd} is given, 
as usual, by the Euclidean path integral of the theory on a thermal circle 
of circumference $\beta$. The free result \eqref{freesykmd}
is obtained by integrating out all fermions at 
at `one loop' (i.e. by computing fermionic determinants -we explain how this 
works in more detail below). Corrections to \eqref{freesykmd} are obtained 
by including the contribution of more general diagrams. 

It was demonstrated in \cite{Witten:2016iux} that, in the strict large $N$ limit 
of interest to this paper, the only graphs that contribute are melonic 
graphs. One way of organising the graphs that contribute to our computation 
is by the number of melons a graph contains. We will refer to a graph 
with $n$ melons as an $n^{th}$ order graph. Such graphs are proportional 
to $J^{2n}$. As in the previous section we will be interested in the effective action as a function of holonomies, $S_{\text{eff}}(U)$. Let the contribution to 
$S_{\text{eff}}(U)$ from graphs of $n^{th}$ order be denoted by $S_n(U)$. We have 
\begin{equation}\label{ssnq}
S_{\text{eff}}(U)= \sum_{n=0}^\infty S_n(U).
\end{equation} 
As in the previous section we will principally be interested in the partition 
function in the scaling limit \eqref{tempscal}. In this limit the temperature is very small and so $\beta$ is 
very large $\beta \sim \ln N$. For this reason it is important to keep track 
of explicit multiplicative factors of $\beta$ 
(as opposed, for instance, to factors of $x=e^{-\beta m}$) 
in our results. Below we will demonstrate that $n^{th}$ order graphs have at 
least one and at most $n$ explicit multiplicative factors of $\beta$. It 
follows 
that the contributions of $n^{th}$ order graphs to the effective action
can be organised in series
\begin{equation}\label{nogs}
S_n(U)=  \frac{J^{2n} \beta^n}{m^n} 
\sum_{a=0}^{n-1} \left( \frac{1}{m \beta} \right)^{a}
f^n_a(x, U) \equiv  -\left(\frac{J }{m}\right)^{2n}F_{2n} (m \beta,x,U). 
\end{equation}
Substituting \eqref{nogs} into \eqref{ssnq}, we can rearrange the sum over 
graphs as 
\begin{equation}\label{ssn}\begin{split}
S_{\text{eff}}(U)&= \sum_{k=0}^\infty \left( \frac{J}{m} \right)^{2k} 
H_k( \frac{J^2 \beta}{m}, x, U),\\
 H_k( \frac{J^2 \beta}{m}, x, U)&= \sum_{n=k}^\infty 
\left( \frac{J^2 \beta }{m}\right)^{n-k} f_k^n(x, U).
\end{split}
\end{equation}
As we are interested in the scaling limit \eqref{tempscal} it follows that: 
\begin{equation}\label{tildeq} \begin{split} 
&H_k\left( \frac{J^2 \beta}{m}, x, U\right)= {\tilde H}_k\left(\frac{J^2 \beta}{m}\right)
 x \prod_{i=1}^{q-1} \Tr U_i, \\
&f_k^n(x, U)=  f_k^n  x \prod_{i=1}^{q-1}\Tr U_i, \\
& H_k\left( \frac{J^2 \beta}{m}\right)= \sum_{n=k}^\infty
\left( \frac{J^2 \beta }{m}\right)^{n-k} {\tilde f}_k^n, \\
&S_{\rm eff}(U)=   x \prod_{i=1}^{q-1} \Tr U_i \times 
\sum_{k=0}^\infty \left( \frac{J}{m} \right)^{2k} 
{\tilde H}_k\left( \frac{J^2 \beta}{m}\right),
\end{split}
\end{equation}
where we will present an argument for the $u$ dependences asserted here below.

\eqref{tildeq} represents an interesting reorganisation of usual perturbation 
theory. This reorganisation is particularly useful at small 
$\frac{J^2}{m^2} \ll 1$ but finite values of $\frac{J^2 \beta}{m}$. 
As $\beta \sim \frac{m}{\ln N}$ in the scaling limit, it follows that 
$\frac{J^2 \beta}{m}$ is fixed only for $\frac{J^2}{m^2} \sim \frac{1}{\ln N}$. 
At these small values of the coupling, $S_{\rm eff}(U)$ is well approximated
by the first term in the expansion in \eqref{tildeq}, i.e. by the term 
proportional to ${\tilde H}_0$. We will explicitly evaluate ${\tilde H}_0$ 
in this section and so reliably determine the partition function when 
$\frac{J^2}{m^2}$ is in the parametric range described above. 
\footnote{Although this is far from guaranteed, it is possible that the approximation 
$S_{\text{eff}} \sim H_0$ has a larger range of validity. Let us consider the 
parametric regime in which $\frac{J^2}{m^2}$ is small compared to unity
but large compared to $\frac{1}{m\beta}$. In this regime $\frac{J^2 \beta}{m}$ 
is effectively scaled to infinity. 
 Let us define 
\begin{equation}\label{rra}
r_k=\lim_{\frac{J^2 \beta}{m} \to \infty} \frac{H_k}{H_0}.
\end{equation}
If it turns out that $r_k$ is bounded (finite) then it follows that 
$H_0$ is in fact also a good approximation to the partition function 
for all values of $\beta$ assuming only that  $\frac{J^2}{m^2} \ll 1$. 
It would be interesting to investigate whether $r_k$ above are actually 
bounded for all $k$; however we leave that to future work. }

In the rest of this section we present the results of our explicit 
perturbative computations. Although we are principally interested in the 
function $H_0$ in the scaling limit, to set notations and for practice 
we first present the results of simpler computations. To start with 
we work out the partition function at level zero and recover the 
free partition function of the previous section. We then work out the 
partition function at level 1 (i.e. including graphs with a single melon). 
Next we present our results at level 2 (i.e. including all graphs with 
two melons). Finally we turn to the problem of principal interest to 
us, namely the sum of the infinite set of graphs that generates $H_0$. 
As preparation for all these computations we first briefly discuss the 
structure of the free Greens function. 

\subsection{Free Greens Function}

Consider the free fermionic Greens function 
$$\langle \psi^a(t) \bar{ \psi }_b(0)   \rangle.$$
We work in a colour basis in which the holonomy $U$ is diagonal. In this 
basis in which the action of holonomies on the fermions is given by  
\begin{equation} \label{holferm} \begin{split} 
U \psi^{a} & = e^{i \theta_a} \psi^a, \\
U {\bar \psi}_a&=  e^{-i \theta_a} \psi_a.\\
\end{split}
\end{equation}
The free fermionic Greens function at finite temperature is given by
\begin{equation}
\label{propexp}
	\begin{aligned}
		\langle \psi^a(t) \bar{ \psi }_b(0)   \rangle = & G_0(t) \delta^{a}_b, \  G_0(t)=f(t,m,\theta_a),   \ \  \textit{for \ $ -\beta \leq t\leq  \beta $ }\\
	\end{aligned}
\end{equation}
where 
\begin{equation} \label{propexpt}
	\begin{aligned}
		f(t,m,\theta_a) = & \frac{1}{2} e^{-(m+i\theta_a)t}\left[\text{sgn}(t)+\tanh\left(\frac{1}{2}(m+i\theta_a)\beta\right)\right]\\
		=& \frac{e^{-(m+i\theta_a)t}}{1+x \ e^{-i\theta_a \beta}}
		\left[ \Theta(t)-\Theta(-t) \ x \ e^{-i\theta_a \beta} \right].
	\end{aligned}
\end{equation}
Note that the function $f$ obeys the identity 
\begin{equation} 
	\begin{aligned}
		f\left(\frac{\beta}{2}+t,m,\theta_a\right)=-f\left(-\frac{\beta}{2}+t,m,\theta_a\right) \ \ \textit{for  \ $0 \leq t<\frac{\beta}{2}$},
	\end{aligned}
\end{equation}
from which it follows that the Greens function is antiperiodic on the circle 
as required on physical grounds. 

Note that we have presented the Greens function only in the `fundamental 
domain' $- \beta < t < \beta$. Our fermionic Greens function is taken 
by definition to be a periodic function of $t$ with period $2 \beta$; 
this property plus the explicit results \eqref{propexp} and \eqref{propexpt}
can be used to define the Greens function at every value of Euclidean time 
as required. The extended Greens function defined in this manner has 
singularities at $t= n \beta$ for every integral value of $n$, and is 
smooth everywhere else. 

Note also that the `reversed' Greens function 
$\langle {\bar \psi}_a (t) \psi^bb(0) \rangle$ is also given in terms of the function $G_0$ by  the formula \footnote{Owing to time translation symmetry.}
\begin{equation}
	\begin{aligned}
		\langle {\bar \psi}_a (t) \psi^b(0)\rangle = -G_0(-t)\delta^b_a.
	\end{aligned}
\end{equation}
This formula is also manifestation of symmetry of mass deformed 
SYK Lagrangian under the simultaneous swaps 
${\bar \psi} \leftrightarrow \psi$, $U \leftrightarrow U^{-1}$, 
$m \leftrightarrow -m$.

\subsection{Level zero: Free theory}

In this brief subsection we compute $S_{\text{eff}}$ at one loop, i.e. in the free 
theory. The result for $S_{\text{eff}}(U)$ was already presented in the previous 
subsection; we obtain that result here from a one loop computation 
as a simple practice exercise. Let 
\begin{equation}\label{omn}
\omega_n=\frac{2 \pi}{\beta} \left(n+\frac{1}{2}\right).
\end{equation}
The fields $\psi^a$ and ${\bar \psi}_a$ can be independently expanded in Fourier space 
as 
\begin{equation}
	\label{frequency space}
	\begin{aligned}
		\psi(t)=\sum_n \psi^n e^{-i \omega_n t}\  ,
\  \bar{\psi}(t)=\sum_n \bar{\psi}_n e^{+i \omega_n t}.
	\end{aligned}
\end{equation}
When substituted the free part of action \eqref{sykqmmd} becomes
\begin{equation} \label{fr}
	\begin{aligned}
		S=\sum_{n,a} \bar{\psi}_{a,n}[\beta(-i \omega_n+m+i\theta_a)]\psi^{a,n}.
	\end{aligned}
\end{equation}
Fermionic integration gives:
\begin{equation}
	\begin{aligned}
		Z_F=& \prod _{a} \prod _{n=-\infty}^{n=+\infty}\left[\beta(-i \omega_n+m+i\theta_a)\right]\\
		   =& \prod _{a} \prod _{n=-\infty}^{n=+\infty}\left[-i (2 \pi n+\pi)+m \beta+i\theta_a \beta\right]\\	
		   =& \prod _{a} \prod _{n=-\infty}^{n=+\infty}\left[-i 2 \pi n+c(\theta_a)\right]\\
		   =& \prod _{a}c(\theta_a)^2 \prod _{n=1}^{n=+\infty}\left[(-i 2 \pi n+c(\theta_a))(+i 2 \pi n+c(\theta_a))\right]\\
		    =& \prod _{a}c(\theta_a)^2 \prod _{n=1}^{n=+\infty}\left[ (2 \pi n)^2+c(\theta_a)^2\right]\\
		    =& \prod _{a}c(\theta_a)^2 \left(\prod _{n=1}^{n=+\infty}(2 \pi n)^2\right) \prod _{n=1}^{n=+\infty} \left[ 1+\left(\frac{c(\theta_a)/2}{ \pi n}\right)^2\right]\\
		    =&  \prod _{a} c(\theta_a)^2 \left(\prod _{n=1}^{n=+\infty}(2 \pi n)^2\right)  \left[ \frac{\sinh  \frac{c(\theta_a)}{2} }{\frac{c(\theta_a)}{2}} \right]\\
		    =& N  \prod _{a}  \left[ \sinh  \frac{c(\theta_a)}{2}  \right] \sim \prod _{a} e^{\frac{c(\theta_a)}{2} } (1-e^{-c(\theta_a)}),\\
		   \end{aligned}
\end{equation}
where $$c(\theta_a)=m \beta+i\theta_a \beta-i \pi$$ and $$N= \prod _{a}  \prod _{n=1}^{n=+\infty}(2 \sqrt{2} \pi n)^2  $$ is an infinite holonomy independent constant. As for every $\theta_a$ there is $-\theta_a$ to be taken into account $\prod _{a} e^{\frac{c(\theta_a)}{2} }$ becomes independent of holonomy. Keeping only holonomy dependent terms \footnote{Note that this also ensures for $\beta \to \infty$ partition function is 1 and for $\beta \to 0$ total number of states for a given $a$ are 2.} 
\begin{equation}
	\label{oth exact}
	\begin{aligned}
		\ln Z= \sum_a \log[1+x e^{-i\theta_a \beta}] \ , \ \ x=e^{-m \beta},
	\end{aligned}
\end{equation} 
In other words 
\begin{equation}
	\label{ot}
	\begin{aligned}
		\ln Z= \Tr  \ln [1+x U]. 
	\end{aligned}
\end{equation} 
Expanding \eqref{ot} in a power series in $x$ we recover \eqref{freesykmd}
at $N_F=1$. In the scaling limit we recover \eqref{pfots}.

\subsection{Level one: single melon graphs}

The contributions of graphs with a single melon to the Free energy is given by 
\begin{equation} \label{smg}
F_2= \frac{1}{2!}\ ^2C_{\frac{2}{2}}\ (-1)^{q/2}q \ m^2 \int \prod_{k=1}^q G_0(t_{1}-t_{2},\theta_{a_k}) \ dt_1 dt_2,
\end{equation} 

\begin{figure}[!h]
   \begin{center}
     \includegraphics[scale=0.7]{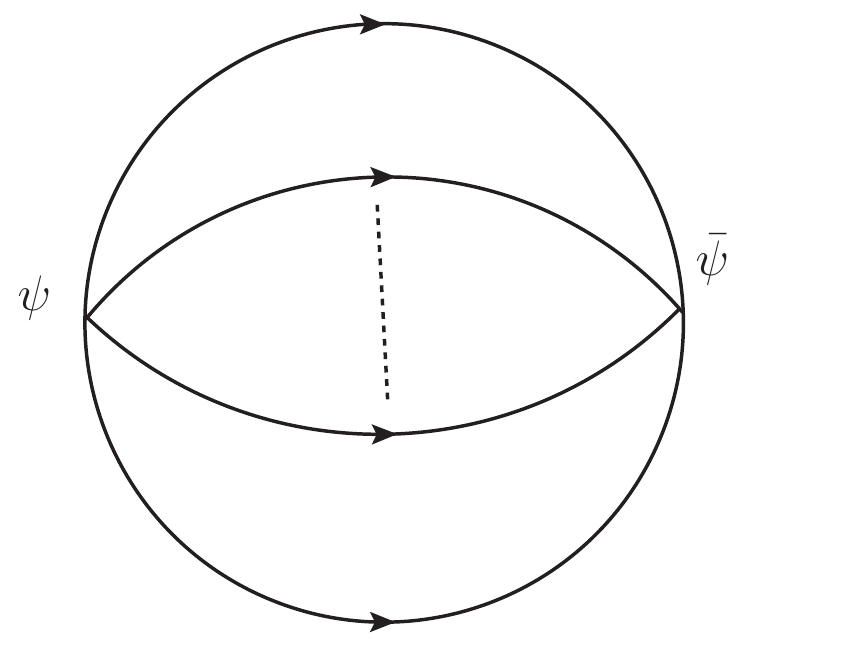}
   \end{center}
   \caption{Single loop contribution to free energy.}
\label{1loop}
   \end{figure} 

In this graph we contract each of fields in the interaction vertex $\psi^q$ with one of the fields in ${\bar \psi}^q$. Consider any 
particular $\psi$ field. This $\psi$ field has to contract with one of the $q$ ${\bar \psi}$ fields in the second interaction vertex. 
It is thus clear that there are $q$ choices for this contraction (the choices of which ${\bar \psi}$ our specified $\psi$ pairs up with).
Once this choice has been made, if we are interested - as we are- in graphs that contribute only at leading order in large $N$ there are no further choices in our contraction. Recall that every one of the remaining $\psi$'s (respectively $\bar{\psi}$'s) has exactly one colour common with the $\psi$ (resp $\bar{\psi}$) that we have just 
contracted together. The leading large $N$ behaviour is obtained only if the $\psi$ that shares any given gauge index with our special contracted $\psi$ is now contracted with the  $\bar{\psi}$ that shares the same gauge index with the special contracted ${\bar \psi}$. This rule specifies a unique contraction structure for the remaining fields. It follows that, up to a sign, the symmetry factor is simply $q$. The sign in question is simply  $(-1)^{(q-1)+(q-2)+..+1}$ Recalling that $q$ is even, it is easy to see that this phase $=(-1)^{q/2}$.
 
The integral in \eqref{smg} is very easy to perform. To see this note that the analytic structure of the integrand as a 
function of $t=t_1-t_2$ takes the form 
$$  e^{-qm t} (A_q \ \text{sgn}(t)+B_q), $$
for various different values of $q$. The integrand is integrated from $-\frac{\beta}{2}$ to $\frac{\beta}{2}$. The integral 
over $t_1+t_2$ produces an overall factor of $\beta$. The integrals are all trivial to do; evaluating them we find the final answer
\begin{equation}
	\label{2nd order}
	\begin{aligned}
		F_2=\frac{(-1)^{q/2}}{2!}\ ^2C_{\frac{2}{2}}\ q\,   m \beta \ I^{(2)}_1(q,x),
	\end{aligned}
\end{equation}
where
\begin{equation}
	\label{2nd order exact}
	\begin{aligned}
		I^{(2)}_1(q,x)= \frac{1-x^q}{q}  {\rm Tr}_F \prod_{k=1}^q \left(   \frac{1}{1+x \tilde{U}_k} \right).
	\end{aligned}
\end{equation}
The expression $\Tr_F(...)$ in the equation above represents the 
trace over an operator built on a particular Auxiliary Hilbert space. 
The operator in question is a function of the elementary operators 
${\tilde U}_k$ that act on this Hilbert space. We will now carefully 
define the relevant Hilbert space and the operators  ${\tilde U}_k$
and so give meaning to \eqref{2nd order exact}.

The operators ${\tilde U}_k$  in \eqref{2nd order exact} have the following meaning. These operators are unitary operators that 
act on a vector space whose dimensionality is $N^{\frac{q}{2} (q-1)}$. The vector space in question is the tenor product 
of $q-1$ factors, each of which has dimension $N^{\frac{q}{2}}$. Each factor described above is associated with one of the 
$q-1$ gauge groups. Let us focus on any one gauge group, say the first. The factor associated with this gauge group
consists of $\frac{q}{2}$ distinct factors of isomorphic $N$ dimensional spaces on which the $N\times N$ holonomy matrices of the 
first gauge group naturally act. 

Recall that each $\psi$ field that appears in an interaction has exactly one gauge 
index contraction with every other $\psi$ field. This means, in particular, that the indices of gauge group 1 are contracted 
between $\frac{q}{2}$ pairs of $\psi$s. This fact is the origin of the $\frac{q}{2}$ distinct factors of the space on which 
the holonomy matrices of the first gauge group act. 

With all this preparation we now explain the form of the operators $\tilde{U}_k$. Each $\tilde{U}_k$ acts as $U_1$ (the holonomy of the first 
$O(N)$ gauge group) on one of the $\frac{q}{2}$ copies of the $N$ dimensional vector space associated with the first $O(N)$, 
and as identity on the remaining $\frac{q}{2} -1$ copies of this space. In a similar fashion it acts as $U_2$ on one of the 
$\frac{q}{2}$ copies of the $N$ dimensional vector space associated with the second $O(N)$, 
and as identity on the remaining $\frac{q}{2} -1$ copies of this space. And so on. 
Exactly two $\tilde{U}_k$s act as $U_1$ on the same Hilbert space. Exactly two $\tilde{U}_k$s act as $U_2$ on the same Hilbert space, etc. 
Finally every two $\tilde{U}_k$s act on the same Hilbert space for one and only one gauge group.\footnote{This means that 
if $U_1$ and $U_3$ act on the same copy of the Hilbert space for gauge group 1, then they necessarily act on 
different copies of the Hilbert space for all the other gauge groups. }
The symbol ${\rm Tr}_F$ in that equation denotes the trace over the full $N^{\frac{q}{2} (q-1)}$ dimensional Hilbert space. 

From a practical point of view it is less complicated to use the 
definitions of the ${\tilde U}_k$ operators than it might at first seem.
We could, for instance, expand the result \eqref{2nd order exact} in a power
series in $x$. The formal looking expressions of traces of sums of products 
of ${\tilde U}_k$ operators that appear as coefficients in this expansion 
can easily be evaluated in terms of traces of powers of the holonomy matrices
$U_1 \ldots U_{q-1}$ of the factors of $O(N)$. 
 
A little thought will allow the reader to convince herself that the rules 
described above imply that, for instance 
\begin{eqnarray}
\label{rules} 
 {\rm Tr} \left( \sum_{k=1}^{q} {\tilde U}_k \right)  
&=&q N^{\frac{(q-2)(q-1)}{2}}  
{\rm Tr} U_1 {\rm Tr} ~ U_2 \ldots {\rm Tr} U_{q-1},  \\
  {\rm Tr} \left( \sum_{k_1 \neq k_2}^{q} {\tilde U}_{k_1}
 {\tilde U}_{k_2} \right)   
&=& q N^{\frac{q^2 -5q +6}{2}}  
\left[\prod_{k=1}^{q-1} {\rm Tr} U_1^2 ~  ({\rm Tr} U_2)^2 \ldots ({\rm Tr} U_{q-1})^2
\right.\nonumber\\&&\left.~~~~~~~~~~~+ (1 \leftrightarrow 2) + (1 \leftrightarrow 3) + \ldots (1 \leftrightarrow 
q-1)  \right], \\ 
 {\rm Tr} \left( \sum_{k=1}^{q} {\tilde U}^2_k \right) 
&=& q N^{\frac{(q-2)(q-1)}{2}}
\left( {\rm Tr} U^2_1 ~ {\rm Tr} U^2_2 \ldots {\rm Tr} U_{q-1}^2 \right).
\end{eqnarray}
As an illustration of these rules let us evaluate the partition function 
in the low energy scaling limit described in the previous section. 
Recall that in the limit of interest $x \sim \frac{1}{N^{q-3}}$ and we are instructed to retain only those contributions 
to $S_{\text{eff}}(U)$ that are linear in $x$; terms of higher order in $x$ can be discarded. It follows that the partition function in this limit may be evaluated 
by Taylor expanding \eqref{2nd order exact} in $x$ and discarding all terms 
that are quadratic or higher order in $x$. Using the first of \eqref{rules} 
we conclude immediately that 
\begin{equation}
\label{2nd order1}
\begin{aligned}
	F_2=\frac{(-1)^{q/2}}{2!}\ ^2C_{2/2}\ q\,   m \beta \ N^{q-1} (-x) \prod_{m=1}^{q-1} \rho^1_m.
	\end{aligned}	
\end{equation}
where $\rho^1_m= \frac{{\rm \Tr } U_m}{N}$ as in the previous section, and 
we have dropped the terms of order $x^0$ which are independent of $U_m$.

\subsection{Level 2: 2 melon graphs}
\label{2-melon}

\begin{figure}[!h]
   \begin{center}
     \includegraphics[scale=0.55]{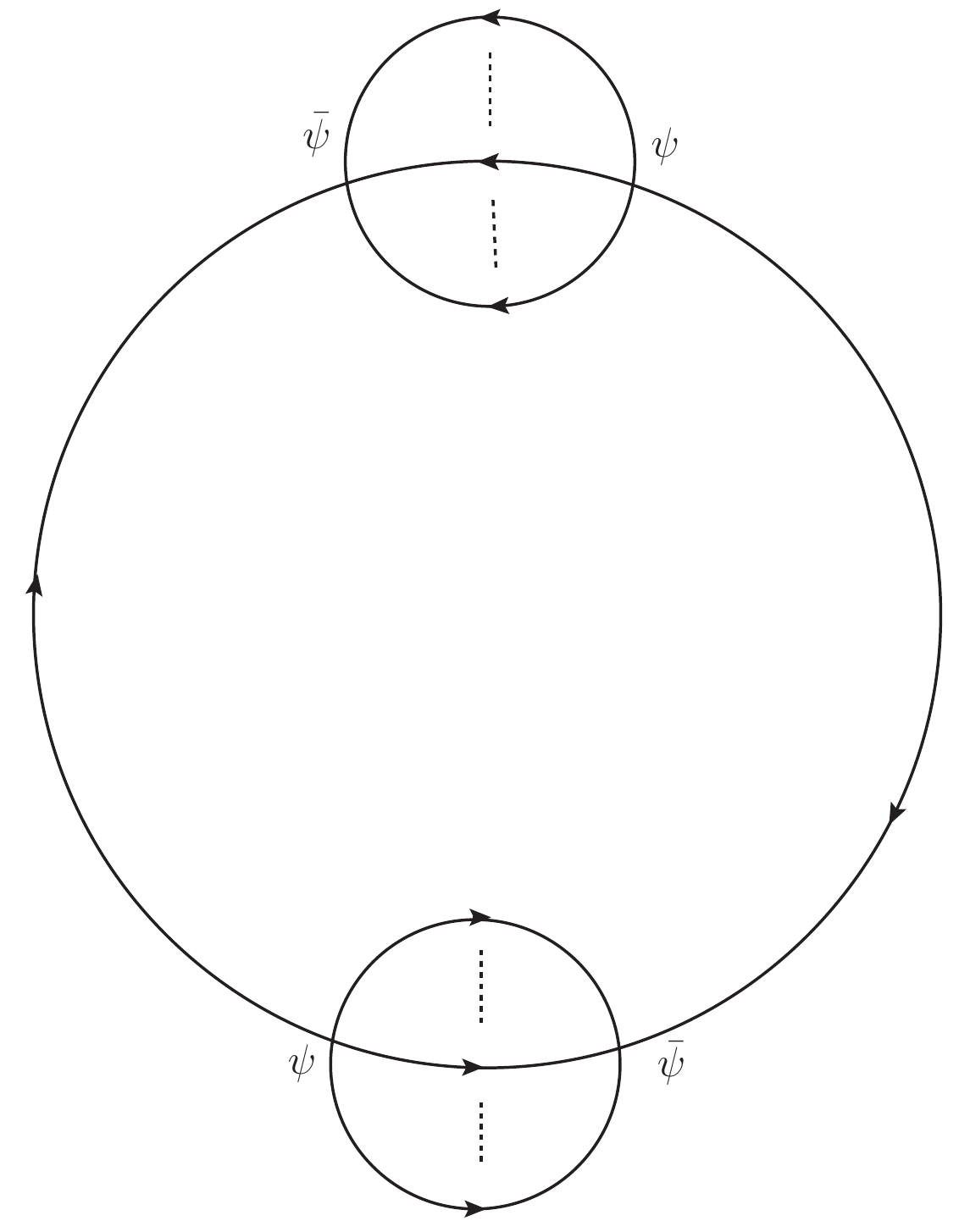}
   \end{center}
     \caption{Two loop contribution to free energy.}
\label{2loop}
   \end{figure}

At level 2 we once again have contributions from a single Feynman diagram 
Fig \ref{2loop}. In order to evaluate this graph we must 
evaluate in integral 
\begin{eqnarray}
	\label{2nd order2}
		F_4&=&\frac{1}{4!}\ ^4C_{4/2}\ (-1)2 (q^2)^2  \int \prod_{i=1}^4 dt_i \  \left( \prod_{i=1}^{q-1} G_0(t_{12},\theta_{a_i}) \right)  
		\left( \prod_{i=1}^{q-1} G_0(t_{34},\theta_{b_i}) \right)  \nonumber\\
		&&~~~~~~~~~~~~~~~~~~~~~~~~~~~~~~~~~~~~~~~~~~~~~~~~~~~~~~~~~~~~~`\times 
		G_0(t_{32},\theta_{c_2})G_0(t_{14},\theta_{c_1}).~~~~~~~~~
\end{eqnarray}

We give some details of this expression and the evaluation of this integral in the Appendix \ref{perturb-melon}. We have completely evaluated this 
integral with the help of mathematica (see Appendix \ref{all-beta} for arbitrary number of melons), but the final result for $S_{\text{eff}}(U)$ in the general case is too complicated to 
transfer to text. As before, however, the answer simplifies dramatically in the low energy scaling limit of the previous section (see Appendix \ref{2-melons})
and we find 
\begin{equation}
	\label{2nd order3}
	\begin{aligned}
		&F_4=\frac{(-1)}{4!}\ ^4C_{4/2}\ 2 (q^2)^2   [m \beta \ I^{(4)}_1(q)+m^2 \beta^2 \ I^{(4)}_2(q)] \ N^{q-1}x \prod_{m=1}^{q-1} \rho^1_m,\end{aligned}
		\end{equation}
		where
		\begin{equation}
	\begin{aligned}	& I^{(4)}_1(q)=-\frac{2}{q}(2q-3),\\
		& I^{(4)}_2(q)=-1.
	\end{aligned}
\end{equation}
Note that the final answer had two terms; one proportional to an overall factor of $\beta$ and the second proportional 
 to $\beta^2$. In the next subsection we will argue that a graph at level $n$, in the low temperature scaling 
 limit, has terms proportional to $\beta^q$ for $q=1 \ldots n$.

\subsection{The infinite sum $H_0$}

We will now turns to a study of the free energy at level $n$. As in the 
previous subsection we will focus on the start at the low energy scaling 
limit of the previous section, and so retain only those terms in all 
graphs that are proportional to $x$. As we will see below, general graphs 
in the scaling limit and at level $n$ break up into different pieces 
that are proportional to $\beta^{k}$ for $k=1 \ldots n$.\footnote{For instance 
the level one graph computed above was proportional to $\beta$ while 
the level two graph was the sum of one term proportional to $\beta$ and 
another term proportional to $\beta^2$.} We will further focus our attention 
on the graph with the largest power of $\beta$, i.e. in this subsection 
we will contribute that piece of the level $n$ answer that scales like 
$\beta^n x$. It turns out that this piece is rather easy to extract 
as we now explain. 

Let us first recall that the propagator in our theory takes the following 
form:\begin{equation} \label{proparg}
	\begin{aligned}
\langle \psi_a(t) {\bar \psi}^b (0) \rangle  
		=& \frac{e^{-(m+i\theta_a)t}}{1+x \ e^{-i\theta_a \beta}}
		\left[ \Theta(t)-\Theta(-t) \ x \ e^{-i\theta_a \beta} \right].
	\end{aligned}
\end{equation}
It will turn out (and we will see explicitly below) 
that the denominator in \eqref{proparg} only contributes at order $\beta^{n-1}$ or lower in free energy linear in $x$. For the purposes of the current subsection, therefore (where we wish to ignore terms at order $x^2$ or higher and only keep highest power of $\beta$)
 this denominator can be dropped, and we can work with the 
simplified propagator\footnote{The role that of the overall holonomy dependent phase factors 
above is quite subtle. Naively these overall factors can be dropped 
in their contribution to free energy diagrams. The naive argument for 
this is that the net contribution to of these phase factors at any 
interaction vertex is proportional to $ \prod_a e^{i (\theta_a)t_1}$ where the 
sum runs over the phases $\theta_a$ of all the $q$ propagators that end 
at that interaction vertex. As the interaction vertex is 
a gauge singlet, $\sum \theta_a$ vanishes, so it might at first seem that 
the contribution of all these phase factors drops out. This is in general 
incorrect. The subtlety is that $t_1$ is not single valued on the circle. 
In diagrams in which propagators `wind' as they go around the circle, 
one of the factors in the product may effectively be evaluated at, e.g. 
$t_1 + \beta$ and so the net contribution of this phase factor could 
turn out to be $e^{i \beta \theta_a}$. While this contribution is constant 
(independent of $t_1$), it is nontrivial in nonzero winding sectors. 
Such a contribution will play an important role in our computation below.}

\begin{equation} \label{propargs}
	\begin{aligned}
\langle \psi_a(t) {\bar \psi}^b (0) \rangle 
		=& e^{-(m+ i \theta_a) t }
		\left[ \Theta(t)-\Theta(-t) \ x \ e^{-i\theta_a \beta} \right].
	\end{aligned}
\end{equation}
In this subsection we assume $m>0$; the case $m<0$ can be argued 
in a completely analogous manner with the role of $\psi$ and ${\bar \psi}$ 
reversed in the analysis below. In the computation of Feynman diagrams 
on the circle we will need to choose a `fundamental domain' on the circle; 
our (arbitrary but convenient) choice of fundamental domain is 
\begin{equation}\label{fdomain}
-\frac{\beta}{2} < t < \frac{\beta}{2}
\end{equation}
Finally some terminology. We will call the part of the propagator 
\eqref{propargs} that is proportional to $\theta(t)$ the `forward' (`normal')  part 
of the propagator, and the part of the propagator proportional to 
$\theta(-t)$ the `reverse' part of the propagator. Note that
the normal part of the 
propagator ranges is modulus from $1$ to $\sqrt{x}$; it is maximum (i.e. unity) 
at $t=0$ and minimum (i.e. $\sqrt{x}$) 
at $t=\frac{\beta}{2}$. The modulus of the 
reverse part of the propagator 
varies in magnitude from $\sqrt{x}$ to $x$. It is minimum (i.e. equal to $x$) 
at $t=0$ and maximum (i.e. equal to $\sqrt{x}$) at $t=-\frac{\beta}{2}$.

With all this preparation we are now ready to isolate the parts of the 
level $n$ diagrams whose contribution is proportional to $x \beta^n$.

\begin{figure}[!h]
\begin{center}
	\includegraphics[scale=0.3]{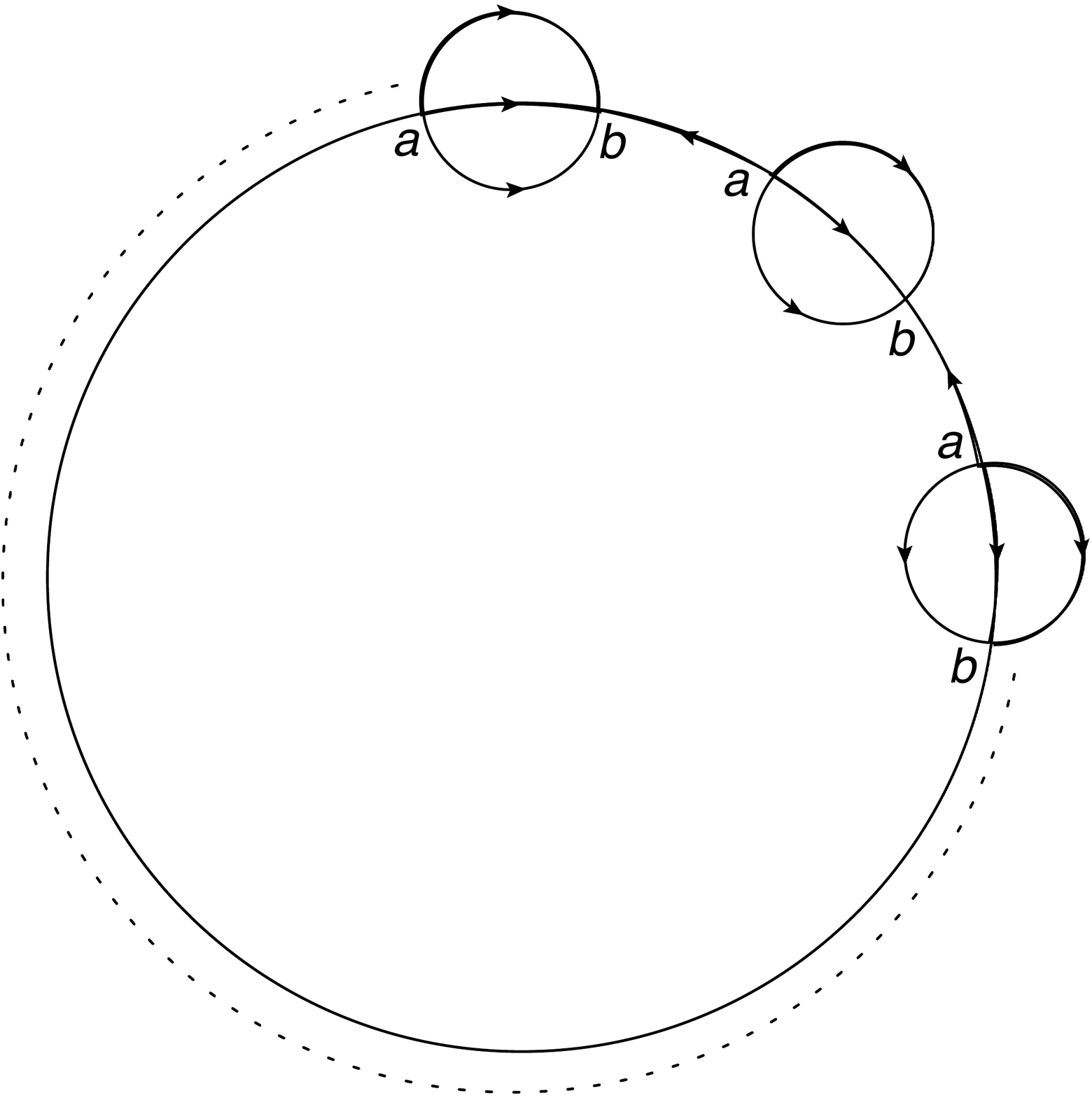}
\end{center}
    \caption{Circle diagram: a, b represents respectively insertions of $\psi$, $\bar{\psi}$. Direction of arrow is from $\psi$ to $\bar{\psi}$. The diagram is drawn for $q=4$.}
\label{nloop}
   \end{figure} 

To start with let us consider the simple $n^{th}$ level ring diagram depicted
in Fig. \ref{nloop}. In this diagram we have $n$ $a$ type vertices and 
$n$ $b$ type vertices. In this graphs we have $q-1$ propagators connecting 
adjacent $a$ and $b$ type vertices, but only a single propagator connecting 
$b$ to $a$ type vertices. 

Consider any propagator between $a$ and $b$ type vortices - which has $a$ type vertex $A$ at time $t_1$ and its adjacent $b$ type vertex 
$B$ at time $t_2$. Depending on whether $t_1>t_2$ or $t_1 <t_2$, all the 
$q-1$ propagators from $A$ to $B$ are either simultaneously all reverse or 
simultaneously all normal. If all propagators are reverse, the modulus of 
these propagators is less than $\sqrt{x}^{q-1} < x$ (recall $q \geq 4$). 
It follows that configurations in which the propagators from $A$ to $B$ 
do not contribute in the scaling limit, and so all propagators from $A$ 
to $B$ must be normal. Given that these propagators are all normal 
their modulus is proportional to $e^{-m (q-1)|t_1-t_2|}$. It is intuitively 
clear that separating $t_1$ from $t_2$ over a finite fraction of the 
circle forces us to pay a high cost in factors of $x$; it can be shown 
(this will be clearer in a bit) that such configurations do not contribute
to the result in the scaling limit. In the scaling limit we only receive 
contributions from configurations in which $|t_1-t_2|$ is of order 
$\frac{1}{m}$. It follows that for parametric purposes, we can simply 
regard $t_1$ and $t_2$ as the same point, replacing the integral over 
$t_1-t_2$ by $\frac{1}{m}$. For parametric purposes, in other words, 
each of the melons in Fig. \ref{nloop} can be thought of as a single 
interaction vertex, inserted at a single `self energy vertex', inserted
at a single time, with effective an effective insertion factor of order 
$\frac{J^2}{m}$.

Now let us turn to the propagators between $b$ and $a$ type vertices. 
These are now $n$ different propagators connecting the effective self 
energy blobs described in the previous paragraph. Let the effective 
times of insertions of these self energy blobs be $T_1$, $T_2 \ldots T_{n}$. 
Our graph is proportional to the product of $n$ propagators, the first 
from $T_1$ to $T_2$, the second from $T_2$ to $T_3$  $\ldots$ and the 
last from $T_{n}$ to $T_1 + w \beta$ where $w$ is an integer.
 As each reverse propagator contributes a 
factor of at least $\sqrt{x}$ to the integrand, no more than two of 
these propagators can be reverse. 

Let us first consider diagrams in which all propagators are forward. 
As all propagators move forward in time, the final propagator in the 
sequence must end not at time $T_1$ but at time $T_1 + w \beta$ where 
$w$ is a positive integer. The modulus of the product of these propagators 
is then easily seen to be proportional to $e^{-w m \beta}= x^w$. In the 
scaling limit of interest to us, the only option is $w=1$. Once we set 
$w=1$, the integrand of the diagram is now independent of the effective 
insertion times $T_i$. The integral over these $n$ insertion times 
thus gives a factor $\beta^n$, and the contribution of the graph 
in question is proportional to $x \beta^n$ as desired. 

Now let us consider diagrams in which one of the propagators between the 
effective self energy vertices is reverse, and the rest are forward. It 
is easy to verify that the modulus of the product of propagators in such 
a graphs is proportional to $x e^{-w m \beta}$ where $w=0, 1, \ldots $. 
In the scaling limit under consideration we are interested only in $w=0$. 
Once again the modulus of these graphs is independent of the insertion 
positions of the effective self energy vertices, and integration over their 
locations produces a result proportional to $x \beta^n$ as required.

Diagrams in which two of the propagators are reverse are kinematically 
very constrained. Similar argument as above shows these graphs are proportional to $x$ only if $w=-1$,i.e, if the two reverse propagators each have length $\frac{\beta}{2}$ 
(up to corrections of order $\frac{1}{m}$) and so all the forward 
propagators have length zero, again up to corrections of order $\frac{1}{m}$. 
These constraints ensure that such graphs are proportional to $\beta$ 
but no higher power of $\beta$ (certainly not $\beta^n$) and so are not 
of interest to the current section. 

In summary, graphs of the form depicted in Fig. \ref{nloop} only contribute
at order $x \beta^n$ if all propagators from $a$ to adjacent $b$ type 
vertices are normal, if the separation between 
$a$ and adjacent $b$ type vertices is of order $\frac{1}{m}$, and 
if the propagators between adjacent melons are either all normal with 
net winding number one or one reverse and the rest normal with net winding 
number zero. Once we have identified the parts of these graphs that 
contribute at order $x\beta^n$, the computation of these contributions 
is very simple (see below). 

Let us now turn to more general graphs than those drawn in Fig. \ref{nloop}. 
All graphs that contribute to the free energy at leading order in the large 
$N$ limit are of the general structure depicted in \ref{nloop}, but with the melons in Fig \ref{nloop} replaced by effective melons or `cactus graphs'. The net effect of this is to replace the bare propagators between $a$ and $b$ type vertices in Fig. \ref{nloop} by exact propagators. Recall that we are only interested in the propagator corrections at times $t=|t_1 -t_2| \sim \frac{1}{m} \ll \beta$. The $k^{th}$ order correction to the forward propagator at short times takes the schematic form 
\begin{equation}\label{kpst}
G(t) \sim \frac{|J|^{2k} t^k}{m^k} \sum_{n=0}^k C_n \left( \frac{m}{t} \right)^n
\end{equation}
As all values of $t$ that contribute to our integrand in the low energy 
scaling limit of interest to this paper are of order $\frac{1}{m}$, 
it follows that all terms on the RHS of \eqref{kpst} are of order 
$\frac{J^{2k}}{m^{2k}}$. As compared to the contribution of the graphs of Fig. \ref{nloop}, in other words, these graphs have extra powers of $J^2$ but no compensating factors of $\beta$. It follows that The contribution of such graphs at level $n$ is always of the form $x \beta^{h}$ with $h$ strictly less than $n$. Consequently all such graphs can be ignored. 

In summary, the only graphs that contribute at terms proportional to $x \beta^n$ at level $n$ are the very simple `necklace' graphs depicted in Fig. \ref{nloop}. We have already explained above that the contribution of each of these graphs is easily evaluated in the scaling limit. It follows that the computation of the sum of these graphs is a relatively simple job. 

Relegating all further details to the Appendix \ref{n-melons} we simply list our results. 
The contribution of order $x \beta^n$ to $S_{\text{eff}}(U)$ from graphs of level 
$n$ is given, for $n\geq 2$ by 
\begin{equation} \label{oin}
	\begin{aligned}
	\frac{J^{2n}}{m^{2n}}F_{2n} =	2x \ N^{q-1} \left( \prod_{m=1}^{q-1} \rho^1_m \right) \ \frac{1}{(n-1)!} \left[ \gamma(q)\ \frac{(-\beta)}{m}|J|^2 \right]^n\left(2-\frac{2^{n-1}}{n}\right)+\mathcal{O}(\beta^{n-1}),
	\end{aligned}
\end{equation}
where
\begin{equation} \label{gde}
	\begin{aligned}
		\gamma(q)=(-1)^{\frac{q}{2}(q-1)} \frac{q}{2}.
	\end{aligned}
\end{equation}
Summing these contributions over all $n=2$ to infinity and adding the 
separate contribution of $n=1$ we find $H_0$.
\begin{equation}\label{parfmo}
\begin{aligned}
	H_0=& 2x \ N^{q-1} \left( \prod_{m=1}^{q-1} \rho^1_m \right) \left[\sum_{n=2}^{\infty} \ \frac{1}{(n-1)!} \left[\gamma(q)\ \frac{(-\beta)}{m}|J|^2 \right]^n\left(2-\frac{2^{n-1}}{n}\right)-\frac{(-1)^{q/2}}{2}q^2 \ \beta \frac{|J|^2}{m} \right] \\ 
 =& 2x \ N^{q-1} \left( \prod_{m=1}^{q-1} \rho^1_m \right)
  \left[ \frac{1}{2}+2\gamma(q)\ \frac{(-\beta)}{m}|J|^2  e^{\gamma(q)\ \frac{(-\beta)}{m}|J|^2}-\frac{1}{2}e^{2 \gamma(q)\ \frac{(-\beta)}{m}|J|^2} -\frac{(-1)^{q/2}}{2} q \ \beta \frac{|J|^2}{m}  \right],
\end{aligned}
\end{equation}
so that the free holonomy effective action takes the form \eqref{nogs} with 
$F_0$ in that equation given by $H_0$ in \eqref{parfmo}. 

Note that $\gamma(q)$ is positive for $q=4, 8, 12 \ldots $ but is negative 
for $q= 6, 10, 14 \ldots$. It follows that the exponential terms in 
\eqref{parfmo} decay at large $\frac{J^2 \beta}{m}$ for the first set of 
values of $q$ but blow up for the second set of values of $q$. It would 
be interesting to better understand the meaning and consequences of this 
observation.

\subsection{Thermodynamics}

At sufficiently weak coupling we have demonstrated in the previous subsection that the free result for 
$S_{\text{eff}}(U)$ in the scaling limit, \eqref{pfots}, is replaced by the formula
\begin{equation}\label{sfa}
-S_{\text{eff}}(U)= N^{q-1} \left(\prod_{m=1}^{q-1} \rho^1_m \right)
x{\tilde H}_0,
\end{equation}
where $H_0$ was computed in the previous subsection. 

Note that \eqref{sfa} has the same structure of $U$ dependence as \eqref{pfots}; it follows that 
the partition function obtained by integrating $e^{-S_{\text eff}(U)}$ over $U$ is simply 
$Z( \tilde{x})$ (where the function $Z(x)$ was defined in \eqref{pfots}). At small enough coupling 
${\tilde x}$ is close to $x$, and the structure of the canonical partition function generated by 
\eqref{sfa} is very similar to the results described in detail for the free theory in the previous section. 

What does consequence does the replacement of $x$ by ${\tilde x}$ have for the micro canonical partition 
function? Let us first recall a simple formal result. Let 
$$e^{-\beta m} \rightarrow e^{-m \beta}(1+\epsilon \ h_0(\beta)).$$
By linearizing the usual thermodynamical formulae it is easy to show that 
this replacement results in the replacement
\begin{equation} \label{repla}
	S(E)=S_0(E)+\epsilon \frac{E}{m}h_0 \left[\frac{\partial S_0(E)}{\partial E} \right]+\mathcal{O}(\epsilon^2),
\end{equation}
(this result holds provided we expand about an analytic point in the phase 
diagram, i.e. away from phase transitions).
Clearly in our context this result applies if $\frac{J^2}{m^2} 
\sim \frac{\alpha }{\ln N}$ and $\alpha$ is taken to be small. 
However our results for the partition function are valid over 
a larger parametric regime; they are definitely valid whenever 
$\frac{J^2}{m^2} \sim \frac{\alpha }{\ln N}$ even at finite values of 
$\alpha$. In order to understand the effect ${\tilde H}_0$ has on the 
entropy as a function of energy at such values of $\alpha$ we take 
a slightly different route.

Define a `particle mass probability function' 
$p(m)$ by the following requirement 
\begin{equation} \label{pde}
\int dm' e^{-\beta m'} p(m')= x H_0. 
\end{equation}
Intuitively $p(m)$ denotes a spread in the mass density 
function (which was a $\delta$ function for the free theory) that mimics
the effects of interactions in thermodynamics. 

A little thought demonstrates that the following ansatz for $p(m')$ 
reproduces the structure of our perturbative expansion for $x H_0$ 
\begin{equation} \label{alli}
	\begin{aligned}
		p(m')=\sum_{k=0}^{\infty} \frac{1}{m}g_k \left( \frac{m'-m}{|J|^2/m} \right) \left( \frac{|J|}{m} \right)^{2k-2},
	\end{aligned}
\end{equation}
where the functions $g_k(y)$ do not depend on $J$. Working with the probability distribution \eqref{alli} is equivalent 
to replacing $x$ by  
	\begin{equation} \begin{aligned} \label{conf}
x \rightarrow  		\int_0^{\infty}e^{-\beta m'}p(m')dm'=x \sum_{n,k=0}^{\infty} \frac{1}{n!} \left(-\frac{|J|^2 \beta}{m}  \right)^n \left(\frac{|J| }{m}  \right)^{2k} \int_{-m^2/J^2}^{\infty}u^ng_k(u)du. 
	\end{aligned}
\end{equation}
The lower  limit of the integration in \eqref{conf} can safely be approximated by $-\infty$. If we want the RHS of 
\eqref{conf} to equal ${\tilde x}$ we must choose
\begin{equation}
	\begin{aligned}
		\int_{-\infty}^{\infty}g_0(u)du =1 \ , \
		\int_{-\infty}^{\infty}u \ g_0(u)du =\frac{2}{q}\gamma(q),\\
		\int_{-\infty}^{\infty}u^n \ g_0(u)du =  4n\left(1-\frac{2^{n-2}}{n}\right)\gamma(q)^n \ , \  \textit{n $\geq$ 2}
	\end{aligned}
\end{equation}
These relationships determine the moments the as yet unknown $g_0$. Inverting these relations we find 
\begin{equation}
\label{leadingmassdistribution}
	\begin{aligned}
		p(m')=& 2\delta(m'-m)- \delta\left(m'-m-2  \gamma(q)\tfrac{|J|^2}{m} \right)  
\\&-4\gamma(q)\frac{|J|^2}{m}\delta'\left(m'-m-\gamma(q)\tfrac{|J|^2}{m}\right)
	\end{aligned}
\end{equation}
Recall that the function $p(m')$ in the free theory was just a $\delta$ function localised at $m'=m$. 
The interaction effects considered in this section split this $\delta$ function into a set of 4 localised 
$\delta$ (or $\delta'$) spikes, distributed in a width of order $\frac{J^2}{m}$ around $m'=m$. As an 
aside we note the striking fact that interaction effects - at least at the order we have computed them - 
do not smoothen the free spectral function out. 

It is not difficult to convince oneself that the function $S(E)$ that follows from  \eqref{leadingmassdistribution}
is qualitatively similar to the entropy as a function of energy derived in detail for the free theory
in the previous section, and in particular displays faster than Hagedorn growth. 
\section{Discussion}

In these notes we have argued that the quantum mechanical model \eqref{sykqm}
- which is known to agree with the SYK model in the strict large $N$ limit - 
displays qualitatively new dynamics at subleading orders in $\frac{1}{N}$. 
We argued that the fluctuation spectrum about the finite temperature 
saddle point in this theory has new light modes - that originate in 
time dependent $O(N)^{q-1}$ transformations - in addition to the 
modes that arise from conformal diffeomorphisms and that were present 
also in the original SYK theory. The total number of new light modes 
is $(q-1)\frac{N^2}{2}$ and so is very large in the large $N$ limit.
We have also proposed that the dynamics of these new modes is governed by 
the sigma action \eqref{effectiveaction}, with a normalisation 
constant ${\cal{A}}$ whose value we have not been able to calculate. 

Assuming that our proposal for the new light modes is correct, it raises 
several interesting questions. It should be possible to check our proposal 
for the structure for the effective action \eqref{effectiveaction} by 
performing an independent computation of the four point function of 
four operators in the theory \eqref{sykqm} (by summing ladder diagrams) 
and comparing the long time behaviour of this computation with 
what one obtains directly from \eqref{effectiveaction}. Such a procedure 
should also permit the direct computation of the as yet unknown constant 
${\cal A}$.

It is also natural to attempt to find a bulk interpretation of our new 
modes. One natural suggestion is that these modes are dual to gauge 
fields in $AdS_2$ \footnote{We thank J. Maldacena for this suggestion.}
 If this is the case it is interesting that the rank 
of the bulk gauge fields diverges in the effectively classical $N \to \infty$
limit. In other words the bulk classical dual of this theory is given 
in terms of a weakly coupled theory of an {\it infinite} number of 
classical fields. The situation is somewhat reminiscent of 
the proliferation of `light states' in the duality of \cite{Gaberdiel:2012uj}, 
and also the situation with ABJ `triality' in the ABJM limit 
\cite{Chang:2012kt} (although in this context the number of 
bulk Vasiliev fields 
is never both parametrically large and parametrically weakly coupled).
It would be very interesting to investigate this further. 

We have also shown that the density of states in an extreme mass deformation
of the model \eqref{sykqm} displays a faster than Hagedorn growth at 
energies of order $N^2$. In our opinion this is also a very striking 
result; the phase that displays this rapid growth is the `thermal graviton' 
or `string gas' phase. The rapid growth in the density of states of this 
phase presumably means it cannot thermally equilibriate with another 
system. It would be interesting to understand what consequences 
this rapid growth has for potential bulk duals of mass deformed 
versions of the theory \eqref{sykqm}. 

Finally we have performed detailed calculations for the holonomy effective 
action of the mass deformed theory \eqref{sykqm} away from the strict 
large mass limit. In a particular scaling limit that zooms in on the 
dynamics of the theory at energies of order $N^2$ we demonstrated 
that the holonomy effective action of our theory, $S_{\rm eff}(U)$ takes 
a simple universal form. We were able to capture 
the leading interaction effects by summing the appropriate infinite 
class of graphs and obtain a very simple effective action that 
captures the leading deviation away from free behaviour. 
It should certainly be possible to generalise our perturbative computation 
of ${\tilde H}_0$ to a computation of ${\tilde H}_1$. More ambitiously, 
it may eventually prove possible to completely sum this perturbative 
expansion. We leave investigation of this possibility to the future.

\section*{Acknowledgement}

We would like to thank
S. Jain, I. Klebanov, C. Krishnan, J. Maldacena, G. Mandal, 
P. Nayak,  S. Sachdev, S. Shenker, D. Stanford, J. Yoon  and E. Witten 
 for useful 
discussions. We would like to thank S. Mazumdar, Y. Dandekar and S. Wadia 
for collaborations at the initial stages of this work. The work of all authors 
was supported in part by a UGC/ISF Indo Israel grant, and the Infosys Endowment for Research into the Quantum 
Structure of Spacetime. Finally we would all like to acknowledge our debt to the steady support of the people 
of India for research in the basic sciences.
\appendix

\vspace{0.2cm}

\section{Conformal Kernel}\label{ConformalKernel}

In this appendix following main result is proved
\begin{equation} 
	\begin{aligned}
	\int dt_3 \ dt_4 \ 	\tilde{K_c}(t_1,t_2;t_3,t_4)(g_c)_a^{\ b}(t_3,t_4)=\frac{1}{|J|^2} \ (g_c)_a^{\ b}(t_1,t_2),
	\end{aligned}
\end{equation}
where relevant quantities are defined by 
\begin{equation} 
	\begin{aligned}
		\tilde{K_c}(t_1,t_2;t_3,t_4)=&-|G_c(t_1,t_2)|^{\frac{q-2}{2}}G_c(t_1,t_3)G_c(t_2,t_4)|G_c(t_3,t_4)|^{\frac{q-2}{2}} ,\\ 
		g_c(t_1,t_2) = &|G_c(t_1,t_2)|^{\frac{q-2}{2}} G_c(t_1,t_2)[H(t_1)-H(t_2)]. 
	\end{aligned}
\end{equation}
Important part of the integration is given by:
\begin{equation}
	\begin{aligned}
		Q(t_1,t_2) \equiv & \int dt_3 \ dt_4 \ G_c(t_1,t_3)G_c(t_2,t_4)G_c(t_3,t_4)^{q-1}[H(t_3)-H(t_4)] \\ =& -\frac{1}{|J|^2} \int dt_3 \ G_c(t_1,t_3) H(t_3) \int  dt_4 \ G_c(t_2,t_4)\ |J|^2 G_c(t_4,t_3)^{q-1} \\
		& -\frac{1}{|J|^2} \int dt_4 \ G_c(t_2,t_4) H(t_4) \int  dt_3 \ G_c(t_1,t_3)\ |J|^2 G_c(t_3,t_4)^{q-1}\\
		=&-\frac{1}{|J|^2} \int dt_3 \ G_c(t_1,t_3) H(t_3) (-\delta(t_2-t_3))  -\frac{1}{|J|^2} \int dt_4 \ G_c(t_2,t_4) H(t_4) (-\delta(t_1-t_4))\\
		=& 	\frac{1}{|J|^2}\left[ G_c(t_1,t_2) H(t_2)+ G_c(t_2,t_1) H(t_1)\right]\\
		=& -\frac{1}{|J|^2} G_c(t_1,t_2) \left[H(t_1)-H(t_2)\right].
		\end{aligned}
\end{equation}
This proves claimed result when multiplied with $-|G_c(t_1,t_2)|^{\frac{q-2}{2}}$.

\section{Details of the perturbative computations}
\label{perturb-melon}
\subsection{Leading Power of $\beta$}
\subsubsection{Two melon graphs}
 \label{2-melons}
\begin{figure}[!h]
\begin{center}
	\includegraphics[scale=0.3]{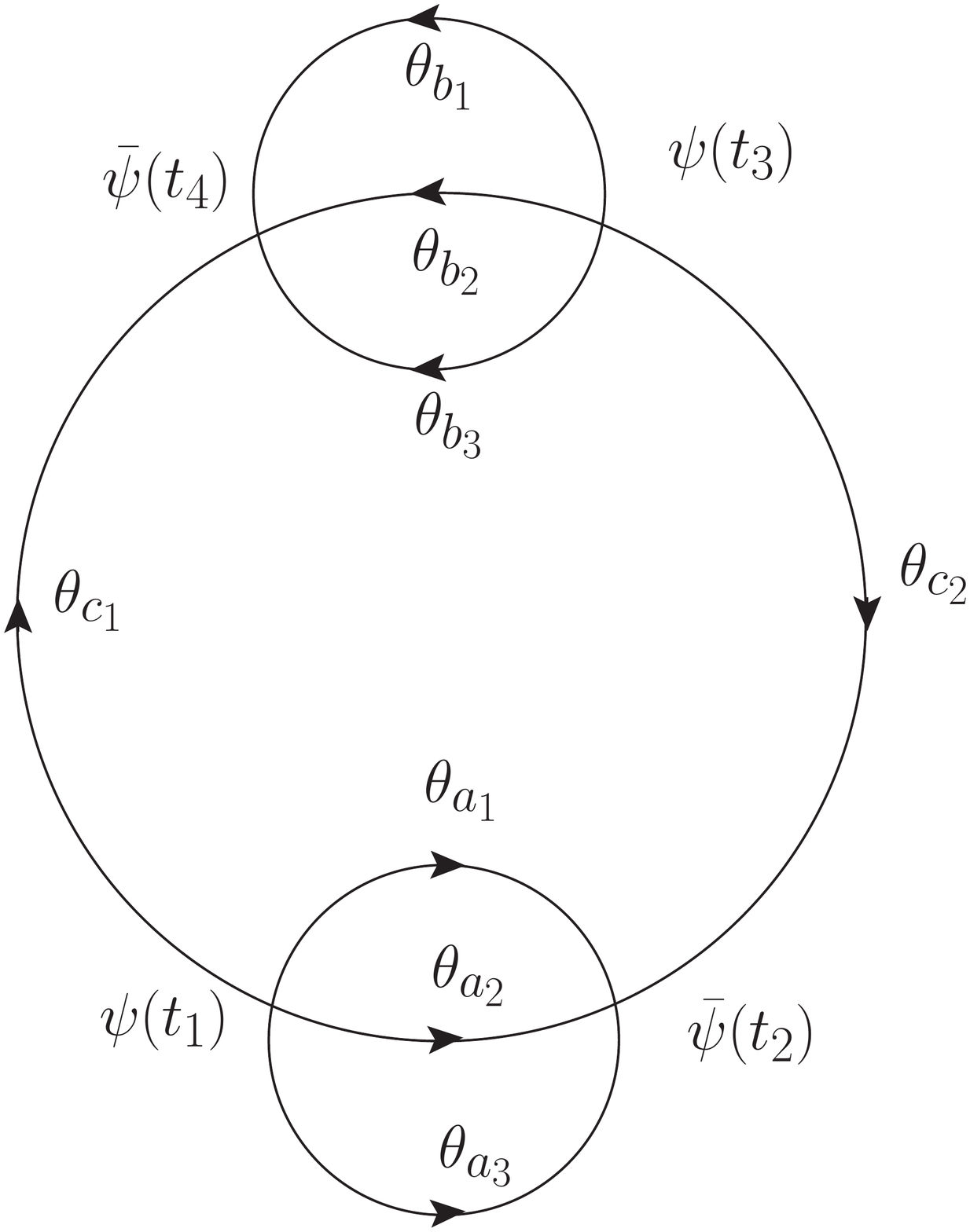}
\end{center}
    \caption{Direction of arrow is from $\psi$ to $\bar{\psi}$. The diagram is drawn for $q=4$.}
\label{2loop_details}
   \end{figure}

 In this subsection we consider the contribution to the free energy given by fig. \ref{2loop_details}. First non-trivial effect of winding is seen at this level as explained below.
 The term whose Wick contraction is calculated is $\frac{1}{4!}(J \psi^4+h.c.)^4$ - where each of $^4C_{4/2}$ terms contribute the same.
   The symmetry factor is calculated as follows. Any one of $q$ number of $\psi$'s of first $\psi$-vortex contracts with any one of $q$ number of $\bar{\psi}$'s of any one of two $\bar{\psi}$-vortex to give a factor of $2 q^2$. Any one of $q$ number of $\psi$'s of second $\psi$-vortex contracts with any one of $q$ number of $\bar{\psi}$'s of remaining $\bar{\psi}$-vortex to give a factor of $q^2$. In large-N only non-suppressed diagram is obtained by joining $\psi$ to $\bar{\psi}$ (of same vortex) of same common colour. Choice of external propagator gives $q-1$ possibilities at each blob. Sign of the symmetry factor comes from noticing as there are two identical 'blobs' sign of contraction of each blob cancel and overall sign is just because of contraction between two 'blobs', it turns out to be -1.  Contribution of symmetry factor at this order becomes
\begin{equation}
	\begin{aligned}
		& F_4=\frac{1}{4!}\ ^4C_{4/2}\ (-1)2 [q^2(q-1)]^2 I^{(4)},\end{aligned}
\end{equation}
where
\begin{equation}
	\begin{aligned}		& I^{(4)}=\int \prod_{i=1}^4 dt_i \  \left( \prod_{i=1}^{q-1} G_0(t_{12},\theta_{a_i}) \right)  
		\left( \prod_{i=1}^{q-1} G_0(t_{34},\theta_{b_i}) \right)  
		G_0(t_{32},\theta_{c_2})G_0(t_{14},\theta_{c_1}).
	\end{aligned}
\end{equation}
Where $\theta$s are holonomies on different propagators.
Here time differences are not necessarily single valued and to satisfy the constraint $$t_{12}+t_{23}+t_{34}+t_{41}=w \beta,$$ where $w=0,\pm 1, \pm 2$ (note that each $t_{ik}$ is in $( -\frac{\beta}{2},\frac{\beta}{2}  ) $, and this restricts allowed values of $n$)  we introduce dimensionless Lagrange multiplier integration
\begin{equation}
	P \equiv \beta \int_{-\infty}^{+\infty} \frac{ds}{2 \pi} e^{i s (t_{12}+t_{23}+t_{34}+t_{41}-w \beta)}=\delta\left(\frac{t_{12}+t_{23}+t_{34}+t_{41}-w \beta}{\beta }\right).
\end{equation}
In the scaling limit (assuming $m>0$), the propagator becomes
\begin{equation}
	\begin{aligned}
		G_0(t)=e^{-(m+i\theta_a)t} \theta(t)-x e^{-i\theta_a \beta} e^{-(m+i\theta_a)t} \theta(t)-x^{1/2} e^{-m\beta/2} e^{-i\theta_a \beta}  e^{-(m+i\theta_a)t} \theta(-t).
	\end{aligned}
\end{equation}
This way of writing ensures in each of three parts of $G_0$ excluding explicit $x$ dependence integration over $-\frac{\beta}{2}$ to $\frac{\beta}{2}$ gives only positive powers of $x$. We will refer to these three parts of $G_0$ as $x^0,x,x^{1/2}$ contributions.

In the scaling limit of interest $\ I^{(4)}$ can receive contribution from 5 different types of integration 
\begin{equation}
\begin{aligned}
	I^{(4)}= & \textit{$x^{0}$ everywhere}+ \textit{$x^{1/2}$ on one of the outer ($\theta_{c_1},\theta_{c_2}$) lines} + \textit{$x^{1/2}$ on both of the outer lines}\\+
	& \textit{$x$ on one of the outer lines}+\textit{$x$ on one of the inner lines ($\theta_{a_1}, \theta_{a_2} \dots  \theta_{a_{q-1}},\theta_{b_1},\theta_{b_2} \dots \theta_{b_{q-1}}$)}.
\end{aligned}
\end{equation}
Note that choosing $x^{1/2}$ on one of the inner propagators will force choosing all the inner propagators in the same blob to be $x^{1/2}$ term due to unit step function. Therefore this choice is ignored in scaling limit calculation.
Here we'll present the calculation corresponding to the first one and mention results for others.

Consider $x^{1/2}$ on $\theta_{c_1}$ say, and on all others we choose x independent part of $G_0$. This ensures following time ordering for non-zero integrand $t_{12}>0,t_{32}>0,t_{34}>0, t_{41}>0$, with which only consistent values of $n$ are $0,1$. Contribution to $I^{(4)}$ becomes, omitting $\beta (-x^{1/2}) e^{-i \theta_{c_1} \beta}e^{+i \theta_{c_1} w \beta} $ (for a contribution like $F_0$ we must have $n=0$ which is shown to be true below)
\begin{equation}
	\begin{aligned}
		I^{(4)} \sim & \  \beta (-x^{1/2}) e^{-i \theta_{c_1} \beta}e^{+i \theta_{c_1} w \beta} \int \frac{ds}{2 \pi} \ dt_{12} \ dt_{32} \ dt_{34} \ dt_{41} \ e^{-i s w \beta} \ e^{-(m(q-1)-is)t_{12}}  \times \\&  \ \ \ \ \ \ \ \ \ \ \ \ \ \ \ \  \ \ \ \ \ \ \ \ \  \ \ \ \ \ \ \ \ \ \ \ \ \ \ \ \ \ \ \ \ \ \ \ \ \ e^{-(m(q-1)-is)t_{34}}   \  e^{+(m+is)t_{41}-m \beta/2}  e^{-(m+is)t_{32}} \\
		= &-\beta (-x^{1/2}) e^{-i \theta_{c_1} \beta}e^{+i \theta_{c_1} w \beta}\int \frac{ds}{2 \pi} \ \frac{(e^{i s \beta /2}-x^{1/2})(x^{1/2}e^{-i s \beta /2}-1)}{(s+i(q-1)m)^2(s-im)^2}  e^{-is w \beta}+ \mathcal{O}(x^{3/2}),
	\end{aligned}
\end{equation}
where we ignored higher order contributions in $x$. Simplifying the numerator gives 3 terms: $x$ independent piece that comes with a non-zero phase factor $e^{i s \beta /2}$ (which will give a factor of $\beta$ upon integration because only $w=0$ will contribute), $x^{1/2}$ term that comes with no non-trivial phase (cannot give a $\beta$ upon integration), $x$ term drops out in scaling limit. Rest of the integration can be done easily choosing proper contour (semi-circle on upper or lower half plane as required by convergence) to ensure only $w=0$ term contributes to give the following result
\begin{equation}
	\begin{aligned}
		\delta_{w,0} x^{1/2} \frac{2}{(qm)^3}\left(-1+\frac{q}{4} m \beta\right).
	\end{aligned}
\end{equation}  
All other integrations can be performed similarly to give leading order contribution to free energy
\begin{equation}
	\begin{aligned}
		F_4 = \frac{1}{4!}\ ^4C_{4/2}\ 2 q^4  \frac{(q-1)^2}{q^2}m^2 \beta^2 \ N^{(q-1)^2} x \prod_{m=1}^{q-1} \rho^1_m+\mathcal{O}(\beta).
	\end{aligned}
\end{equation}

\subsubsection{n melon graphs}
  \label{n-melons}

 Here a circle diagram with $n \geq 2$ blobs is considered and leading term in $\beta$ is calculated using methods demonstrated in previous sub-section. 
 
 Symmetry factor for the diagram in large N limit is \footnote{Here an extra factor of $(n-1)!$ comes as compared to $n=2$ case because of freedom of joining $n$ blobs with one another.}
 \begin{equation}
 \label{symmetryfactor}
 	 	\frac{(-1)^{\frac{nq}{2}+n+1}}{(n!)^2} \  \ n! \ (q^2)^n \ (n-1)!.
 \end{equation}
 The leading order contribution in $\beta$ comes from two distinct choices - i) considering $x^{1/2}$ in any one of the $n$ external propagators (with holonomy $\theta_{a}$ say) with $x^0$ part of the free propagator in all others and ii) $x^0$ part of the free propagator in all propagators.

Contribution from the integral due to choice (i) is easily seen to be
 \begin{equation}
 	\begin{aligned}
 		& -x^{1/2}|g|^n \ e^{-i\theta_a \beta+i w \theta_a  \beta} \ \beta \int \frac{d s}{2 \pi} \ e^{-i(w-\frac{1}{2}) s \beta} \ \frac{1}{(-i s +m(q-1) )^n(is+m )^n}\\
 		&= -2x \ |g|^n\ e^{-i \theta_a \beta} \frac{1}{(n-1)!} \left( \frac{\beta}{2 m q} \right)^n \ \delta_{w,0},
 	\end{aligned}
 \end{equation}
 where we have kept only highest power of $\beta$. 
 Note that extra powers of beta $\beta^{n-1}$ came from the integration because of evaluation of residue around a pole of order $n$. This contribution is to be multiplied with a factor of $n$ due to freedom in choosing one external propagator on which $x^{1/2}$ is considered.
 
 Now we turn to the choice (ii). In this case contribution to the integral is 
 \begin{equation}
 	\begin{aligned}
 		& |g|^n \ e^{i w \theta_a  \beta} \ \beta \int \frac{d s}{2 \pi} \ e^{-iw s \beta} \ \frac{(1-x^{1/2}e^{-is \frac{\beta}{2}})^n}{(-i s +m(q-1) )^n(is+m )^n}\\
 		&= 2x \ |g|^n\ e^{-i \theta_a \beta} \frac{1}{(n-1)!} \left( \frac{\beta}{2 m q} \right)^n \left(\frac{2^{n-1}}{n}-1 \right) \ \delta_{w,1}.
 	\end{aligned}
 \end{equation}
As before we have kept only highest power of $\beta$. Note that this contribution vanishes for $n=2$.

After summing over the holonomies, and canceling loop N's with that of scaling of $g$, contribution to free energy becomes
\begin{equation}
	\begin{aligned}
	F_{2n} =	2x \ N^{q-1} \left( \prod_{m=1}^{q-1} \rho^1_m \right) \ \frac{1}{(n-1)!} \left[ \gamma(q)\ \frac{(-\beta)}{m}|J|^2 \right]^n\left(2-\frac{2^{n-1}}{n}\right)+\mathcal{O}(\beta^{n-1}),
	\end{aligned}
\end{equation}
where
\begin{equation}
	\begin{aligned}
		\gamma(q)=(-1)^{\frac{q}{2}(q-1)} \frac{q}{2}.
	\end{aligned}
\end{equation}

\subsection{All powers of $\beta$ in a circle diagram}
\label{all-beta}

\begin{figure}[h!]
\centering
\includegraphics[scale=0.4]{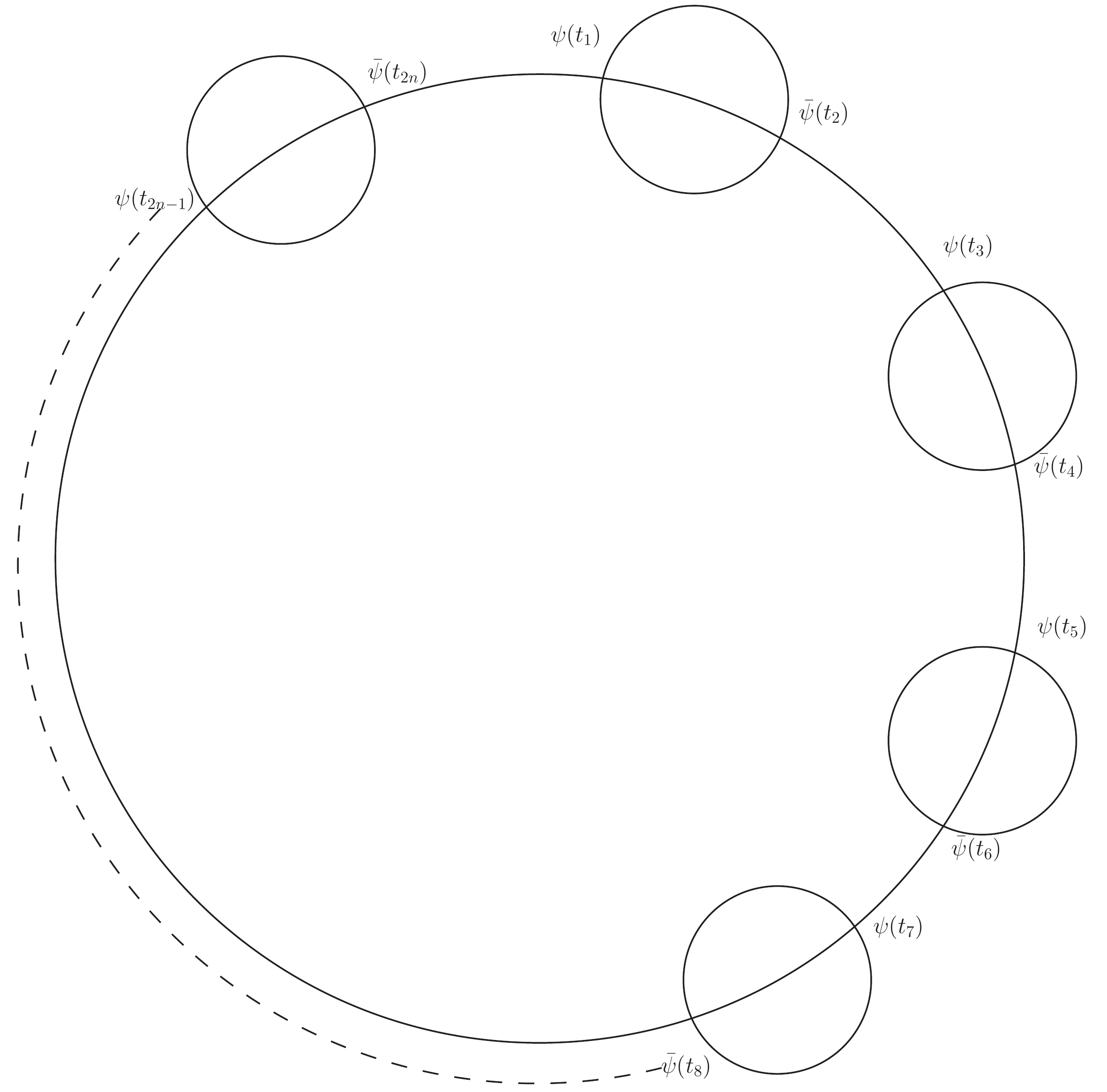}
\label{n-melon}
\end{figure}

In this subsection we shall compute explicitly the integral involved in computing the  contribution to the free energy in the scaling limit linear in $x=e^{-m\beta}$.
\newline\newline
The free fermionic Green's function at any finite temperature is given by,
\begin{eqnarray}
\langle \psi(t) \bar{\psi}(0) \rangle &\equiv&  G_0(t)  \nonumber \\
&=& \frac{1}{2} e^{-(m+i\alpha_j)t}\Big[\textrm{sgn}(t)+\tanh\Big(\frac{\beta}{2}(m+i\alpha_j)\Big)\Big]\nonumber \\
 &=& e^{-(m+i\alpha_j)t}\Big[\theta(t)-x e^{-i \alpha_j \beta}\Big]\, ,
\end{eqnarray}
where, $ x=e^{-m \beta}<<1$ (scaling limit).
Hence, one can also write the `reversed' Green's function at finite temperature as,
\begin{eqnarray}
\langle \bar{\psi}(0) \psi(t) \rangle = G_0^*(-m)&=&  \frac{1}{2} e^{(m+i\alpha_j)t}\Big[\textrm{sgn}(t)-\tanh \Big(\frac{\beta}{2}(m+i\alpha_j)\Big)\Big].
\end{eqnarray}
Here $\alpha_j$ are holonomies, satisfying the following constraint
\begin{eqnarray}
  \sum\limits_{j=1}^{q} \alpha_j=0\, .
  \label{holoconstraint}
\end{eqnarray}
Now in the computation we use discrete representation of the delta function
\begin{eqnarray}
 &&\delta(t_{21}+t_{32}+t_{43}+t_{54}+t_{65}+ \ . \ . \ .-t_{\overline{2n-1}\ \overline{2n}}+t_{1\ \overline{2n}})\nonumber \\
 &&={1\over 2\pi \beta}\sum\limits_{\omega=-\infty}^{\infty}e^{-2\pi i {\omega\over \beta}(t_{21}+t_{32}+t_{43}+t_{54}+t_{65}+ \ . \ . \ . \ -t_{\overline{2n-1}\ \overline{2n}}+t_{1\ \overline{2n}}\ )}.
\end{eqnarray}

\subsubsection{Evaluating the integral}
Let us focus on the diagram which can be computed as using the integral,
\begin{eqnarray}
  I^{(2n)}&=&\tfrac{1}{2\pi\beta}\left(\tfrac{J}{4}\right)^{2n}\sum\limits_{\omega=-\infty}^{\infty}\left[\int_{-\beta/2}^{\beta/2}  dt_1 \ e^{-t_1 \left(m+i \alpha _q\right)} e^{-2\pi i \frac{\omega}{\beta } t_1} \left( \textrm{sgn}(t_1)+\tanh(\tfrac{m\beta+i\alpha_q\beta}{2})\right)  \right. \nonumber \\
    &&\hspace{3cm}\left.\int_{-\beta/2}^{\beta/2}  dt_q \ e^{t_q \left((q-1)m-i \alpha _q\right)} e^{-2\pi i \frac{\omega}{\beta } t_q}  \left(A \ \textrm{sgn}(t_q)-B)\right)\right]^{n}\, .
    \label{Fn}
\end{eqnarray}
Here the first integral inside the sum is a single propagator while the second one represents the melon with $q-1$ propagatrs, where $A$ and $B$ are defined as
\begin{eqnarray}
\prod\limits_{j=1}^{q-1}\left[ \textrm{sgn}(t_q)-\tanh(\tfrac{m\beta+i\alpha_j\beta}{2})\right]=(-1)^q\left(A \ \textrm{sgn}(t_q)-B\right)\, .
\label{AndB}
\end{eqnarray}
We integrate over the time intervals of these propagators in (\ref{Fn}) and since there are $n$ of them we raise it to the power $n$. However, we would also have to implement the constraint that the times add up to an integral of $\beta$. This is achieved by repesenting the delta function on a circle of length $\beta$ as an infinite sum. This contributes a factor of $e^{2\pi i \tfrac{\omega}{\beta}t_i}$ in each of the propagators as shown in (\ref{Fn}).
  
Now we would like to focus on the integrals within the box brackets in (\ref{Fn})
\begin{eqnarray}
F^{(2n)}&=&\sum\limits_{\omega=-\infty}^{\infty}\left[I^{(q)}_\omega\right]^n.
\end{eqnarray}
Upon integrating over $t_1$ and $t_q$ one finds that 
\begin{eqnarray}
I^{(q)}_\omega &=& \frac{f_1+f_2}{((q-1)m-i\alpha_q+z)(m+i\alpha_q-z)},\,\,\,z =-\frac{2\pi i\omega}{\beta}
\end{eqnarray}
Here $f_1$ consists of terms with $e^{\pm k z \beta}$ where $k\in{\mathbb{Z_{\rm even}}}$ while $f_2$ consists of terms $e^{\pm kz\beta/2}$ where $k\in{\mathbb{Z_{\rm odd}}}$.
Its is evident that upon raising $I^{(q)}_\omega$ to $n$ one would have to evaluate sums in $z$ of the form 
\begin{eqnarray}
S_1&=&\sum\limits_{\omega=-\infty}^{\infty}\frac{e^{\pm kz\beta}}{(((q-1)m-i\alpha_q+z)(m+i\alpha_q-z))^n},\,k\in{\mathbb{Z_{\rm even}}}\cr
S_2&=&\sum\limits_{\omega=-\infty}^{\infty}\frac{e^{\pm kz\beta/2}}{(((q-1)m-i\alpha_q+z)(m+i\alpha_q-z))^n},\,k\in{\mathbb{Z_{\rm odd}}}.
\end{eqnarray}
Since $z=-\tfrac{2\pi i \omega}{\beta}$ we see that these reduce to
\begin{eqnarray}
S_1&=&\sum\limits_{\omega=-\infty}^{\infty}\frac{1}{(((q-1)m-i\alpha_q+z)(m+i\alpha_q-z))^n},\cr
S_2&=&\sum\limits_{\omega=-\infty}^{\infty}\frac{e^{z\beta/2}}{(((q-1)m-i\alpha_q+z)(m+i\alpha_q-z))^n}.
\label{sums}
\end{eqnarray}

We will use the technique of Matsubara summation to evaluate the above, where a weighting function is included
to replace the sum by a contour integral.
So, first let us evaluate $S_1$.
With a weighting function $f(z)=\frac{1}{1-e^{z\beta}}$, one can replace the above summation with the following contour integral,
\begin{equation}
 S_{1}= \oint \frac{dz}{(1-e^{z\beta})(((q-1)m-i\alpha_q+z)(m+i\alpha_q-z))^n}
\end{equation}
Notice that the integrand has two poles at $z \equiv z_a=-(q-1)m+i \alpha_q$ and $z \equiv z_b=m+i \alpha_q$ and both are of $n$-th order. Using the residue theorem, one can evaluate the above integral as,
\begin{eqnarray}
 S_{1}=\lim_{z \to z_a} {1\over (n-1)!}  \partial_z^{(n-1)}\frac{1}{(1-e^{z\beta})(z-z_b)^n} +\lim_{z \to z_b} {1\over (n-1)!}  \partial_z^{(n-1)}\frac{1}{(1-e^{z\beta})(z-z_a)^n}
 \label{Sevensum}
\end{eqnarray}
Now, it is very easy to verify that for any function $f(z)$,
\begin{eqnarray}
 \partial_z^{(n-1)}\Big[f(z) \frac{1}{(z-z_a)^n}\Big]=\sum\limits_{k=0}^{n-1}(-1)^k \  
 \Comb{(n-1)}{k} \ {(n+k-1)!\over (n-1)!} \ \frac{\partial_z^{(n-k-1)}f(z)}{(z-z_a)^{n+k}}
 \label{nthderv}
\end{eqnarray}
In the present case, taking $f(z)={1\over (1-e^{\beta z})}$, one can evaluate
\begin{eqnarray}
 \partial_z^{(n)}\Big[\frac{1}{1-e^{\beta z}}\Big]={\beta^n e^{-\beta z}\over (e^{-\beta z}-1)^{n+1}}A(n)
 \label{nthderv1}
\end{eqnarray}
where, $A(n)$ is the Eulerian polynomial in $e^{-\beta z}$, given by,
\begin{eqnarray}
 A(n)=\sum\limits_{m=0}^{n-1}\sum\limits_{k=0}^{m+1}(-1)^k \ \Comb{n+1}{k}(m+1-k)^n \ e^{-\beta z m}
 \label{EulerianPolyA}
\end{eqnarray}
Using equation (\ref{nthderv1}) and (\ref{EulerianPolyA}), one can easily obtain,
\begin{eqnarray}
 \partial_z^{(n-k-1)}f(z)&=&\partial_z^{(n-k-1)}\Big[\frac{1}{1-e^{\beta z}}\Big]\nonumber \\
 &=&{\beta^{n-k-1}\over (e^{-\beta z}-1)^{n-k}} 
 \sum\limits_{m=0}^{n-k-2}\sum\limits_{l=0}^{m+1} (-1)^l \ \Comb{n-k}{l} \ (m+1-l)^{n-k-1}e^{-\beta(m+1)z}\nonumber \\
 &+&{1\over 1-e^{\beta z}}\delta_{n-k-1,0}
 \label{nthderv1final}
\end{eqnarray}
Finally Substituting equation (\ref{nthderv1final}) into equation (\ref{nthderv}), we have,
\begin{eqnarray}
 \partial_z^{(n-1)} \Big[{f(z)\over (z-z_a)^n}\Big]&=&\partial_z^{(n-1)} \Big[\frac{1}{(1-e^{\beta z})(z-z_a)^n}\Big]\nonumber \\
 &=&\sum\limits_{k=0}^{n-1}{(-1)^k\over (z-z_a)^{n+k}} \  \Comb{(n-1)}{k} \ {(n+k-1)!\over (n-1)!} \Big[
 {\beta^{n-k-1}\over (e^{-\beta z}-1)^{n-k}} \sum\limits_{m=0}^{n-k-2}\sum\limits_{l=0}^{m+1} (-1)^l \nonumber \\ 
 &&\ \Comb{n-k}{l} \ (m+1-l)^{n-k-1}e^{-\beta(m+1)z}+{1\over 1-e^{\beta z}}\delta_{n-k-1,0}\Big] 
\end{eqnarray}
Evaluating the above expression at both the poles $z=z_a$ and $z_b$, one can compute $S_{1}$ as expressed in equation (\ref{Sevensum}).

Now let us discuss about evaluating the summation $S_2$ as given in equation (\ref{sums}).
With a weighting function $f(z)=\frac{\ e^{\beta z/2}}{1-e^{\beta z}}$, one can replace the above summation with the following contour integral,
\begin{equation}
 S_{2}= \oint \frac{ e^{\beta z/2} dz}{(1-e^{\beta z})(((q-1)m-i\alpha_q+z)(m+i\alpha_q-z))^n}
 \label{oddint}
\end{equation}
Notice that we encounter the same $n$-th order poles in the contour integral as we had with $S_1$. The residue computation for evaluating this
contour integral needs to evaluate the following term as before,
\begin{eqnarray}
 \partial_z^{(n)}f(z)=\partial_z^{(n)}\Big[\frac{e^{\beta z/2}}{1-e^{\beta z}}\Big]={ \beta ^n e^{-\beta z/2}\over 2^n(e^{-\beta z}-1)^{n+1}}B(n)
 \label{nthderv2}
\end{eqnarray}
where, $B(n)$ is the Eulerian polynomial of type-B in $e^{-\beta z}$, given by,
\begin{eqnarray}
 B(n)=\sum\limits_{m=0}^{n}\sum\limits_{k=0}^{m}(-1)^{m-k} \ \Comb{n+1}{m-k}(2k+1)^n \ e^{-\beta z m}
 \label{EulerianPolyB}
\end{eqnarray}
Finally, using equation (\ref{nthderv}), (\ref{nthderv2}) and (\ref{EulerianPolyB}), one can obtain,
\begin{eqnarray}
 \partial_z^{(n-1)} \Big[\frac{f(z)}{(z-z_a)^n}\Big]&=&\partial_z^{(n-1)} \Big[\frac{e^{\beta z/2}}{(1-e^{\beta z})(z-z_a)^n}\Big]\nonumber \\ &=&
 \sum\limits_{k=0}^{n-1}{(-1)^k\over (z-z_a)^{n+k}} \  \Comb{(n-1)}{k} \ {(n+k-1)!\over (n-1)!} 
 {\beta^{n-k-1}\over 2^{n-k-1}(e^{-\beta z}-1)^{n-k}} \nonumber \\ 
 &&\sum\limits_{m=0}^{n-k-1}\sum\limits_{l=0}^{m} (-1)^{m-l} \ \Comb{n-k}{m-l} \ (2l+1)^{n-k-1}e^{-(2m+1)\beta z/2}
\end{eqnarray}
Now using the above equation one can compute the residue and hence the integral (\ref{oddint}). This finishes the computation of $S_{2}$ as given in
equation (\ref{sums}).
\newline\newline
One finds that $S_1$ depends only linearly on $x=e^{-m\beta}$ while  $S_2$ depends as $\sqrt{x}$. Further noting that the difference in  $A$ and $B$ in (\ref{AndB}) behaves as $A-B={\mathcal{O}}(x^{q-1})$ we find that $f_1=(A-B){\mathcal{O}}(x^{\tfrac{-q}{2}+1})={\mathcal{O}}(x^{q/2})$ and  $f_2=(A-B){\mathcal{O}}(x^{\tfrac{-q+1}{2}})={\mathcal{O}}(x^{\tfrac{q}{2}-1})$. Therefore in the scaling limit one can take $A=B=2^{q-2}\left(1-x\sum\limits_{j=1}^{q-1}e^{-i\beta\alpha_j}\right)$. 
\newline\newline
Therefore evaluating $(f_1+f_2)^N\approx F_1+F_2$ in the scaling limit- where once again $F_1$ consists of terms with $e^{\pm k z \beta}$ where $k\in{\mathbb{Z}}$ while $F_2$ consists of terms $e^{\pm kz\beta/2}$ where $k\in{\mathbb{Z_{\rm odd}}}$, $F^{(n)}=S_1 \bar{F}_1+S_2 \bar{F}_2$. Here $\bar{F}_{1,2}=F_{1,2}(z=0)$\footnote{Since their $z$ dependences were where taken into account in evaluating $S_1$ and $S_2$}.
\newline\newline
The fact that only these two type of summations contribute for any integer value of $k$, makes it easier to evaluate equation (\ref{Fn})
in the scaling limit as,
\begin{eqnarray}
I^{(2n)}\!\!\!&=&\!\!\!J^{2n}\sum\limits_{k=0}^{n-2}\frac{2^{(q-1)n}x\beta^{n-k}}{(mq)^{n+k}\Gamma(n)^2}(2^n-2^{2+k}n)(^{n-1}C_k)\Gamma(n+k)\prod_{m=1}^{q-1}\rho^1_m
\end{eqnarray}
which can be re-written as 
\begin{eqnarray}
I^{(2n)}\!\!\!&=&\!\!\!\left(\frac{J^2\beta}{mq}\right)^{n}\,\,\sum\limits_{k=0}^{n-2}\left(\frac{J^2}{m^2}\right)^k\left(\frac{m}{qJ^2\beta}\right)^k\frac{2^{(q-1)n}x}{\Gamma(n)^2}(2^n-2^{2+k}n)(^{n-1}C_k)\Gamma(n+k)\prod_{m=1}^{q-1}\rho^1_m\cr
&&\hspace{11cm}+{\mathcal{O}(\beta)}
\end{eqnarray}
or equivalently keep all orders in $\beta$ as
\begin{eqnarray}
I^{(2n)}={J^{2n}\over m^{2n}q^n}\frac{(-1)^{q(n-1)}}{ (n-1)! }  \bigg[ && \frac{(2n-2)!}{(n-1)! \ q^{n-1}}m\beta  \bigg(1  -n  \Big(q -2 n+3 \Big)
   \bigg) \\ && + \sum _{k=0}^{n-2}   \frac{(n+k-1)!}{k!(n-k-1)! \ q^k}(1-2^{k+2-n} n)(m\beta)^{n-k} \ \bigg ] \ x  
 \left( \prod_{m=1}^{q-1} \rho^1_m \right).   \nonumber 
\end{eqnarray}
This multiplied with \eqref{symmetryfactor}$\times N^{q-1}$ gives contribution of a circle diagram with n melons.

\subsection{Evaluating the subleading correction}

We end this appendix by presenting a technical result which we do not use in the main text of the paper, but record here 
anyway, just in case this  result finds application subsequent work. 

\begin{figure}[h!]
\centering
\includegraphics[width=6in,height=4in]{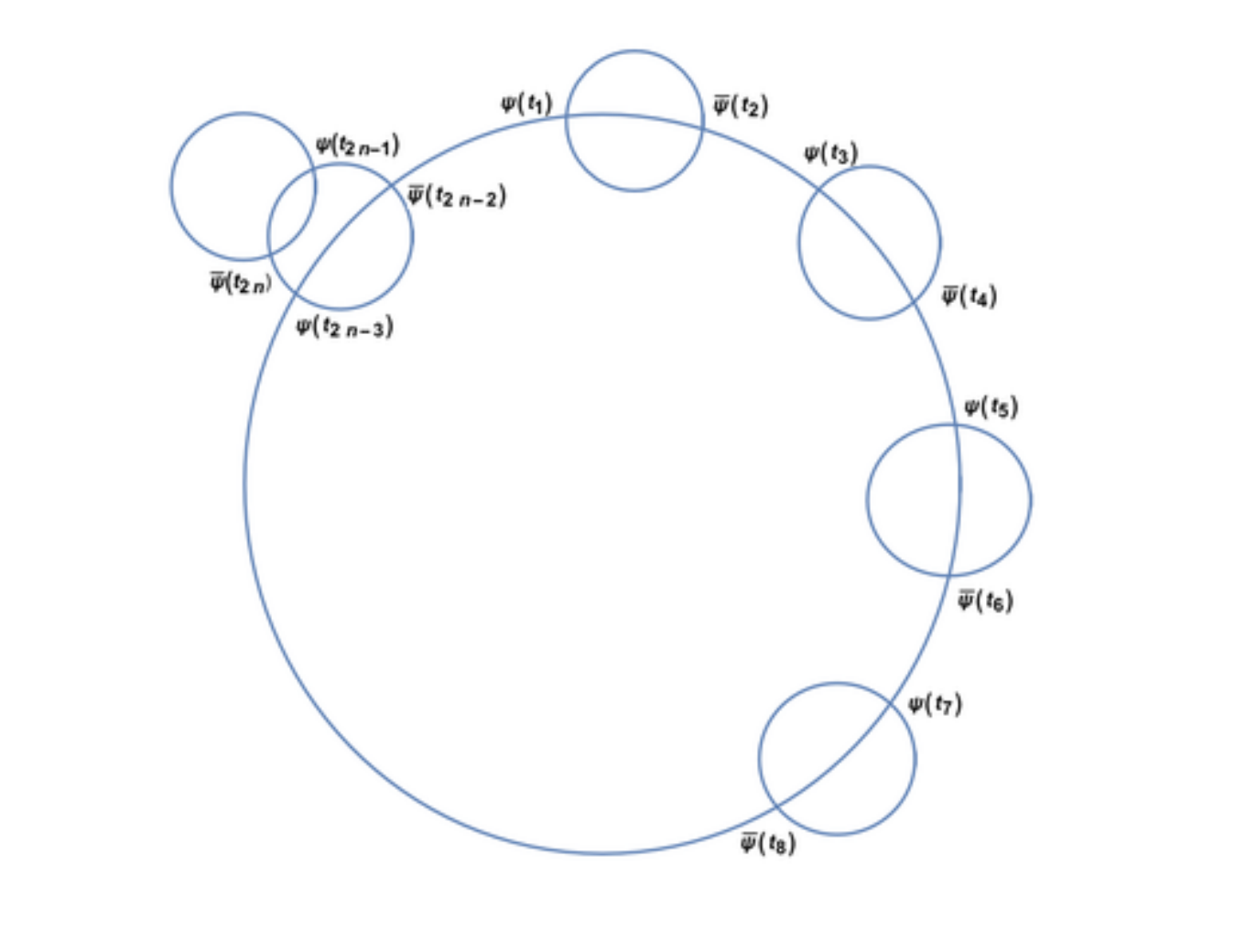}
\caption{subleading diagram}
\label{subleading_diagram}
\end{figure}

\begin{figure}[h!]
\centering
\includegraphics[scale=0.7]{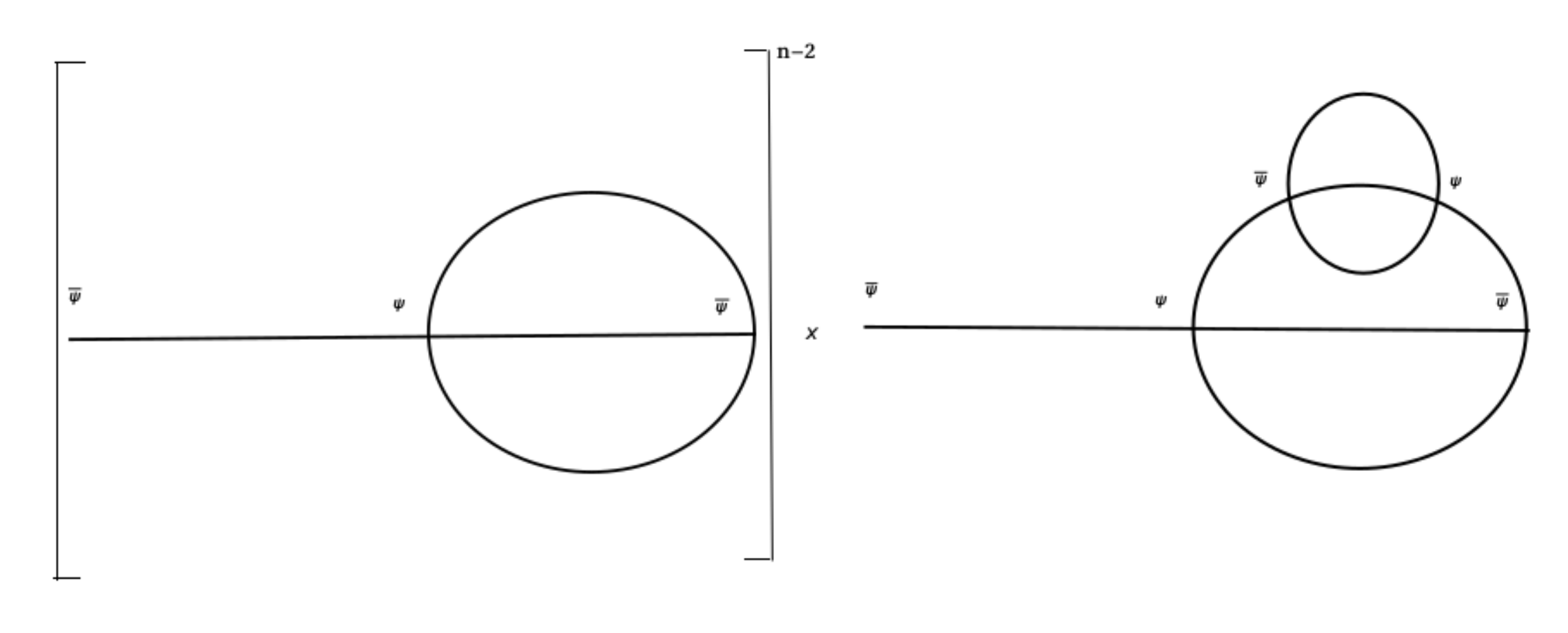}
\caption{parts of subleading diagram}
\label{subleading_diagram_parts}
\end{figure}

The technical result we report here is the evaluation of the Feynman integral for diagram Fig. (\ref{subleading_diagram}) (the figure is drawn 
for $q=4$ but we present the evaluation in general), which is one of the diagrams that 
would contribute to the generalization of the results presented in this paper to subleading orders in $\frac{1}{\beta}$. We present the 
result for the Feynman diagram ignoring the symmetry factor (which can easily be independently evaluated). We evaluate the diagram of Fig. (\ref{subleading_diagram}) as follows.
In order to get the integrand of the diagram we first multiply together all the propgators that make it up, keeping careful track of 
holonomy factors and making use of the fact that holonomies at any interaction vertex sum to zero. The integrand is the term in 
the big square bracket in \eqref{bigstuff} with $\epsilon_1$ and $\epsilon_2$ temporarily set to zero. The first two lines on the RHS 
of \eqref{bigstuff} are the $n-2$ factors on the in the diagram Fig.(\ref{subleading_diagram_parts}).  \footnote{$t_1$ in this term is the length of the straight
line in these factors, while $t_2$ is the length of the 3 (or more generally $q-1$) melonic lines in the part Fig.(\ref{subleading_diagram_parts}) that is 
enclosed in the square bracket. Really there are $n-1$ different $t_1$s and $n-2$ different $t_2$. As $t_1$ and $t_2$ are dummy variables 
that we integrate over, we have used the same symbol for all of them.} 

The next four lines on the RHS of 
\eqref{bigstuff} represent the second factor in Fig.(\ref{subleading_diagram_parts}). Lines 3-6 on the RHS of \eqref{bigstuff} are the 
remaing factors (the propagators outside the square bracket) in  Fig.(\ref{subleading_diagram_parts}).  \footnote{The third line in \eqref{bigstuff} is the 
straight line in this part of Fig.(\ref{subleading_diagram_parts}). The last and secondlast lines in \eqref{bigstuff} are, respectively, the blobs of $q-1$ and 
$q-2$ propagators in this part of Fig.(\ref{subleading_diagram_parts}). Finally the fourth line in \eqref{bigstuff} is the product of the two proagatogrs that run 
between the `$q-1$ blob' and the `$q-2$ blob'. The times in all these terms represent the lengths of the corresponding propagators. }

\newpage
\begin{eqnarray} 
I^{(2n-2)}&=&\left(\tfrac{1}{2\pi\beta}\right)^2\sum\limits_{\omega_1,\omega_2=-\infty}^{\infty}
\left[\left(\int_{-\beta/2}^{\beta/2}dt_1 e^{-(m+i\alpha_1+i\tfrac{\epsilon_1}{\beta})t_1}\left({\rm sgn}(t_1)+\tanh(\tfrac{m\beta+i\alpha_1\beta}{2})\right)\right.\right.\cr
&&\left.\hspace{3.5cm}\int_{-\beta/2}^{\beta/2}dt_2e^{((q-1)m-i\alpha_1-i\tfrac{\epsilon_1}{\beta})t_2}\left({\rm sgn}(t_2)A_1-B_1\right)\right)^{n-2}\cr
&&\hspace{3.5cm}\int_{-\beta/2}^{\beta/2}dt_1e^{-(m+i\alpha_1+i\tfrac{\epsilon_1}{\beta})t_1}\left({\rm sgn}(t_1)+\tanh(\tfrac{m\beta+i\alpha_1\beta}{2})\right)\cr
&&\hspace{3.5cm}\left(\int_{-\beta/2}^{\beta/2}dt_3e^{(m+i\alpha_2+i\tfrac{\epsilon_2}{\beta})t_3}\left({\rm sgn}(t_3)-\tanh(\tfrac{m\beta+i\alpha_2\beta}{2})\right)\right)^2\cr
&&\hspace{3.5cm}\int_{-\beta/2}^{\beta/2}dt_4e^{((q-2)m-i(\alpha_1+\alpha_2)-i\tfrac{(\epsilon_1+\epsilon_2)}{\beta})t_4}\left({\rm sgn}(t_4)A_{1,2}-B_{1,2}\right)\cr
&&\left.\hspace{3.5cm}\int_{-\beta/2}^{\beta/2}dt_5e^{(-(q-1)m+i\alpha_2+i\tfrac{\epsilon_2}{\beta})t_5}
\left({\rm sgn}(t_5)A_2+B_2\right)\right]
\label{bigstuff}
\end{eqnarray}

After evaluating the integrand we need to perform the integrals. Roughly speaking we must integrate all propagator lengths in the integrand 
above from $-\frac{\beta}{2}$ to $\frac{\beta}{2}$. However we need to do this subject to the constraint that as we go round either of the 
two circles in the diagram Fig.(\ref{subleading_diagram}) we come back to the same time as we started out, modulo $\beta$. This is where the parameters 
$\epsilon_1$ and $\epsilon_2$ in \eqref{bigstuff}  come in. $\epsilon_1$ couples to the sum of lengths of propagators in units of $\beta$ 
around the big circle in Fig.(\ref{subleading_diagram}), while $\epsilon_2$ multiplies the sum of the lengths of all the propagators as we go around 
the small circle - again in units of $\beta$ in Fig.(\ref{subleading_diagram}). The constraint that these lengths evaluate to an integral multiple of $\beta$ can then be implemented by setting 
$\epsilon_{1,2}=2\pi \omega_{1,2}$ and then summing $\omega_i$ over all integral values, as we have done in \eqref{bigstuff}. 

In order to proceed we perform the time integrals in an unconstrained manner. The result can be rearranged (according to its $\omega_i$ 
dependence) as a sum of four types of terms. 
\begin{enumerate}
\item[1.]Terms containing $e^{k(z_1+z_2)\beta}$ where $k\in {\mathbb{Z}}$
\item[2.]those with $e^{kz_1\beta/2}$ where $k\in {\mathbb{Z}_{\rm odd}}$
\item[3.]with $e^{kz_2\beta/2}$ where $k\in {\mathbb{Z}_{\rm odd}}$
\item[4.]and $e^{k(z_1+z_2)\beta/2}$ where $k\in {\mathbb{Z}_{\rm odd}}$;
\end{enumerate}
where $z_i = -\frac{2 \pi i}{\beta}\omega_i.$

We deal with these four classes of terms spearately; for each class we 
explicitly perform the sum over $\omega_i$ (by reducing it to a contour integral as in the previous subsection) and 
expand the resultant expression in a Taylor series in $x$ (again as in the previous subsection), keep only the terms that are linear in $x$. 
Combining together the results from each of the four classes we obtain our final result 
\newpage
\begin{eqnarray} \label{poi}
I^{(2n-2)}\!\!&=&\!\!-\left(\frac{J^2\beta}{mq}\right)^n\sum\limits_{k=0}^{n-4}\frac{x(q-1)}{(mq\beta)^{k+1}}\frac{2^{(q-1)n}(2^n+(n-1)2^{3+k})(2n+k-2)\Gamma(n+k-1)}{\Gamma(n-k-1)\Gamma(n)\Gamma(1+k)}\prod_{m=1}^{q-1}\rho^1_m\cr
&&\hspace{11cm}+{\mathcal{O}(\beta^2)}
\end{eqnarray}
(the terms ${\mathcal{O}(\beta^2)}$ that we have omitted to list in \eqref{poi} are the terms with $k=n-3$ and $k=n-2$ which exist in the 
final answer but the values of whose coefficients do not follow the uniform rule of the other terms).

Note that \eqref{poi} scales like $\frac{1}{\beta}$ in coordinated large $\beta$ small $J$ limit in which $J^2 \beta$ is held fixed.

\section{The holonomy effective action from the sigma model } \label{sec5}

In this section we ask the following question: what is the contribution to $S_{\text{eff}}(U)$ - the effective action for holonomies - 
resulting from integrating out the new light degrees of freedom discovered in the massless tensor model in early sections in this paper?
In the bulk of this section we address this question at the technical level. At the end of the section we turn to a quick discussion 
of its physical import. 

Turning on holonomy is equivalent to putting appropriate boundary condition on fermion fields. This translates into boundary condition on $V_l$, given by $V_l(-\frac{\beta}{2})=UV_l(+\frac{\beta}{2})$, $U \in O(N)$. This boundary condition is equivalent to the computation of the 
partition function 
\begin{equation}\label{tpfo}
Z= e^{-S_{\text{eff}}(U)}= \Tr e^{-\beta H} {\hat U},
\end{equation}
where $H$ is the Hamiltonian of the quantum mechanical system \eqref{effectiveaction} and 
$U$ is the quantum mechanical operator that implements left rotations on the sigma model by
the $O(N)^{q-1}$ group rotation $U$. 
The partition function \eqref{tpf} is the product of $q-1$ factors, associated with the sigma models 
on the $q-1$ gauge groups. It follows that the effective action $S_{\text{eff}}(U)$ that follows from this computation 
takes the form 
\begin{equation}\label{seffu}
S_{\text{eff}}(U) = \sum_i  S(U_i).
\end{equation}
In the rest of this section we compute the functions $S(U_i)$

Let us first note that the Hilbert $H$ space on which  any one of the factors of $q-1$ distinct factors the sigma model 
\eqref{effectiveaction} acts is given as follows. The Hamiltonian acts on the Hilbert space $H$
\begin{equation}\label{hspace}
H= \sum_{R_i} {\tilde R_i} \otimes {\tilde R_i}.
\end{equation}
The sum $R_i$ runs over all genuine (as opposed to spinorial) representations of $O(N)$. 
${\tilde R_i}$ denotes the vector space on which $O(N)$ acts in the $i^{th}$ representation. 
The space ${\tilde R_i} \otimes {\tilde R_i}$ transforms in the representation $R_i \times R_i$ 
under $O(N)_L \times O(N)_R$; the operator ${\hat U}$ acts as an $O(N)$ rotation on the first 
${\tilde R_i}$ but as identity on the second ${\tilde R_i}$. The Hamiltonian corresponding to  action 
\eqref{effectiveaction} is diagonal under the decomposition \eqref{hspace}; the energy of the 
$i^{th}$ factor of the Hilbert space is $\frac{J  C_2(R_i)}{2 {\mathcal A} 
N^{q-2} }$. 

Representations of $O(N)$ are conveniently labeled by the highest weights $(h_1, h_2, h_3 \ldots )$, 
the charges under rotations in mutually orthogonal two planes. 
Let $h = \sum_i h_i$. At leading order in the large $N$ limit the
dimensionality of the representation $R_i$ depends only on $h$  and is given by 
$$ d(R_i)= \frac{N^h}{h!}.$$
Moreover the Casimir $C_2(R_i)$ of 
representations of $O(N)$ also depends only on $h$ at leading order in the large $N$ limit and is given by 
$$ C_2(R_i)= N h.$$

Let $\chi_{R_i}(U)$ denote the character in the $R_i$ representation of $O(N)$ and let 
\begin{equation}\label{pro}
\chi_n(U)= \sum_{R_i \in {\hat n}} \chi_{R_i}(U), 
\end{equation}
where ${\hat n}$ denotes the collection  of all representations of $O(N)$  with $h=n$. 
In other words $\chi_n(U)$ is the sum over the characters of all representations with $h=n$. 
 
Note that all representations with $h =n$ can be constructed - and can be constructed 
exactly once - from the direct products of $n$ vectors of $O(N)$ (this is true when $N \gg n$ as we assume). 
\footnote{Note, however, that not every representation of $n$ vectors has $h=n$; the product space 
includes representations (formed by contracting 2 vector indices) with $h=n-2$, and representations 
(formed by contracting 4 vector indices) with $h=n-4$ \ldots.}
Let $P_n$ denote the projector onto representations with 
$h=n$
\begin{equation}\label{defpn} 
P_n \left[ f(U)) \right] = \int dU' \sum_{R_i \in {\hat n} }  \chi_{R_i}(U) \chi^*_{R_i}(U') f(U').
\end{equation}
It follows that 
\begin{equation} \label{chn}
\chi_n(U) = P_n \left[ (\Tr U)^n \right].
\end{equation}
where $U$ on the RHS of \eqref{chn} represents the group element in the vector representation of $O(N)$. 

Finally we define
\begin{equation}\label{zdef}
z= e^{- \frac{J}{2 A N^{q-3}} }.
\end{equation} 

It follows immediately from all the facts and definitions presented above that 
\begin{equation}\label{fsu}
e^{-S(U_i)} = \sum_{n=0}^\infty  \frac{(z N)^n}{n!} \chi_n(U_i).
\end{equation}
Using \eqref{chn}, \eqref{fsu} can be rewritten in the (perhaps deceptively) elegant form
\begin{equation} \label{decelf}
e^{-S(U_i)} = P_{z \partial_z} e^{ N z \Tr (U_i) }.
\end{equation}
Note that
\begin{equation}\label{proj}
\int dU e^{-S(U)} =1.
\end{equation}
This is an immediate consequence of the fact that the vacuum is the only representation in the spectrum of the 
group sigma model that is a singlet under $O(N)_L$. It follows that the partition function generated by 
$S(U)$ by itself is trivial. However $S(U)$ is only one piece of the effective action for $U$ in the massless 
tensor model \eqref{sykqm}; we get other contributions to the effective action by integrating out the fermionic 
fields themselves (as was explicitly done earlier in this paper for the case of massive fermions). When 
put together with other contributions the effective action \eqref{decelf} could have a significant impact on 
the partition function, especially at temperatures scaled to ensure that the matter contribution to the
effective action  - like the contribution of the sigma model considered in this section - is of order $N^2$.

\bibliographystyle{JHEP}
\bibliography{tri-fun.bib}
\end{document}